\documentclass{emulateapj}
\usepackage{color}	
\usepackage{multirow}
\newcommand{\HI}{\ion{H}{1}~}
\newcommand{\HInospace}{\ion{H}{1}}
\newcommand{\kms}{km~s$^{-1}$ }
\newcommand{\Lya}{Lyman~$\alpha$~}

\newcommand{\Lyanospace}{Lyman~$\alpha$}

\usepackage{lscape} 

\slugcomment{March 5, 2013}

\shorttitle{Impact of Starbursts on the CGM}
\shortauthors{Borthakur et al.}

\begin{document}

\title{The Impact of Starbursts on the Circumgalactic Medium}

\author{Sanchayeeta Borthakur, Timothy Heckman, David Strickland}
\affil{Department of Physics \& Astronomy, Johns Hopkins University, Baltimore, MD, 21218, USA}
\email{sanch@pha.jhu.edu}

\author{Vivienne Wild}
\affil{School of Physics \& Astronomy, University of St. Andrews, St. Andrews, Fife, KY16 95S, UK; The Scottish Universities Physics Alliance (SUPA)}

\author{David Schiminovich}
\affil{Department of Astronomy, Columbia University, New York, NY 10027, USA}

\begin{abstract}
We present a study exploring the impact of a starburst on the properties of the surrounding circum-galactic medium(CGM): gas located beyond the galaxy's stellar body and extending out to the virial radius($\sim$200 kpc). We obtained ultraviolet spectroscopic data from the Cosmic Origin Spectrograph(COS) probing the CGM of 20 low-redshift foreground galaxies using background QSOs. 
Our sample consists of starburst and control galaxies. The latter comprises normal star-forming and passive galaxies with similar stellar masses and impact parameters as the starbursts. We used optical spectra from the Sloan Digital Sky Survey(SDSS) to estimate the properties of the starbursts, inferring average ages of $\sim$200 Myrs and burst fractions involving $\sim$10\% of their stellar mass. The COS data reveal highly ionized gas traced by \ion{C}{4} in 80\%(4/5) of the starburst and in 17\%(2/12) of the control sample. 
The two control galaxies with \ion{C}{4} absorbers differed from the four starbursts in showing multiple low-ionization transitions and strong saturated Lyman-alpha lines. 
They therefore appear to be physically different systems. We show that the \ion{C}{4} absorbers in the starburst CGM represent a significant baryon repository. The high detection rate of this highly ionized material in the starbursts suggests that starburst-driven winds can affect the CGM out to radii as large as 200kpc. This is plausible given the inferred properties of the starbursts and the known properties of starburst-driven winds. This would represent the first direct observational evidence of local starbursts impacting the bulk of their gaseous halos, and as such provides new evidence of the importance of this kind of feedback in the evolution of galaxies.
\end{abstract}


\section{INTRODUCTION\label{intro}}

The evolution of galaxies is regulated by how and when they accrete gas, produce stars and supermassive black holes, and by how different feedback processes may regulate this cycle. Similarly, the evolution of the intergalactic medium (IGM) will be regulated by these same feedback processes which can photoionize, collisionally heat, and chemically enrich the IGM. The interface between the region where star formation has taken place and the IGM is the galaxy halo, or circum-galactic medium (CGM). We take this to be the region extending from the main stellar body of the galaxy outward to the virial radius at a few hundred kpc. It is through the CGM that the two-way communication between galaxies and the IGM occurs.

One of the major ways in which feedback occurs is via outflows of gas driven from strongly star-forming galaxies by the energy and/or momentum injected by massive stars \citep[e.g.][and references therein]{veilleux05}. These outflows are ubiquitous in high redshift galaxies \citep{shapley03, adelberger05,weiner09}. At low redshifts the conditions needed to drive these large-scale galactic winds seem to be present only in starburst galaxies \citep[][and references therein]{heckman90,lehnert96, veilleux05,martin05, rupke05b}. However, while less common, these low redshift outflows can be studied in far greater detail than their high-z counterparts. 

While it is clear that intensely star-forming galaxies at all redshifts drive outflows, the effect of these outflows on the evolution of galaxies and the IGM has not yet been well-quantified. Are they sufficiently strong to propagate out into the CGM and IGM? In so-doing can they eject baryons from low-mass dark matter halos (dwarf galaxies)? Can they shut down the delivery of gas through the CGM in more massive systems, leading to the cessation of star-formation and the migration of galaxies from the blue star-forming sequence onto the red quiescent sequence? What role do they play in establishing the galactic characteristic stellar mass of $\sim10^{10.5}$ M$_{\odot}$ which separates the low-mass mainly-star-forming and high-mass mainly-quiescent galaxy populations?

Unfortunately, the investigations of outflows in local starbursts cannot directly address these questions because they have been limited to regions in the inner most part of the CGM (corresponding to radii of a few tens-of-kpc or less). Moreover, the outflow speeds that are directly measured in the molecular, atomic, and warm ionized gas are typically comparable to or less than the galaxy escape velocity, so that it is not clear whether this material is truly a wind or a fountain \citep[e.g.][]{chen10,grimes09}. The very high specific energies implied by the high temperatures of the X-ray emitting gas suggest that this gas is likely to escape at least from the less massive starbursting galaxies \citep{grimes05}, but again these measurements pertain to locations far inside the virial radius.

 \begin{figure*}
 \hspace{2cm}
\includegraphics[trim = 0mm 120mm 0mm 20mm, clip,scale=.6,angle=-0]{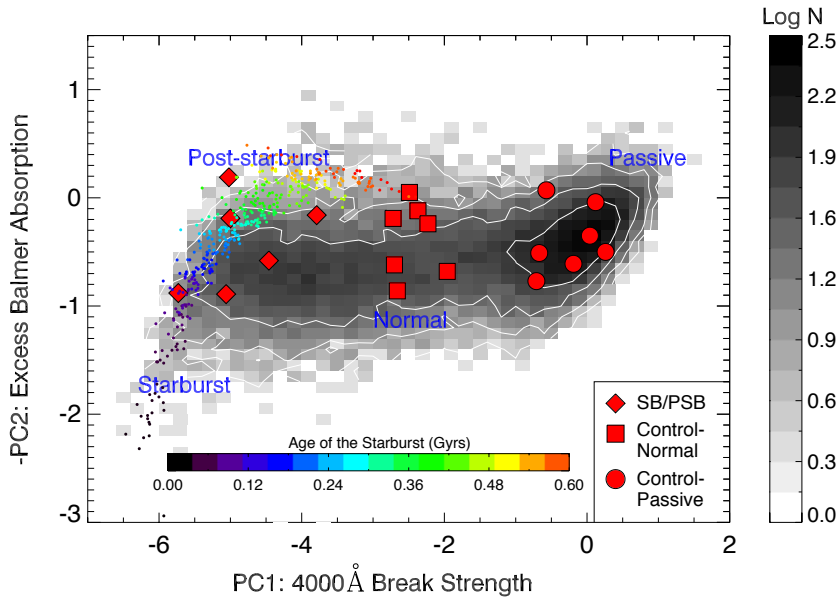}
\caption  {Figure shows the first and the second principle components, PC1 and PC2 respectively, from the Principal Component Analysis (PCA) study of SDSS spectra (rest-frame 3750-4150~$\rm\AA$) by \citet{wild07}.  
Contours show the distribution of SDSS galaxies in the PC1 versus PC2 diagram and the greyscale shows their numbers in logarithmic units. 
The first principal component (PC1) provides the same information as the traditional D$_n$(4000) index whereas the second principal component (PC2) provides additional information on any excess (or deficit) of Balmer absorption, over that expected for the 4000~\AA\ break strength of the galaxy. 
Therefore, a combination of the first two components can identify young stellar populations and hence starburst and post-starburst galaxies. 
The trajectory of starbursts from Figure~1 of \citet{wild10} is shown as colored dots indicative of their ages. The regions of the plot associated with various classes of galaxies with different star-formation histories are labeled in violet on the diagram. The galaxies from our sample are over-plotted as red symbols with their shapes representing their classification- diamonds for starburst and post-starburst, squares and circles for the control sample consisting of normal star-forming galaxies and passive galaxies respectively.}
\label{fig-sample_pc}  
\end{figure*}

Direct imaging of the bulk of the CGM to probe the effects of winds is difficult at high-z \citep[but see][]{steidel11}, and is impossible with existing facilities at low-redshift. The best available probe of the properties of the CGM comes from QSO absorption-line systems (QALS), which can detect this tenuous material with great sensitivity (albeit along single sight-lines). 
Various authors have suggested the connection between QALS and starburst driven winds. \citet{zibetti07} argued that models of metal-enriched outflows from star-forming/bursting galaxies could explain the origin of strong \ion {Mg} {2} absorbers. 
\citet{bouche07} found that 2/3 of their strong \ion{Mg}{2} absorbers (with rest-frame equivalent width W$\rm _{\lambda 2796} > 2 \AA$) at $z \sim 1$ had starburst galaxies (with SFR=1-20~$\rm M_{\odot}~yr^{-1}$) around them.
This result was also later confirmed by  \citet{nestor11} with the discovery of bright, starburst/post starburst galaxies in the field around 2 ``ultrastrong" \ion{Mg}{2} QALs.
Another indication of this connection comes from the study by \citet{menard11}, where they found a strong correlation between the equivalent width of the \ion{Mg}{2} absorbers and the [OII] emission-line luminosity of the host galaxy. In a slightly different experiment, \citet{steidel10} have used more distant galaxies located behind a sample of intensely star-forming galaxies at $z \sim$ 2 to 3 to trace outflowing gas in absorption to radii of $\sim$100 kpc.

With the installation of the Cosmic Origins Spectrograph (COS; \citet{green12}) on the Hubble Space Telescope (HST) it has now become feasible to probe the CGM of low-redshift galaxies via QALS. Most of these transitions lie in the ultraviolet and the superb sensitivity of COS allows us to directly detect and characterize the CGM in a representative samples of low-redshift galaxies that can be studied in detail \citep[e.g.][]{tumlinson11b}.
 In this paper we utilize COS to directly probe the CGM of starburst galaxies and to learn the effect of starburst-driven winds on the circumgalactic gas. To that end, we have selected a small sample of galaxies from the SDSS that have undergone a significant starburst event within the last few $\times 10^8$ year, and we have also selected an appropriate control sample of typical star-forming and quiescent galaxies.  For all these galaxies we have obtained ultraviolet (UV) spectroscopic data with the COS, which can directly detect gas in the CGM via absorption lines seen in the spectra of background UV-bright QSOs

Detailed descriptions of our sample selection criterion, the COS observations, and data reduction are presented in Section~2.  The results and their implications are discussed in Section~3 and 4. And finally, we summarize our findings and conclude in Section~5. The cosmological parameters used in this study are $H_0 =70~{\rm km~s}^{-1}~{\rm Mpc}^{-1}$, $\Omega_m = 0.3$, and $\Omega_{\Lambda} = 0.7$. 
Atomic data used throughout our analysis was obtained from the published work by \citet{morton03}.

\section{OBSERVATIONS  \label{sec:observations}}

\subsection{Sample \label{sec:sample}}

\begin{figure*}
\hspace{1cm}
\includegraphics[trim = 15mm 60mm 15mm 60mm, clip,scale=1,angle=-90]{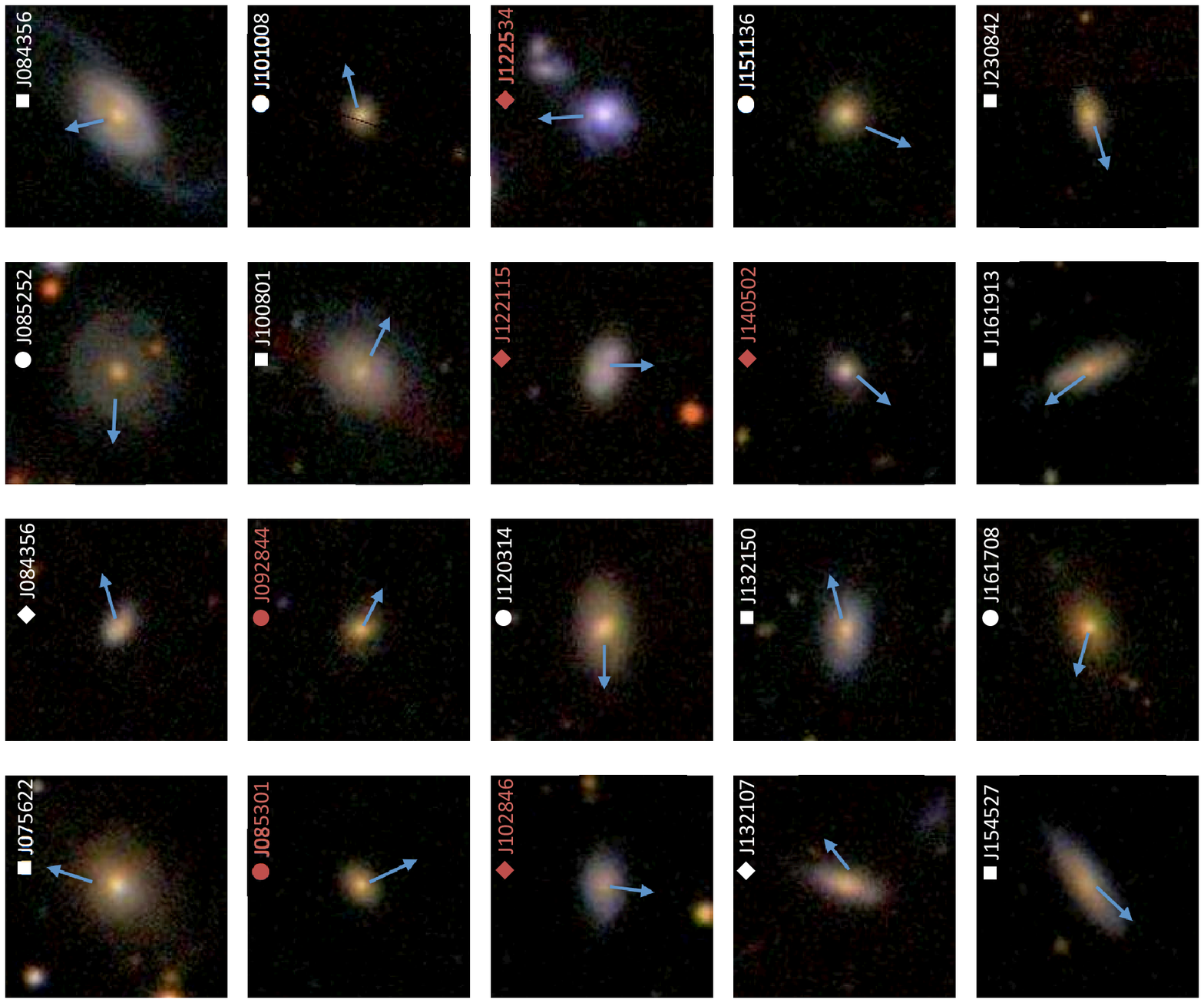} 
{\caption{SDSS composite image of $\rm 37.5^{\prime\prime}\times37.5^{\prime\prime}$ for our sample of 20 foreground galaxies with stellar mass, M$_*$ between $ \rm 10^{10.1} -  10^{11.4}M_{\odot}$. At the median redshift of our sample of $z=0.09$, the scale of the image corresponds to 63~kpc. The galaxies are identified by their name on the top right corner and the symbol represents their subclass: diamond for SB/PSB, square for normal star forming, and circle for passive galaxies. The red label indicate galaxies where  \ion{C}{4} was detected. The blue arrow marks the direction of the background QSO. Details of our sample are presented in Table~\ref{tbl-sample}. }\label{fig-sample_sdss}}
\end{figure*}

Our sample was selected in a two-phase process. The first step was to identify those galaxies in the SDSS Data Release 7 (DR7) ``Main” galaxy sample \citep{strauss02} that are under-going or have recently undergone a strong starburst. To do so we followed the procedure described in \citet{wild07}. They used a Principal Component Analysis (PCA) approach to improve spectroscopic diagnostics for bursts of star formation for a bulge dominated sample. 
In this analysis, SDSS spectra from rest-frame 3750-4150$\rm \AA$ were decomposed in multiple eigenspectra and the resulting principal component amplitudes represent the amount of each eigenspectrum present in the galaxy spectrum. 
The first principal component (PC1) provides the same information as the traditional D$_n$(4000) index \citep{balogh99}. It is a good proxy for the specific star formation rate of the galaxy, averaged over the past few Gyr. The second principal component (PC2) provides additional information on any excess (or deficit) of Balmer absorption, over that expected for the 4000~\AA\ break strength of the galaxy. This is useful in identifying galaxies with young stellar populations  of ages up to 1~Gyr \citep[see also][]{wild10}\footnote{PCA catalogs for SDSS DR7 are available for download at \url{http://www-star.st-and.ac.uk/$\sim$vw8/downloads/DR7PCA.html} }.

 \begin{figure*}
\begin{tabular}{l l l l }
\includegraphics[trim = 00mm 0mm 0mm 0mm, clip, scale=.45,angle=-0]{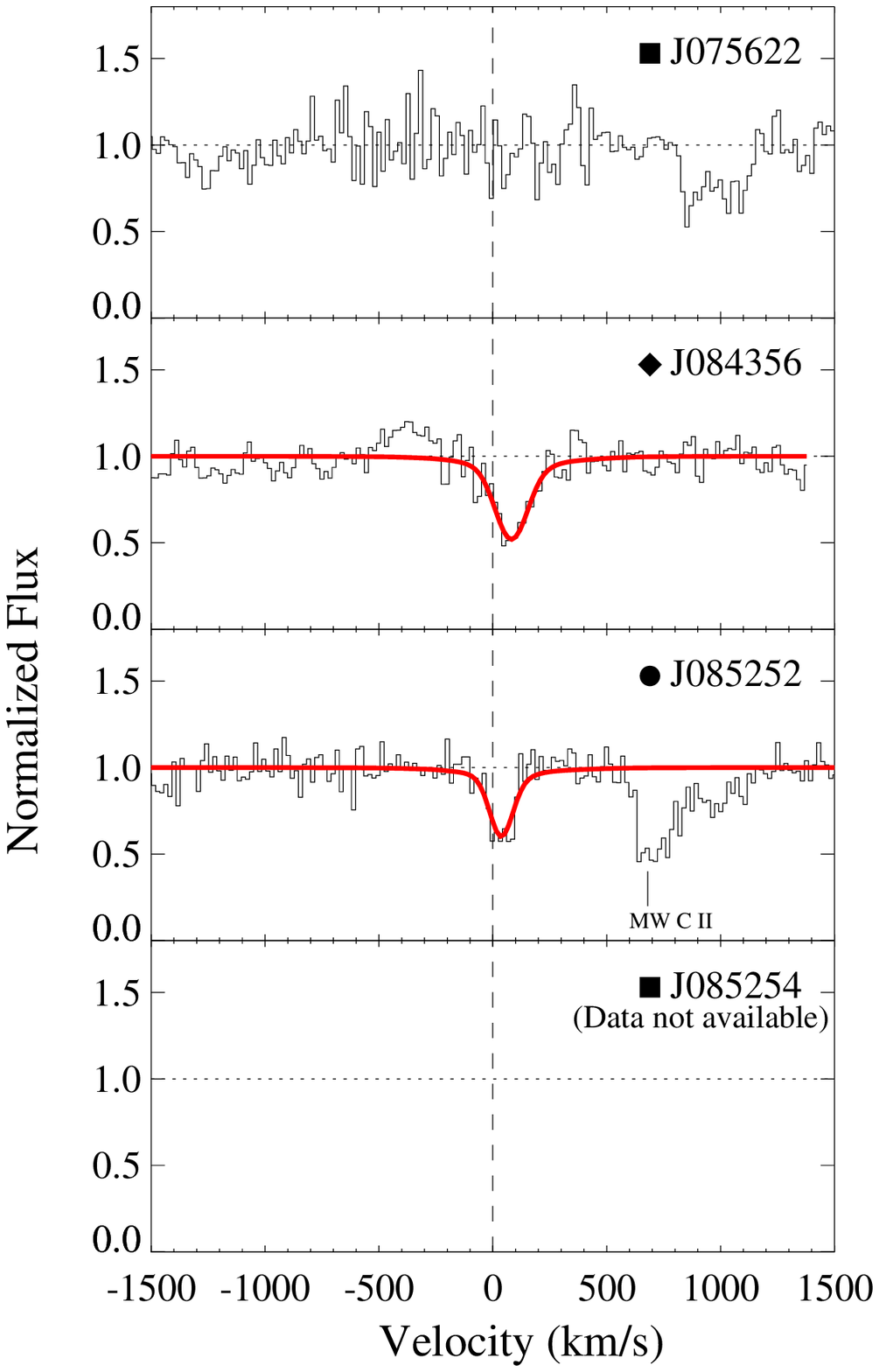} & \includegraphics[trim = 15mm 0mm 0mm 0mm, clip, scale=.45,angle=-0]{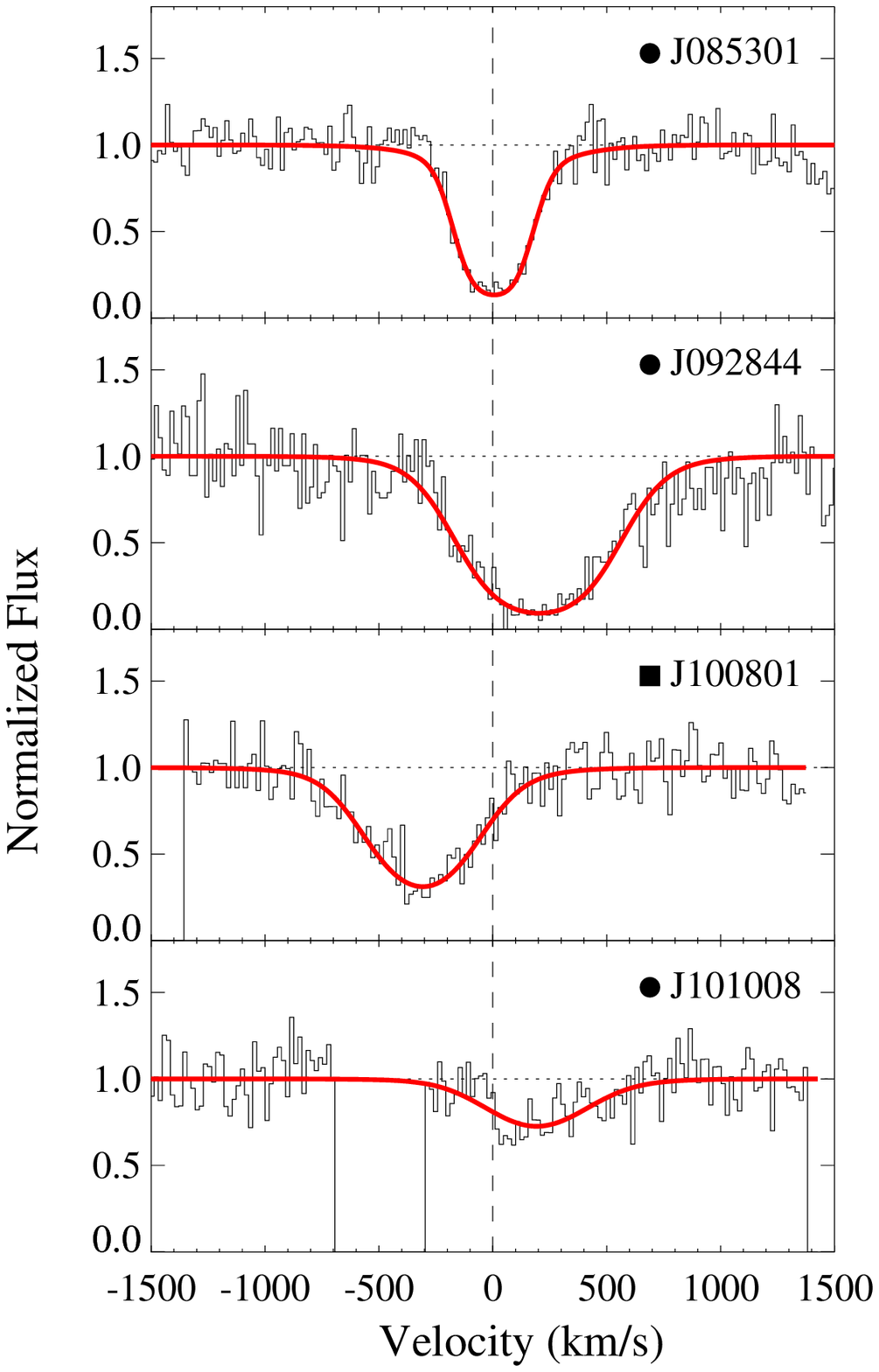} & \includegraphics[trim = 15mm 0mm 0mm 0mm, clip, scale=.45,angle=-0]{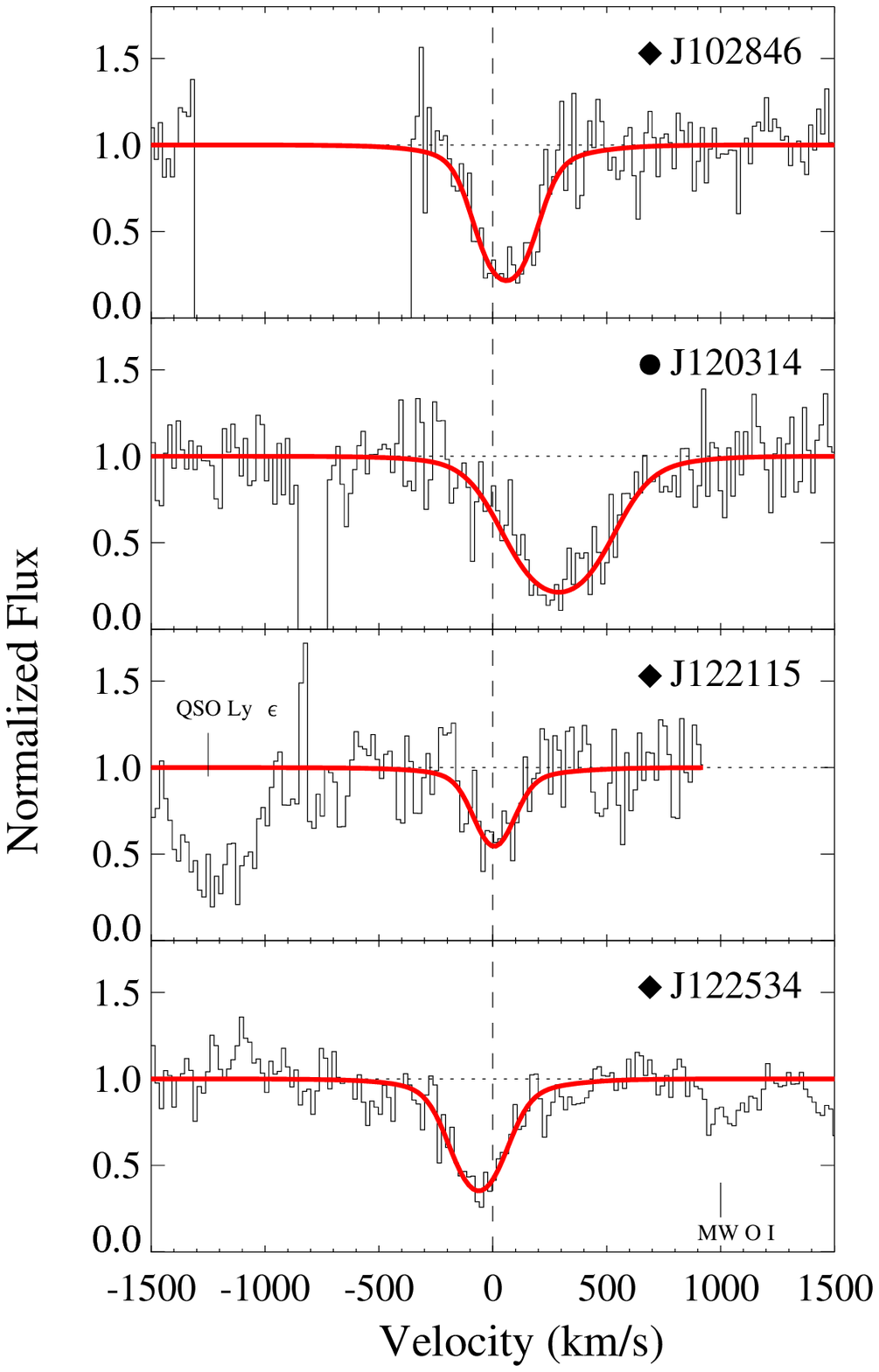} \\
\includegraphics[trim = 00mm 0mm 0mm 0mm, clip, scale=.45,angle=-0]{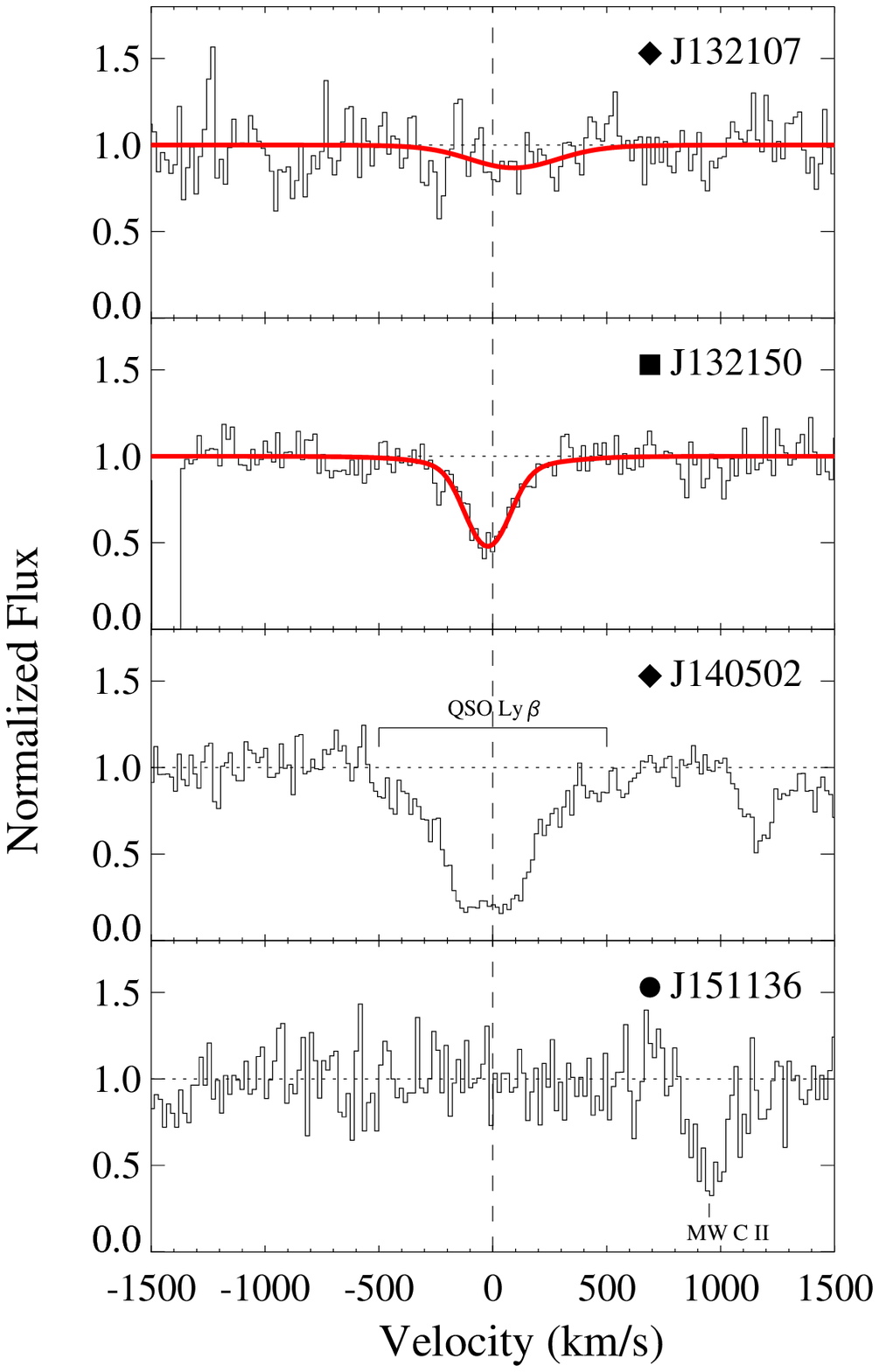} & \includegraphics[trim = 15mm 0mm 0mm 0mm,clip, scale=.45,angle=-0]{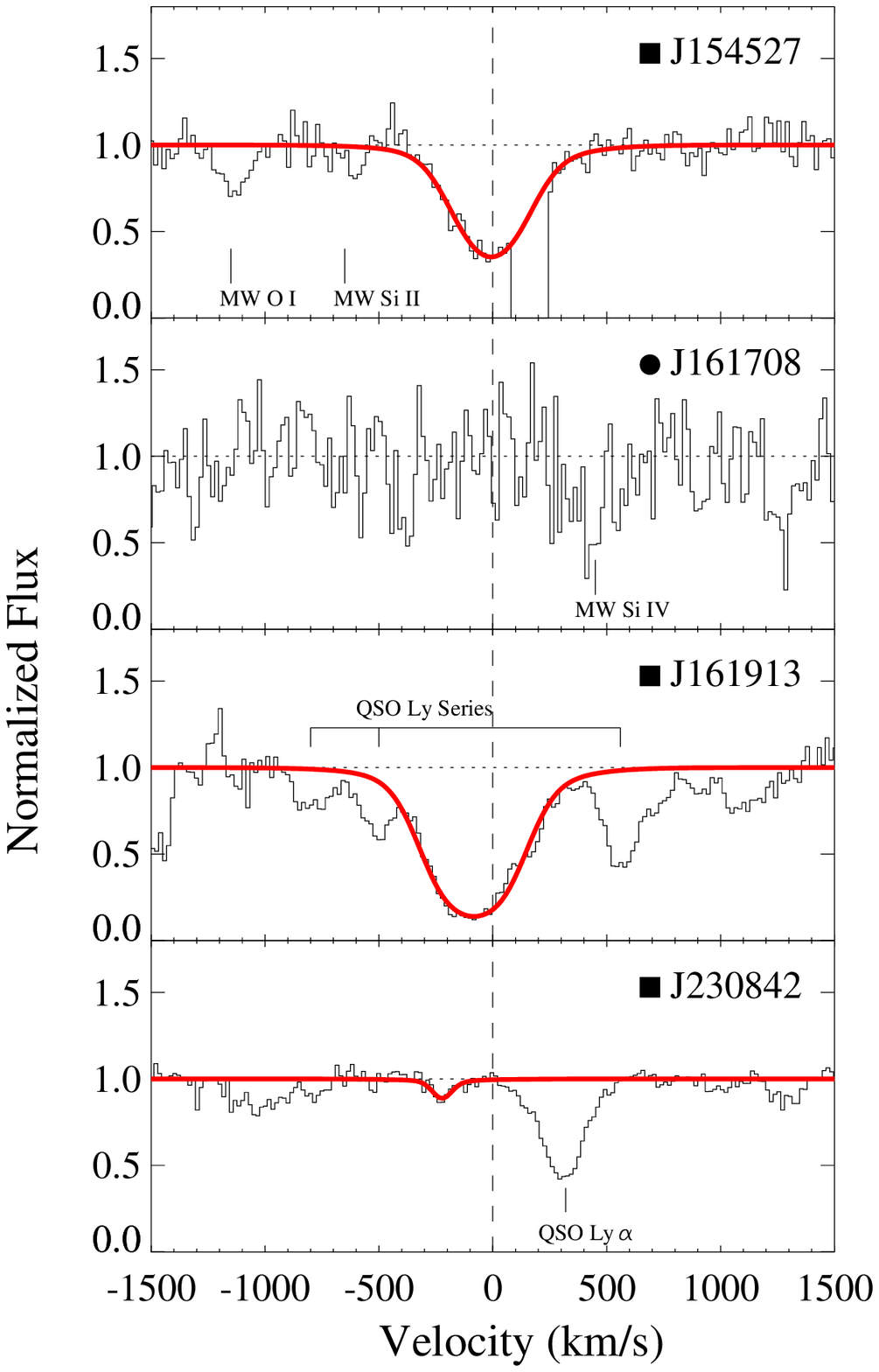}  \\
\end{tabular}
\caption{Rest frame \Lya absorbers associated with 18/20 foreground galaxies for which we have data at the observed wavelength of the \Lya . The dotted line shows the redshift of the target galaxy. The data are labeled with the target galaxy name in the upper right corner and its type is shown as a symbol next to the name: diamonds represent the 5 starburst galaxies, squares and circles represent the control sample of 6 normal star forming galaxies and 7 passive galaxies respectively. The best fit Voigt profile to the absorption feature is shown in red.}
\label{fig-lya}
\end{figure*}

 \begin{figure*}
\begin{tabular}{l l l   }
\includegraphics[trim = 00mm 0mm 0mm 0mm, clip, scale=.46,angle=-0]{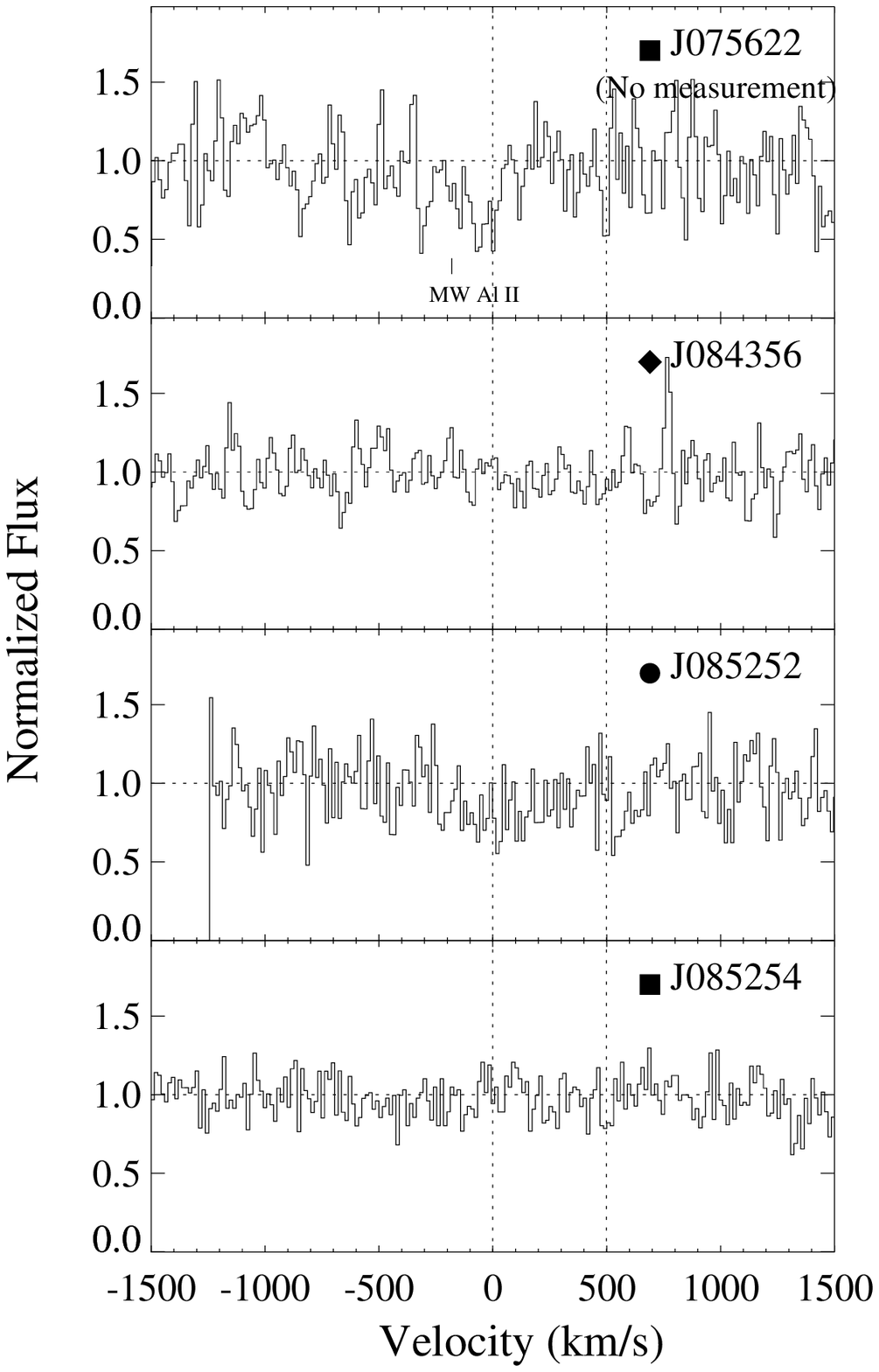} & \includegraphics[trim = 15mm 0mm 0mm 0mm, clip, scale=.46,angle=-0]{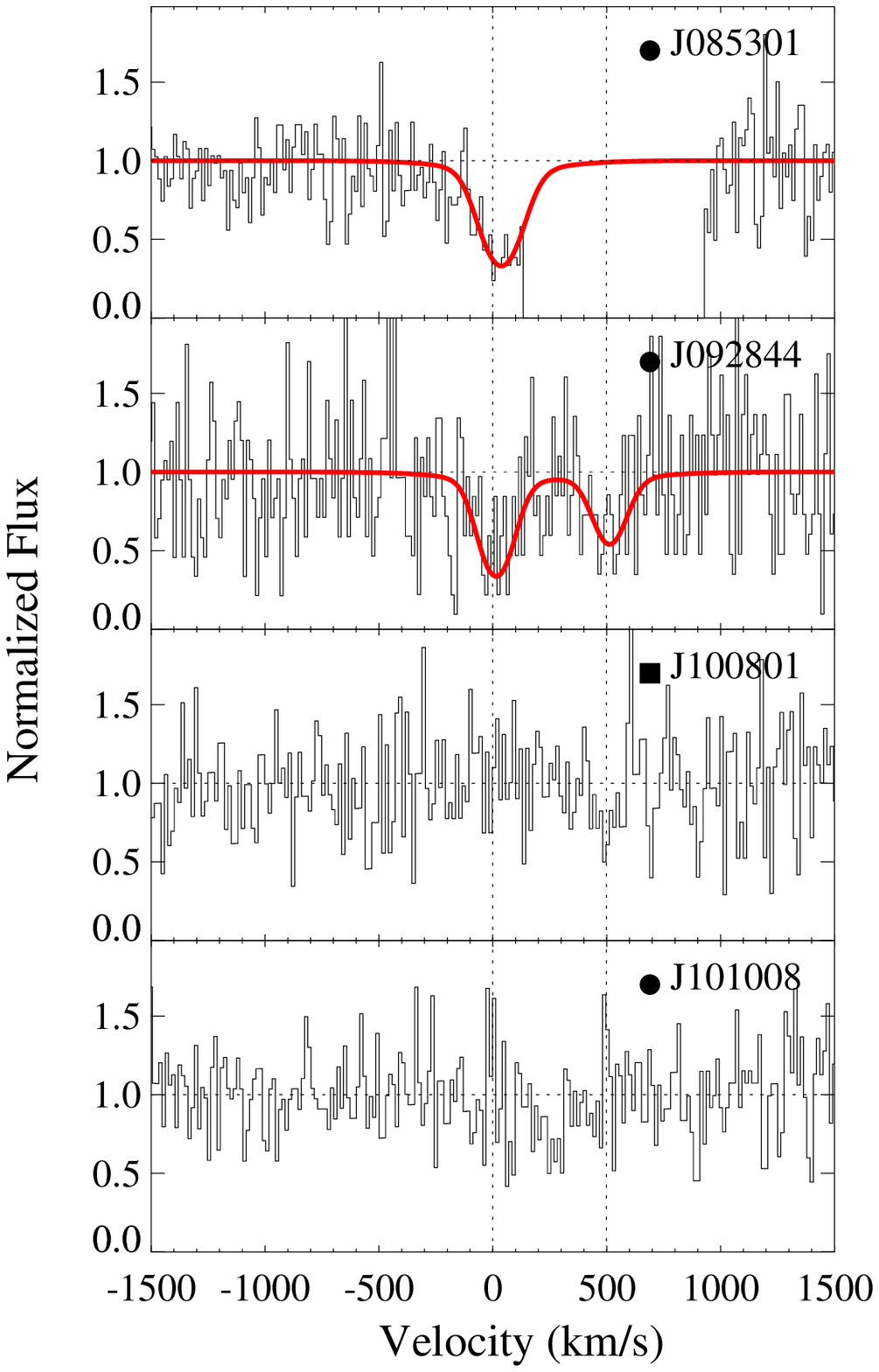} & \includegraphics[trim = 15mm 0mm 0mm 0mm, clip, scale=.46,angle=-0]{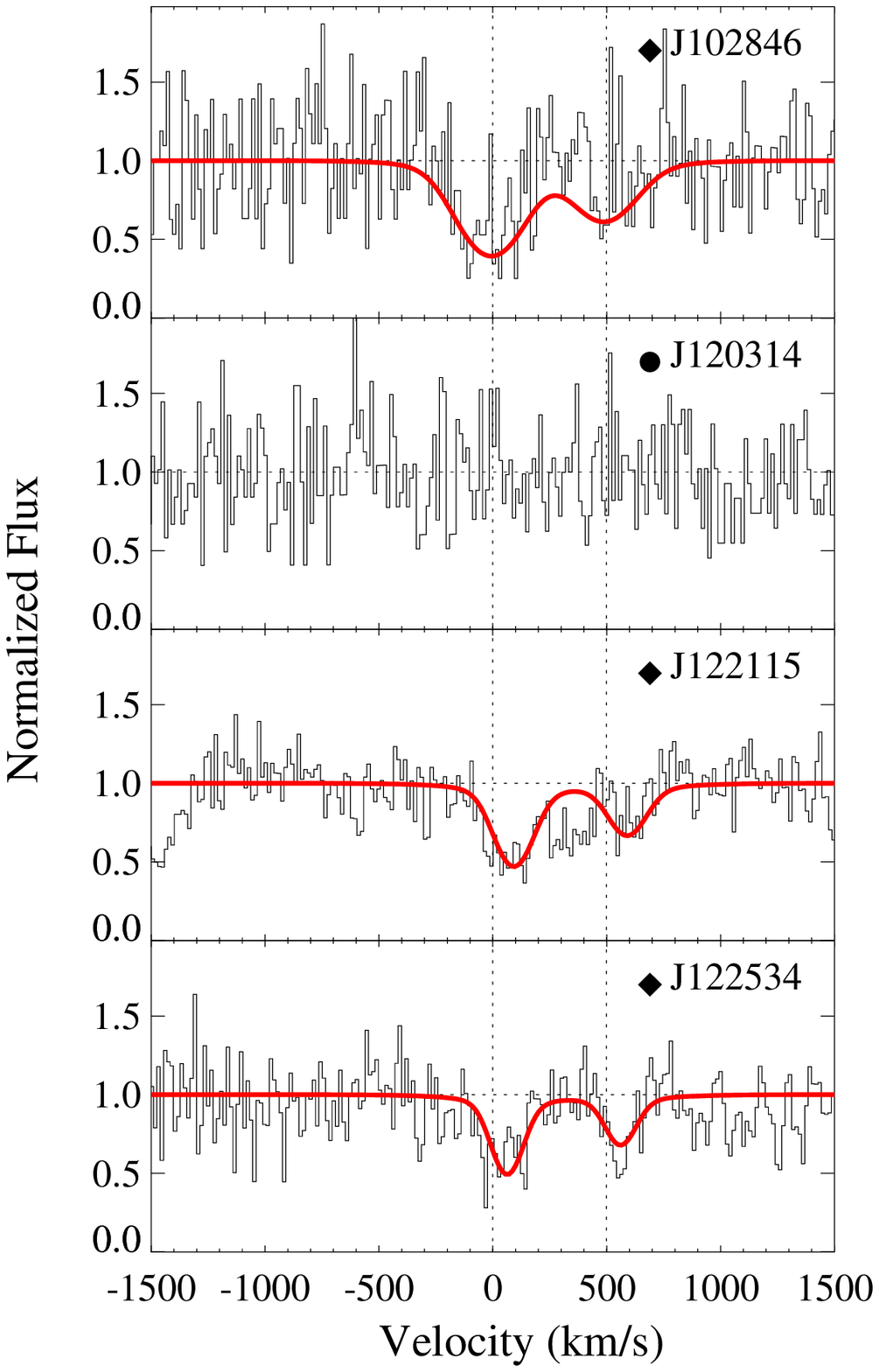} \\
\includegraphics[trim = 00mm 0mm 0mm 0mm, clip, scale=.46,angle=-0]{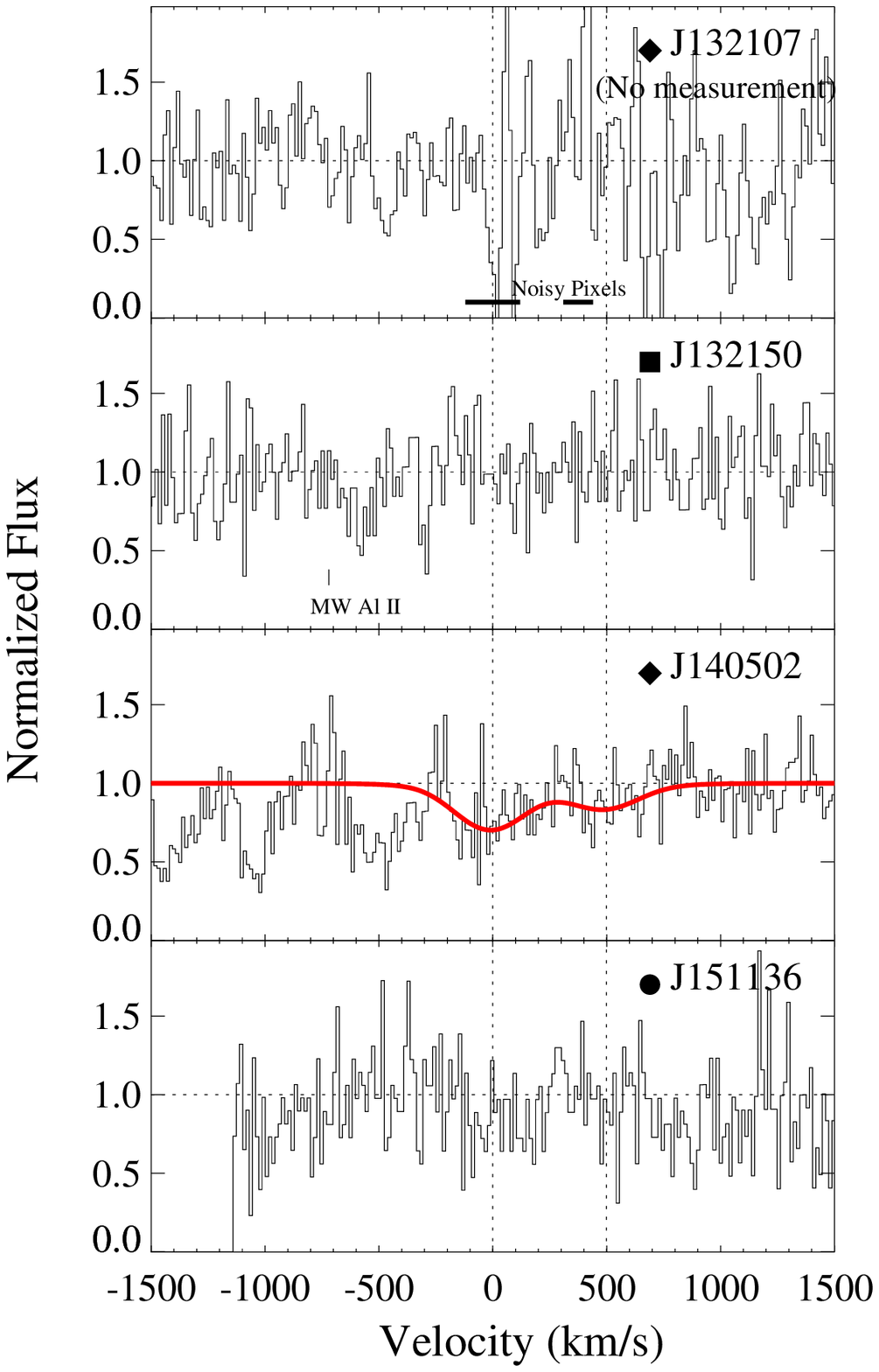} & \includegraphics[trim = 15mm 0mm 0mm 0mm, clip, scale=.46,angle=-0]{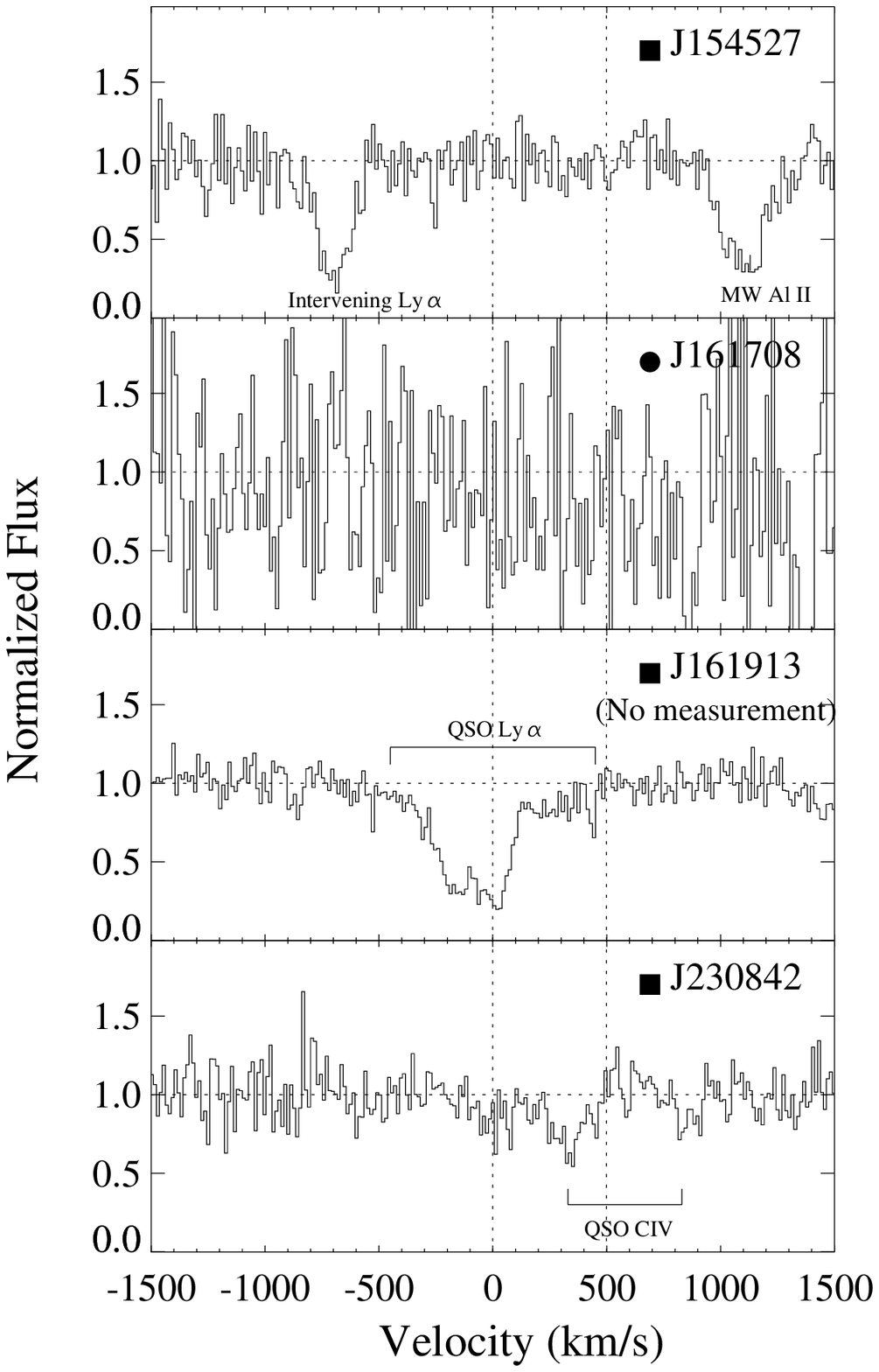}  \\
\end{tabular}
\caption{Rest frame \ion{C}{4} absorption features associated with our sample of 20 foreground galaxies. The data are labeled with the name of the foreground galaxy in the upper right corner and the galaxy-type is shown by the symbol. The fits to the \ion{C}{4} $\lambda \lambda$1548,1550~$\rm\AA$ doublet are plotted in red. The dotted lines show the position of the \ion{C}{4} doublet at the rest-frame of the target galaxy. The targets where no measurements could be made are indicated.}
\label{fig-ci4}
\end{figure*}

In Figure~\ref{fig-sample_pc} we show the distribution of the SDSS main galaxy sample in greyscale as well as contours in the PC1 vs. PC2 plane. The locations of the starburst, post-starburst, passive, and typical star-forming galaxy populations are indicated. The colored dots traces out a time sequence for strong starbursts \citep[details described by][]{wild10}. These bursts can be fit with a model having an exponentially declining star formation rate with an e-folding time of 200-350 Myr and typical burst mass fractions of 10 to 25\% of the stellar mass. To find a sample of recent/on-going starbursts we have then searched the SDSS DR7 Main galaxy sample for galaxies lying along or near this sequence.

The second step in generating our sample was cross-correlating this starburst sample with the Galaxy Evolution Explorer (GALEX) far-ultraviolet (FUV) bright QSO catalog (GR3) to find QSO-starburst pairs. We used the following criteria:  
\begin{itemize}
\item[1.] The background QSO is brighter than FUV = 19.5 mag or $\rm F_{\lambda=1300\AA}=1.02\times10^{-15}~erg~cm^{-2}~s^{-2}\AA^{-1}$;
\item[2.] Signal-to-noise ratio $>$ 8 in the SDSS g-band spectrum of the galaxy;
\item[3.] Galaxy stellar mass, M$_* > 10^{10} \rm M_{\odot}$\footnote{Stellar mass catalogs, derived from 5 band SDSS photometry by \citet{brinchmann04} following the prescription by \citet{kauffmann03} and \citet{salim07}. The catalogs were obtained from  \url{http://www.mpa-garching.mpg.de/SDSS/}.}. We are interested in galaxies in the mass range where the galaxy population transitions from predominantly low-mass star-forming to high-mass quiescent systems \citep[e.g.][]{kauffmann03};
\item[4.] Impact parameter to nearest UV bright QSO, $\rho \le 200$~kpc. This distance is approximately the virial radius, $\rm R_{vir}= \Big[\frac{GM_h}{9\pi^2H_0^2}\Big]^{	1/3}~\sim 215~kpc$ for galaxies with halo masses, $\rm M_h\sim10^{12}~M_{\odot}$ (approximate halo mass for galaxy of stellar mass $\rm10^{10.5}~M_{\odot}$);
\item[5.] The foreground galaxy does not belong to a cluster as defined by the SDSS galaxy cluster catalog \citep[see][]{miller05}. We are interested in the CGM in typical ``field" galaxies.
\end{itemize}

\begin{figure}
\begin{tabular}{l l l }
\hspace{-0.5cm}
\includegraphics[trim = 5mm 0mm 0mm 0mm, clip,scale=.36,angle=-0]{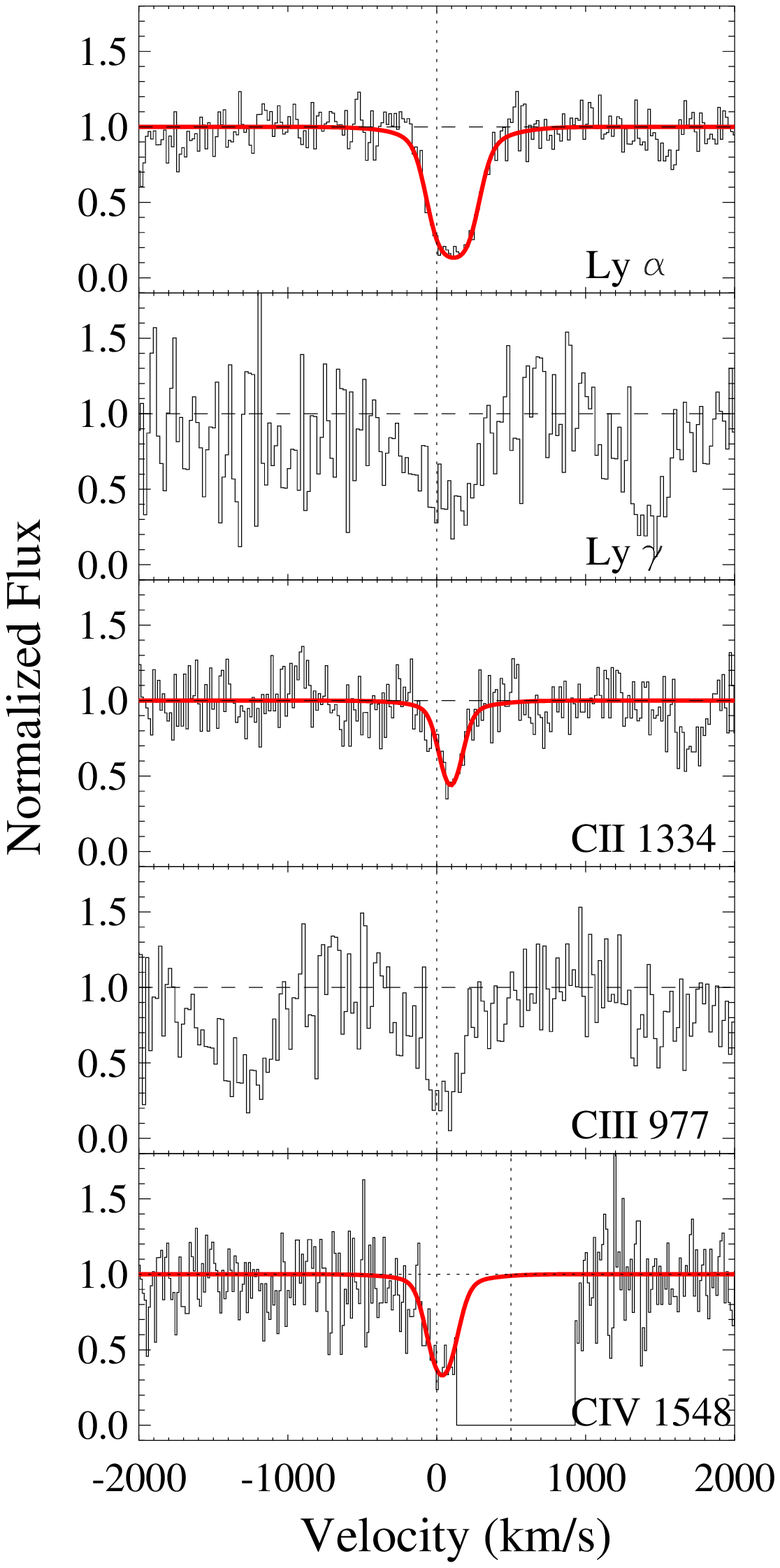} & \includegraphics[trim = 15mm 0mm 0mm 0mm,scale=.36,angle=-0]{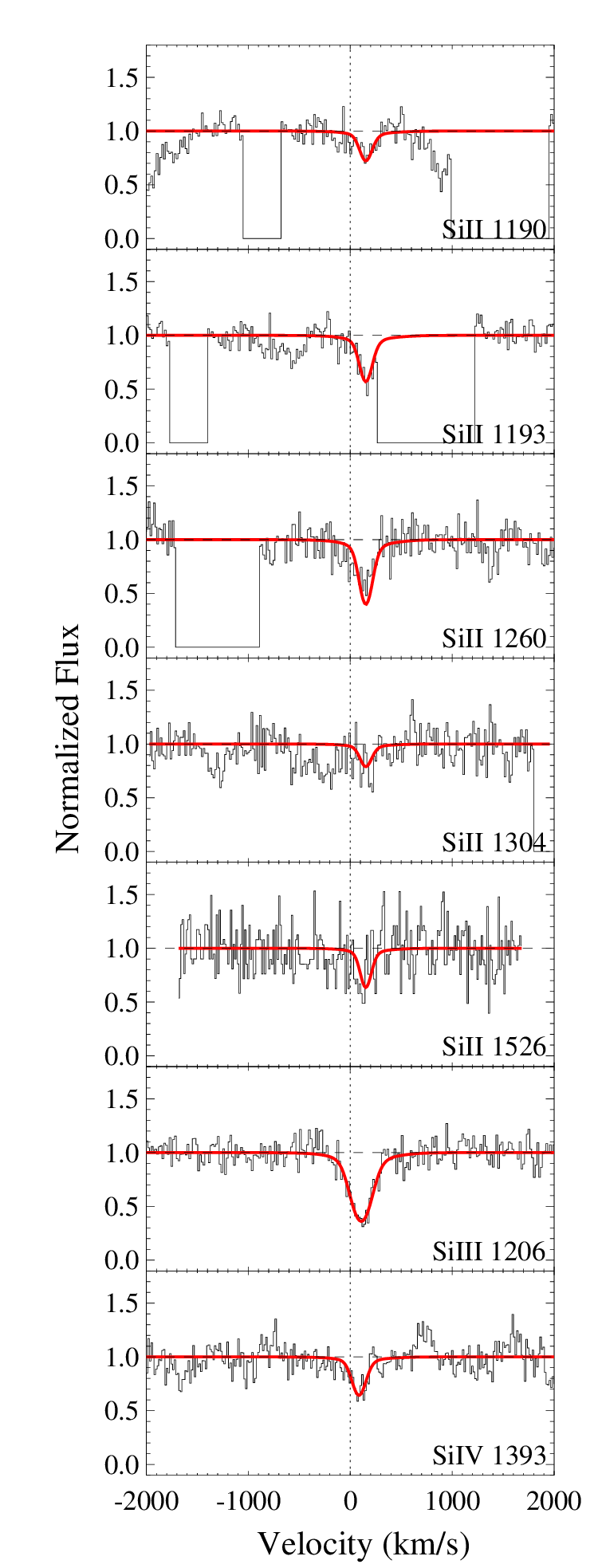}  \\
\end{tabular}
\caption{Plots showing various transitions detected in the foreground galaxy J085301. Lines include \HInospace, low-ionization metal transitions such as \ion{C}{2}, and \ion{Si}{2}, intermediate-ionization transition like \ion{C}{3} and \ion{Si}{3}, along with hot gas traced by \ion{Si}{4} and \ion{C}{4} respectively. The Voigt fits to the profile are shown in red. We do not fit \ion{C}{3} and Lyman~$\gamma$ as they lie in the segment B of G140L grating where wavelength calibration is highly uncertain.}
\label{fig-J085301}
\end{figure}

This resulted in six QSO-starburst pairs (hereafter the``starburst sample - SB/PSB"). We then repeated the above selection procedure for non-starbursts or ``control" galaxies in SDSS DR7. Since there were many eligible QSO-galaxy pairs, we selected our targets so that we sampled the rest of the PC1 vs. PC2 plane uniformly with galaxies that roughly has similar stellar masses and impact parameters as the  starburst sample. For multiple pairs with similar properties, the one with the brightest background QSO in FUV was selected. This resulted in 14 control galaxies with half of them passive (red) galaxies with PC1$>$ -1 and other half consisting of normal star-forming galaxies with PC1 ranging between -3 and -1. The starburst and control samples are overplotted in the PC1 versus PC2 plane in Figure~\ref{fig-sample_pc}, while Figure~\ref{fig-sample_sdss} shows the SDSS multicolor image of the 20 target galaxies from our sample. The target galaxy is at the center of the image and the direction of the QSO is shown with a blue arrow. The galaxy names are labeled in the top right corner and the ones where we have \ion{C}{4} detections are labeled in red. The symbols before the labels refer to the category of galaxies. We use diamonds to identify the SB/PSB sample and circles and squares for the control sample of passive and normal star-forming galaxies respectively.

The properties of the galaxies and QSOs in our sample are summarized in Table~\ref{tbl-sample}. The name, position, redshift of the foreground galaxy and the background QSOs are presented in columns 1, 2, 3, 4, 11, 12, 13, and 14 respectively. The impact parameter of the QSO sightline is presented in the last column. The PC1, PC2 and the corresponding class of the foreground galaxy are presented in columns 5-7. Column 8 presents the stellar masses obtained from the MPA-JHU catalog, These masses were estimated using methods described by \citet{brinchmann04}.
The orientations of the sightlines are presented in column 9, which represents the angle of the QSO sightline with respect to the major axis of the galaxy.  For example, any sightline along the major axis has $\rm \Theta = 0^{\circ}$, whereas a sightline probing the halo perpendicular to the disk has $\rm \Theta = 90^{\circ}$. The orientation was estimated using SDSS photometric i-band measurements. We considered all galaxies where the ratio of the major to minor axis (deVAB) $\rm >$ 0.7 to be face-on, and defined   $\rm \Theta$ as $0^{\circ}$. 

The star formation properties of the galaxies are presented in Table~\ref{tbl-sfr}. The H$\alpha$ fluxes, luminosities, and ionizing photons per second, Q$_*$ are presented in columns 2, 3, and 4 respectively. These values were derived by applying dust corrections \citep[similar to that used by][for emission lines]{wild07} to the fluxes from the MPA-JHU catalog\footnote{Line fluxes were obtained from the MPA-JHU catalog.} and then converting to L($\rm H\alpha$) \citep[see Eq.9.5 from][]{conti_crowther_leitherer08}. In this paper we wanted to evaluate the energy input to the CGM over the lifetime of the starburst (rather than using the instantaneous value). For this reason we do not measure the star formation rates for the starbursts using H$\alpha$. Instead, for the starburst sample we use the burst mass or burst mass fraction and the burst age to estimate a SFR or specific SFR (respectively) averaged over the over the duration of the burst (column 5). For the control sample we use the SFR and specific SFR from \citet{brinchmann04} (column 6). Both these estimates are for the central 3$^{\prime\prime}$ region of the galaxies which is covered by the SDSS fiber.

 \begin{figure}
\hspace{-0.5cm}
\begin{tabular}{l l l }
\includegraphics[trim = 0mm 0mm 0mm 0mm, clip,scale=.36,angle=-0]{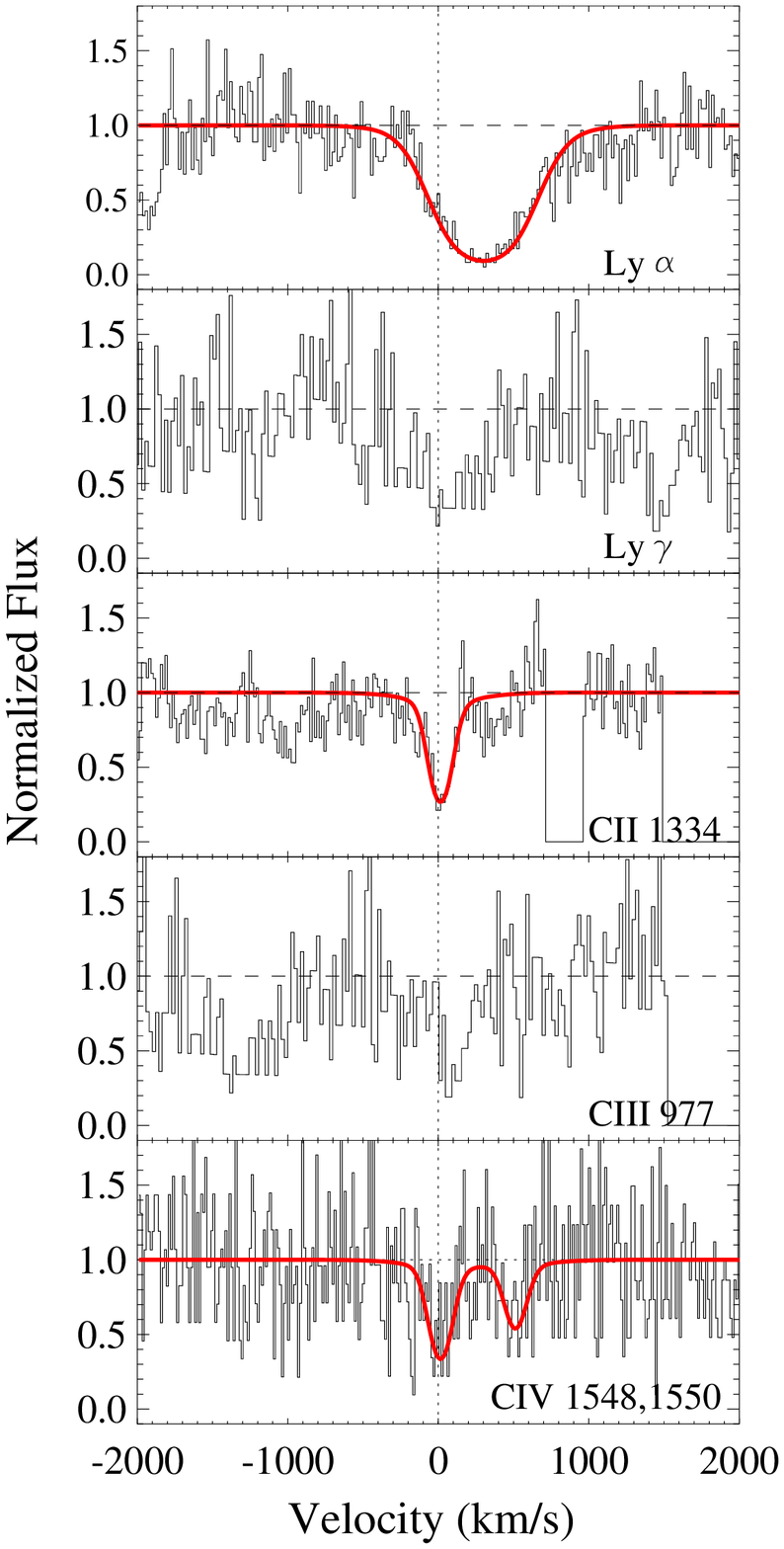} & \includegraphics[trim = 15mm 0mm 0mm 0mm,scale=.36,angle=-0]{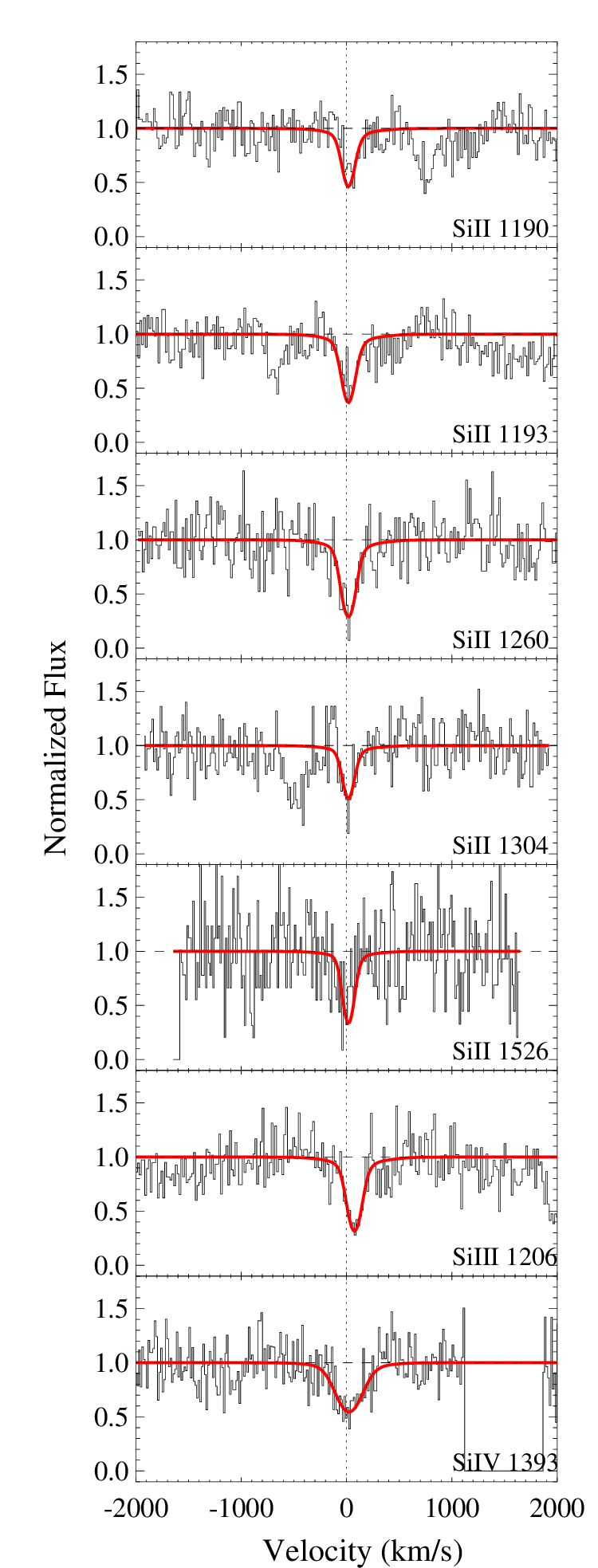}  \\
\end{tabular}
\caption{Similar plots as in Figure~\ref{fig-J085301}, but for the foreground galaxy J092844. }
\label{fig-J092844}
\end{figure}

\subsection{ Cosmic Origin Spectrograph Observations \label{sec:cos}}

The sample was observed with COS using segments B and A of the grating G140L, providing a resolution, $\rm R \sim~2000-3000~(\sigma = 53\pm 10$ km s$^{-1}$). The data were reduced using the standard COS pipeline and covered the wavelength range between rest-frame Lyman~$\alpha$ to \ion{C}{4} with the exception of one control galaxy, J085254. Part of its spectrum (up to $\lambda_{rest} < 1245 \rm~ \AA$) was missed as a result of an error in the setup of the grating. For the remaining 19 targets the data covered the entire range. The wavelength covered includes transitions of various other species such as \ion{C}{2}, \ion{Si}{2}, \ion{Si}{3}, and \ion{Si}{4}. This allows us to probe both neutral gas via low-ionization transitions like \ion{Si}{2} and \ion{C}{2} as well as highly ionized gas via \ion{C}{4}.

Although we have COS coverage for all the other sightlines, we do miss a few transitions due to intervening absorbers or hot pixels at the position of the expected transition. For example, in the case of J140502, the saturated Lyman~$\beta$ transition of the background QSO completely absorbed the QSO flux at the expected position of the \Lya line associated with the foreground galaxy. Similarly, we have three galaxies namely J075622, J161913, and J132107, where the \ion{C}{4} is blended with the Milky Way's \ion{Al}{3}, \Lya of the background QSO, and lies at the position of noisy (hot) pixels respectively. No measurements could be made in these cases. This leaves us with a total of 18 targets with useful data at the expected wavelength of \Lya and 17 targets for \ion{C}{4}.

\begin{figure*}
\hspace{1cm}
\includegraphics[trim = 0mm 130mm 0mm 20mm, clip,scale=.70,angle=-0]{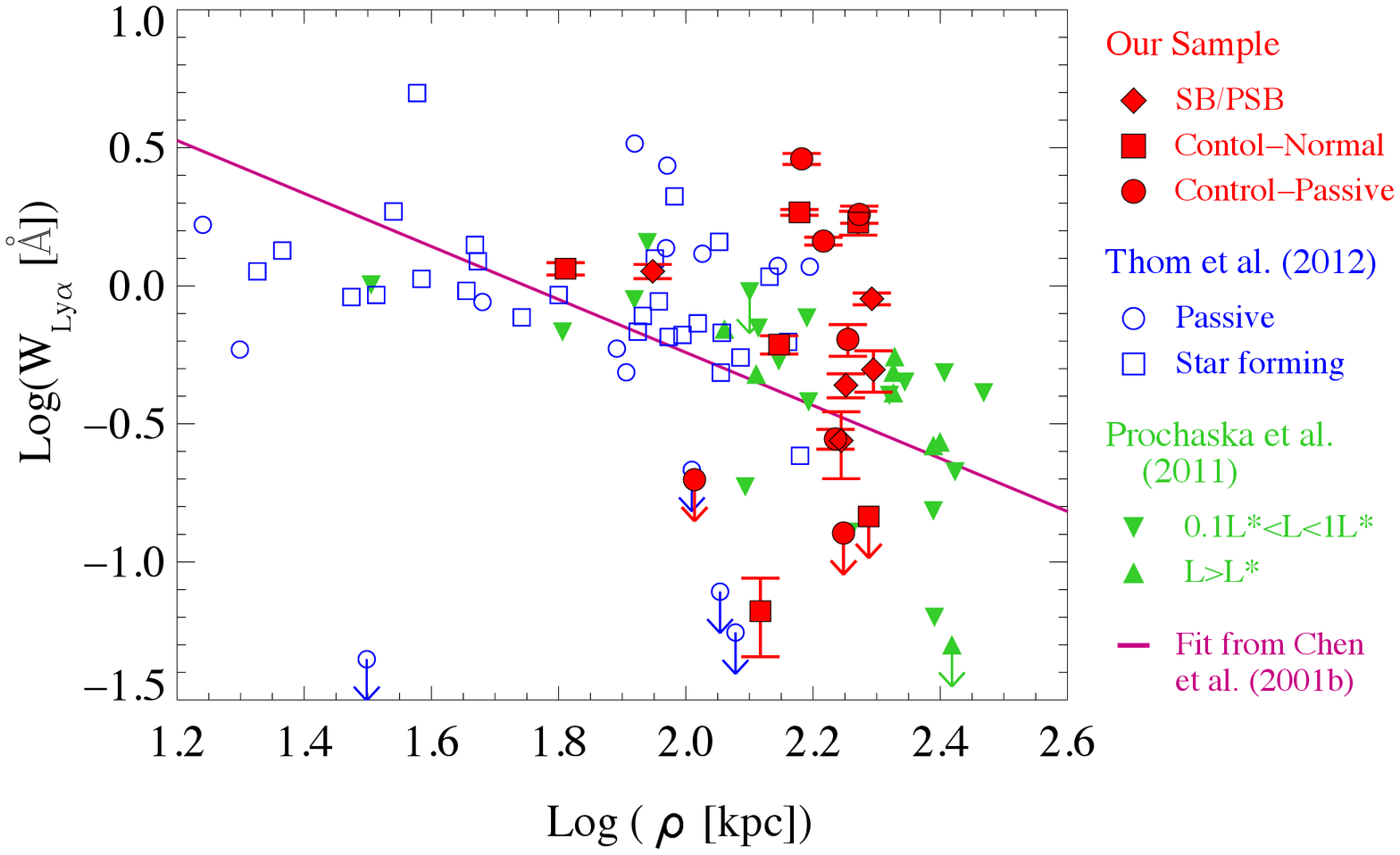}  \\
\caption{Plot showing the variation of the \Lya equivalent widths as a function of the impact parameter from the foreground galaxies. \Lya data for our sample are shown in red with symbols indicating galaxy class (same as in Figure~\ref{fig-sample_pc}). The data in blue and green are from the recently published work by \citet{thom12} and \citet{prochaska11} respectively. The magenta line shows the best fit from \citet{chen01b} to their \Lya absorber sample.  }
\label{fig-rho_eqw}
\end{figure*}

 \begin{figure}
 \center{\includegraphics[trim = 0mm 0mm 00mm 0mm, clip,scale=.55,angle=-0]{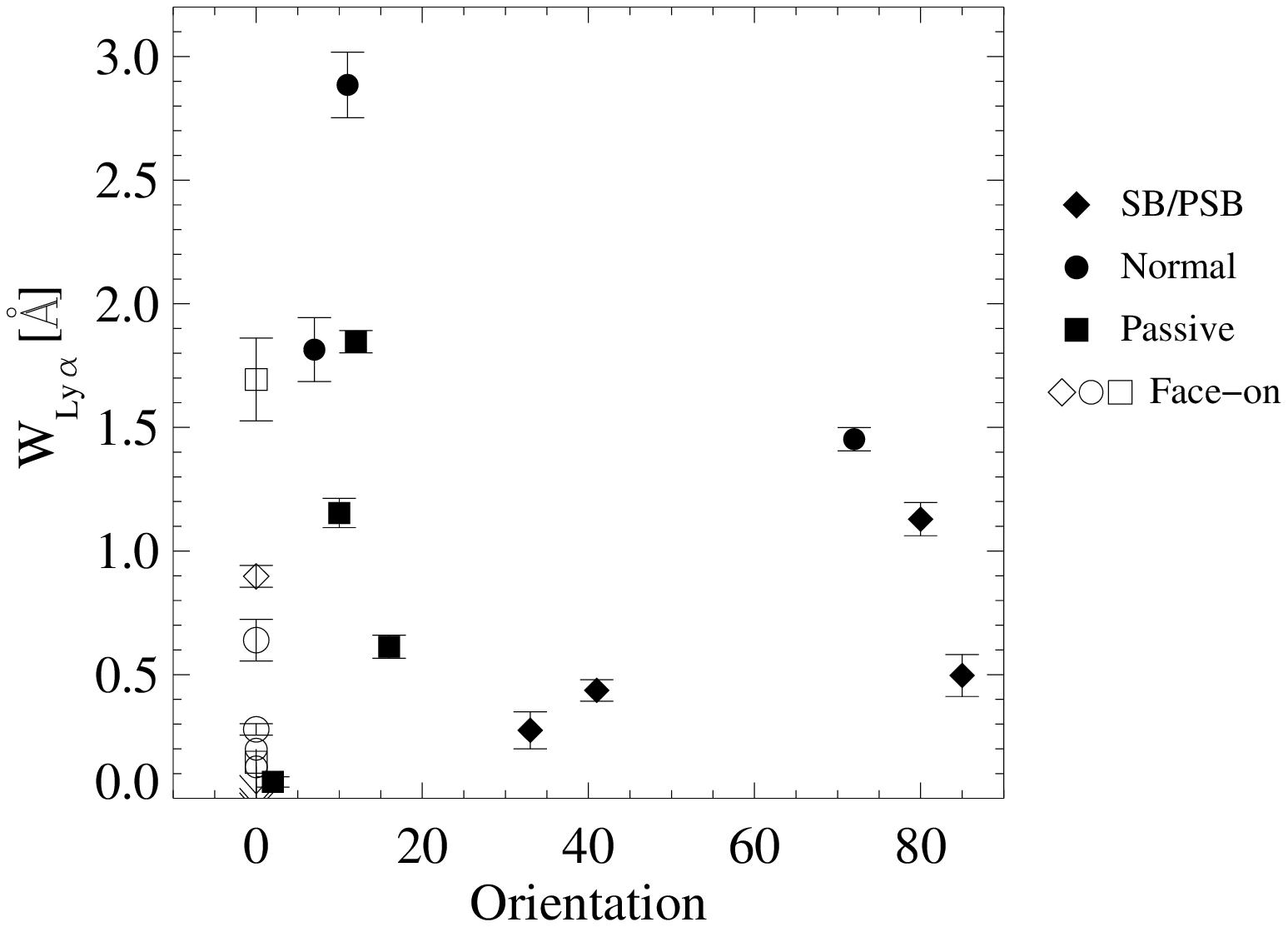}}\\
\caption{\Lya equivalent width is plotted against the orientation of the sightlines, defined as the angle between a sightline and the major axis of the target galaxy. The orientation is 0$^\circ$ when the sightline probes the CGM along the major axis and 90$^\circ$ when the sightline is along the minor axis. All data points with an orientation of exactly 0$^\circ$ are for face-on systems. The symbols indicate galaxy types as in previous figures.}
\label{fig-orient_eqw}
\end{figure}

We detected \HI\ \Lya in 15/18 targets. We also have unambiguously detection of \ion{C}{4} in 6/17 systems.
Information on the properties of the \Lya and \ion{C}{4} absorbers such as equivalent widths, full width at half max (FWHMs), and column densities are presented in Table~{\ref{tbl-absorptiondata}. 
Figures~\ref{fig-lya} and \ref{fig-ci4} show the \Lya and \ion{C}{4} features at the rest frame of the target galaxy, also marked with a dotted line.
The \Lya lines are resolved and unsaturated barring a few cases. Two cases where the \Lya lines are unresolved are marked in Table~{\ref{tbl-absorptiondata}.
The \ion{C}{4} doublets are also resolved and are unsaturated, based on the \ion{C}{4} $\lambda$1548 to $\lambda$1550 doublet ratio (which ranges from 1.54 to 1.87). The corresponding optical depths in the (weaker) \ion{C}4}$\lambda$1559 lines are $\sim$ 0.2 to 1. Apart from \Lya and \ion{C}{4}, we do not detect any other major metal transitions except in the cases of J085301 and J092844. 

The absorption systems associated with target galaxies J085301 and J092844 showed multiple Lyman series and metal transitions. The absorbers are shown in Figures~\ref{fig-J085301} \&  \ref{fig-J092844} and the measurements are presented in Table~\ref{tbl-J085301_J092844}. Besides multiple Lyman series transitions, these absorbing systems show the presence of low-ionization transitions such as \ion{Si}{2} and \ion{C}{2} as well as higher ionization species as traced by \ion{C}{4}. In J085301, the \ion{C}{4} $\lambda$1548 transition lies at the edge of the COS spectral grating and the \ion{C}{4} $\lambda$1550 transition falls in the gap. Hence in this case, the measurements presented in Table~\ref{tbl-J085301_J092844} were derived from the data with incomplete wavelength coverage and may also suffer from flux inaccuracies sometimes seen near the edge of the gratings.

 \begin{figure*}
 \center{\includegraphics[trim = 15mm 20mm 00mm 0mm, clip,scale=0.85,angle=-0]{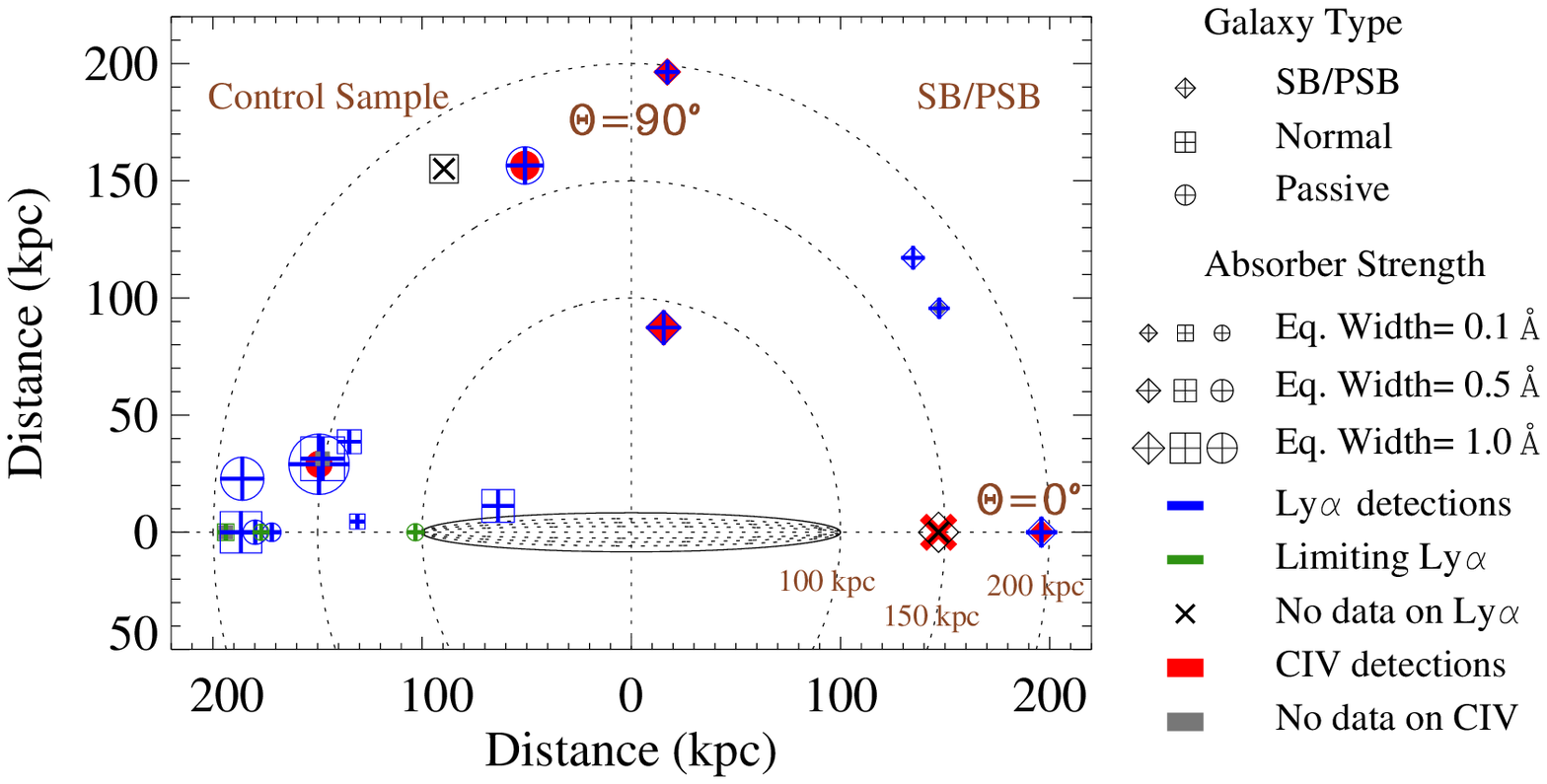} }
\caption{A graphical representation showing our sightlines with respect to a model galaxy. The plot shows impact parameters, orientations, and galaxy-types, as well as the strength of associated \Lya and \ion{C}{4} absorbers for each galaxy in our sample. All the data points with an orientation of exactly 0$^\circ$ are for face-on systems. For easy comparison the starburst galaxies are plotted on the right side of the plot and the control sample consisting of the normal and passive galaxies on the left. The dotted circles mark the loci of impact parameter of 100, 150, and 200~kpc from the center of the galaxy. The strength of the absorber is coded as the size of the symbol. Note that the symbol size scale is not linear.  }
\label{fig-eqw_rho_orient}
\end{figure*}

We fitted Voigt profiles to the data to estimate all the parameters. The associated uncertainties were estimated using the error analysis method published by \citet{sembach92}. Errors estimated using this procedure include continuum placement uncertainties, Poisson noise fluctuations, and zero point uncertainties. The continuum was established using absorption free regions of the spectrum within $\pm$2000-4000~kms$^{-1}$ of the absorption feature and subsequently fitted with a Legendre polynomial of order between 1 and 5, similar to the procedure used by \citet{sembach04}. This ensures that the continuum is described in the neighborhood of the absorption feature precisely and nullifies any slow variations in the shape of the QSO's spectrum. The measurement also takes into account the appropriate line spread functions (LSFs; \citet{osterman11}) for the aperture of the spectrograph from the COS Instrument Handbook \citep{dixon10}. 
For the non-detections, we present the limiting equivalent width, which represents the 3$\sigma$ error in the equivalent width over 200~\kms i.e. $+$100 to $-$100~\kms in the rest frame of the foreground galaxy. Limiting column densities are estimated from the limiting equivalent widths using the relationship between the two for the linear part of the curve of growth.

We have corrected for inaccuracy in the absolute wavelength calibration of the COS data using Milky Way's strong ISM transitions. Since we can estimate the offset of these transitions from the expected velocity with reasonable accuracy, we can then use them to correct for any offsets in the COS wavelength calibration. In order to find the intrinsic velocity offset of the ISM transitions, we use \HI\ 21~cm measurements from the Leiden/Argentine/Bonn (LAB) Survey of galactic \HI\ \citep{kalberla05} from the same region of the sky. Although the beam-size of the \HI\ measurements is enormous as compared to the COS aperture, nevertheless we do not expect this would cause a large error as our velocity resolution is of the order $\sim 100$~\kms (full width at half maximum, FWHM). In most cases for the segment A, we found velocity corrections to be of the order of 100~\kms. However, in the absence of ISM transitions in the segment B, we could not apply the same corrections and the velocity calibration remains highly uncertain. In cases of multiple Lyman series transitions located in both the segments, the ones in segment B (e.g. Lyman~$\gamma$) show a velocity offset of $\sim-700$~\kms relative to those in segment A (\Lya).

\section{RESULTS\label{sec:results}}

\subsection{The Strength of Lyman~$\alpha$ Absorbers\label{sec:Lya}}

 \begin{figure*}
\includegraphics[trim = 0mm 105mm 52mm 20mm, clip,scale=0.45,angle=-0]{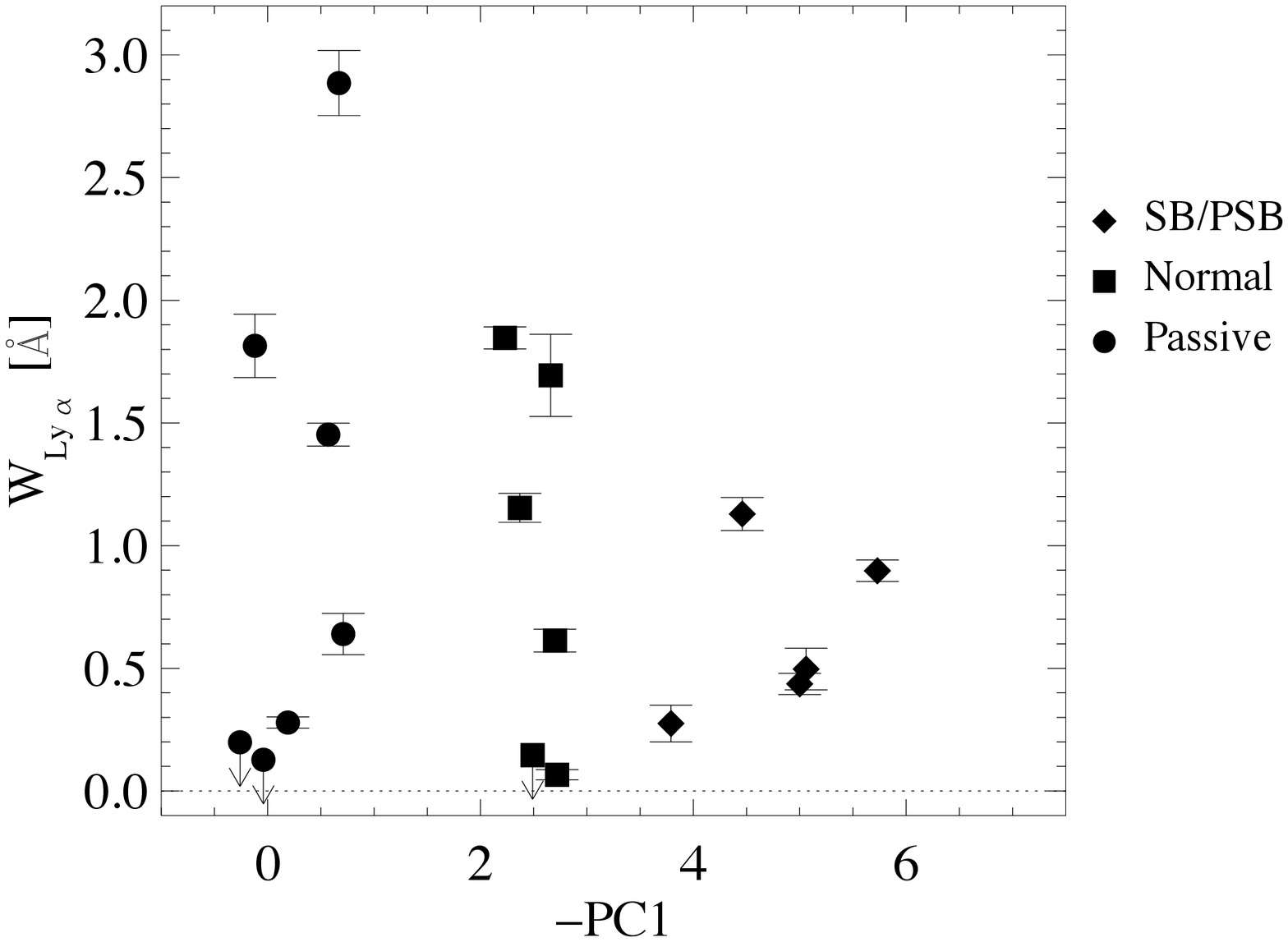}  
\includegraphics[trim = 0mm 105mm 00mm 20mm, clip,scale=0.45,angle=-0]{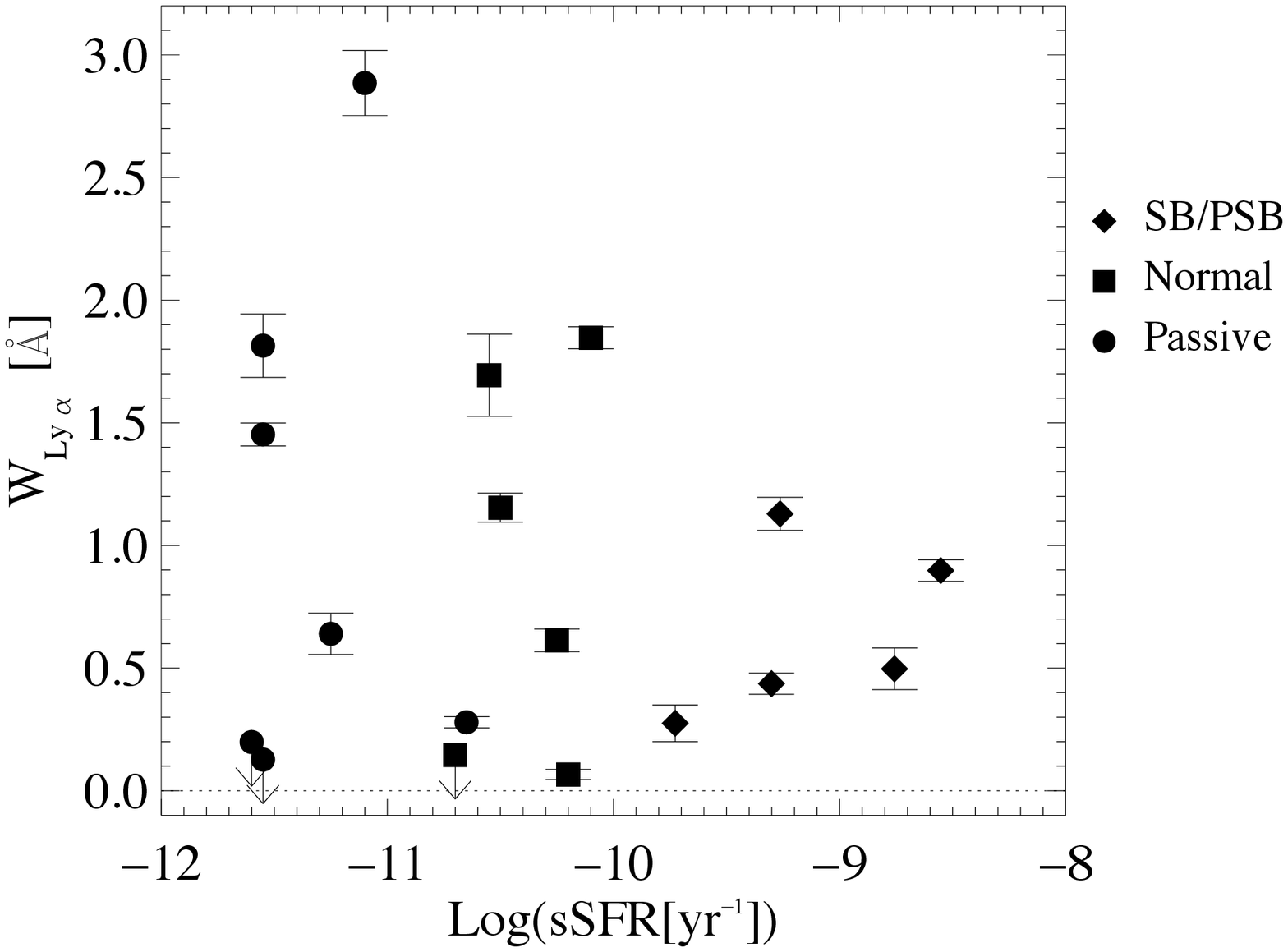}    \\
\caption{Plots showing the variation of \Lya equivalent width as a function of PC1, which represents the 4000~$\rm \AA$ break ($\rm D_n(4000)$) and as a function of the specific star-formation rate (SFR/M$_*$ in inverse years). The galaxy type is represented by the symbols as in previous figures.}
 \label{fig-pc1_pc2_eqw} 
\end{figure*}

The properties of the CGM as a function of properties of the host galaxy are not yet clearly understood. Studies have detected gas traced by \Lya out to radii of a few hundred kpc \citep[][and other authors]{chen98,chen01b,bowen02,steidel11} and found a weak inverse correlation between the strength of the \Lya line and the impact parameter. 

Figure~\ref{fig-rho_eqw} shows \Lya equivalent width versus impact parameter. Our data are shown as red symbols with their shapes representing their spectral classes based on the indices PC1 and PC2. 
The magenta line shows the best fit obtained by \citet{chen01b} for their data. We also overplot data from recent studies by \citet{prochaska11} and \citet{thom12} for a visual comparison. 
We do not see a trend of equivalent width with impact parameter in our sample.  This is not surprising since our sample spans a rather narrow range in impact parameter.
Furthermore, these objects cover the full range of equivalent width seen in other surveys. Interestingly, the top five strongest absorbers are associated with the CGM of the control sample. 

Some authors have found a correlation between the strength of \Lya absorption and the luminosity of the host galaxy \citep{chen01b, bowen02}.
We see no correlation between the \Lya equivalent width and the galaxy stellar mass, which is a proxy for its luminosity. This may be because we span a rather narrow range in stellar mass ($\rm 10^{10.1}-10^{11.4}~M_{\odot}$), although the implied range in the dark halo mass is considerably larger (few $\times 10^{11}$ to a few $\times10^{13}$ M$_{\odot}$).

We also investigate the strength of the \Lya transition as a function of orientation of the sightline with respect to the galaxy major axis (Figure~\ref{fig-orient_eqw}). 
The strongest absorbers are aligned close to the major axis, although some of the weak absorbers can also be found there. However, due to small number statistics of our data a correlation cannot be verified with certainty (the Spearman's rank analysis implies that there is a $\sim$20\% chance that there is no correlation).

A more intuitive way of understanding the combined effect of orientation and impact parameter on the strength of the absorbers is presented in Figure~\ref{fig-eqw_rho_orient}. The figure shows a cartoon of a generic galaxy at the center. The equidistant regions from the center are marked with dotted semi-circles. It is to be noted that this model is just for illustration and does not represent the range of morphologies and sizes of our target galaxies. In order to distinguish between the two sub-samples, the starburst galaxies are plotted on the right side of the figure and the control galaxies containing the normal and passive galaxies are plotted on the left. It is evident that the \Lya absorbers with equivalent width $\rm W_{Ly\alpha}>1\AA$ are only found in the control sample.

To further emphasize the difference in the average \Lya strength between the starburst sample and the control sample, we study the variation in terms of both the value of PC1 and the star-formation rate per unit stellar mass (the specific star-formation rate see Table~\ref{tbl-sfr}). Figure~\ref{fig-pc1_pc2_eqw} suggests that the scatter in the \Lya equivalent width is larger in the passive galaxies and smallest in the starbursts. 

\subsection{Highly Ionized Gas in the CGM in Starburst Galaxies \label{sec:CIV}}

 \begin{figure}
 \hspace{-0.75cm}
\includegraphics[trim = 5mm 0mm 5mm 0mm, clip,scale=0.55,angle=-0]{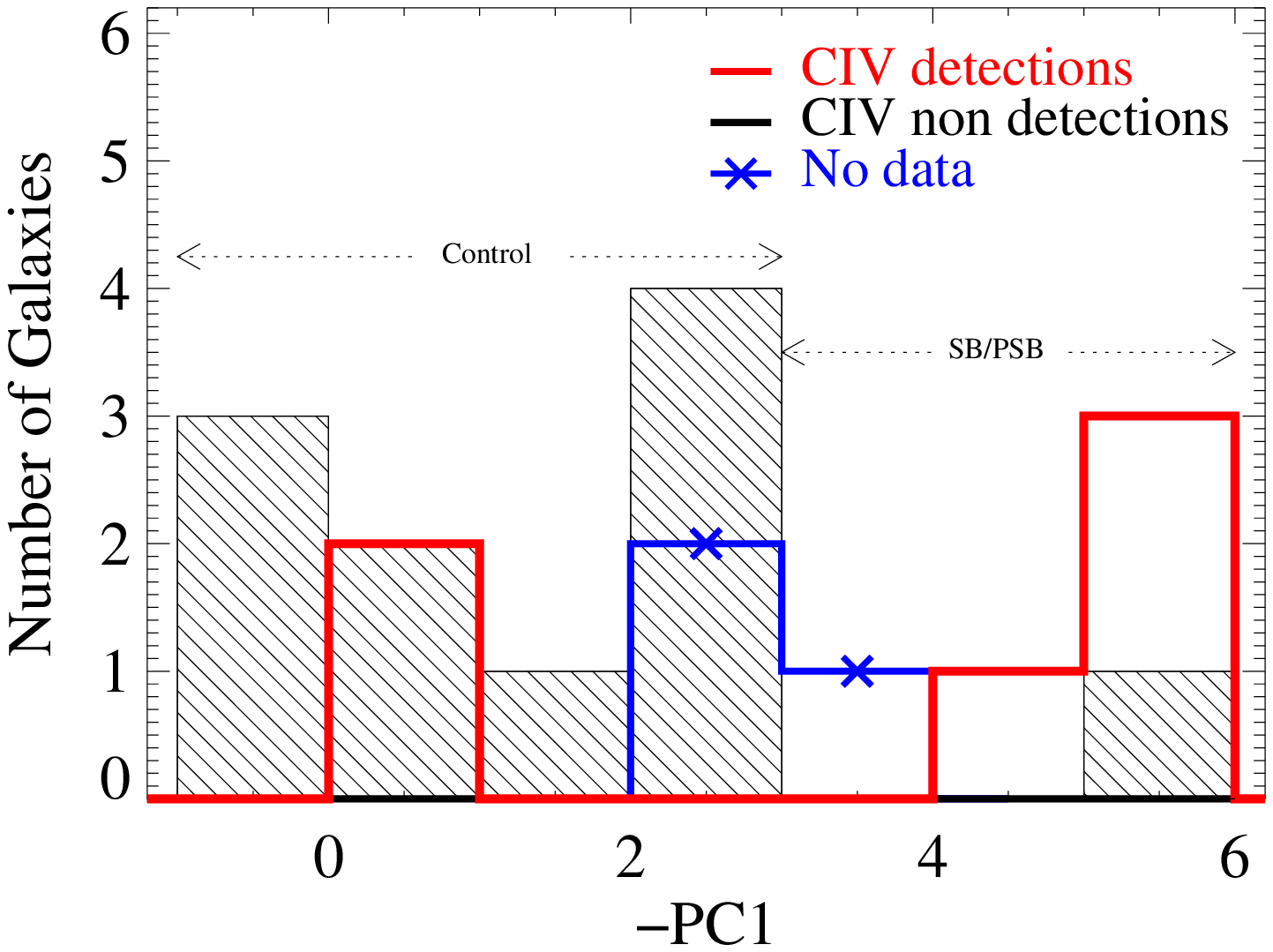}   \\

 \hspace{-0.75cm}
\includegraphics[trim = 5mm 0mm 5mm 0mm, clip,scale=0.55,angle=-0]{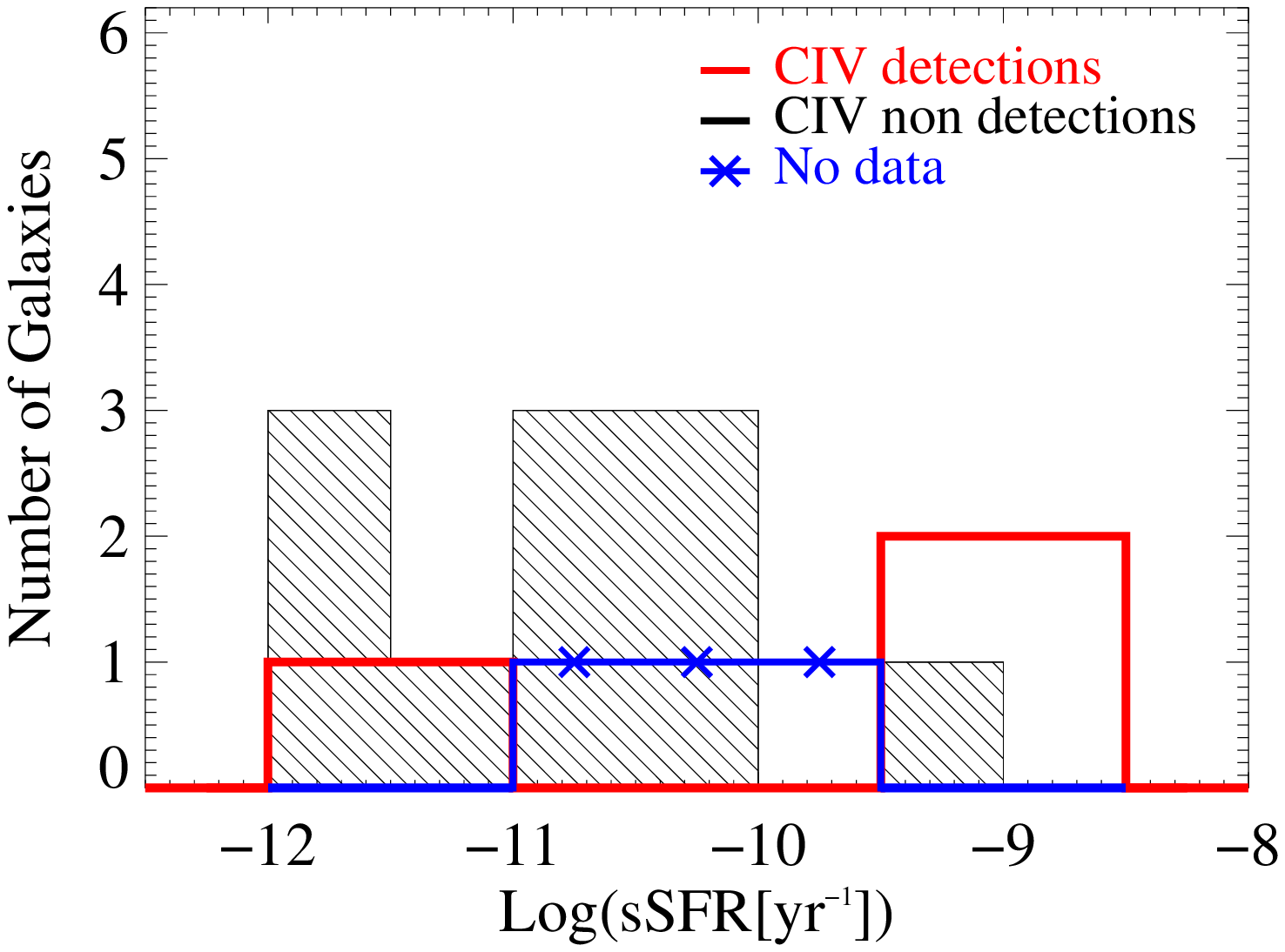}   \\
\caption{Histogram showing the rate of \ion{C}{4} detection (and non detections) as a function of PC1 and the specific SFR respectively. The SB/PSB and the control samples have been labeled. These figures show the higher detection fraction for SB/PSB sample as compared to the control sample. Chi-squared contingency test confirms that the two samples have different \ion{C}{4} detection rates with 99.9\% confidence.  }
\label{fig-civ_hist}
\end{figure}

 \begin{figure*}
\hspace{-1.5cm}
\includegraphics[trim = 5mm 110mm 33mm 55mm, clip,scale=0.60,angle=-0]{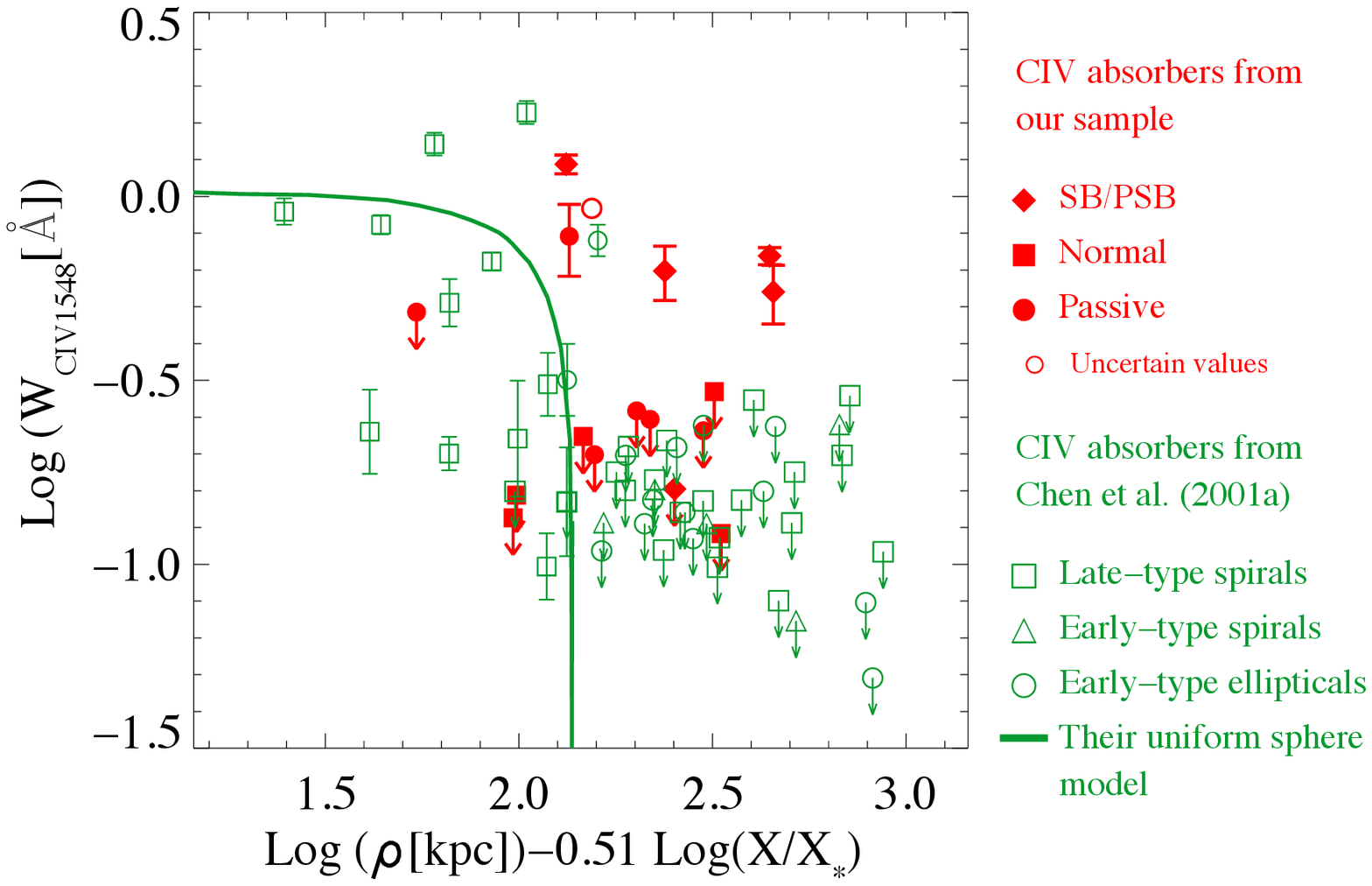}   
\includegraphics[trim = 7mm 110mm 20mm 55mm, clip,scale=0.60,angle=-0]{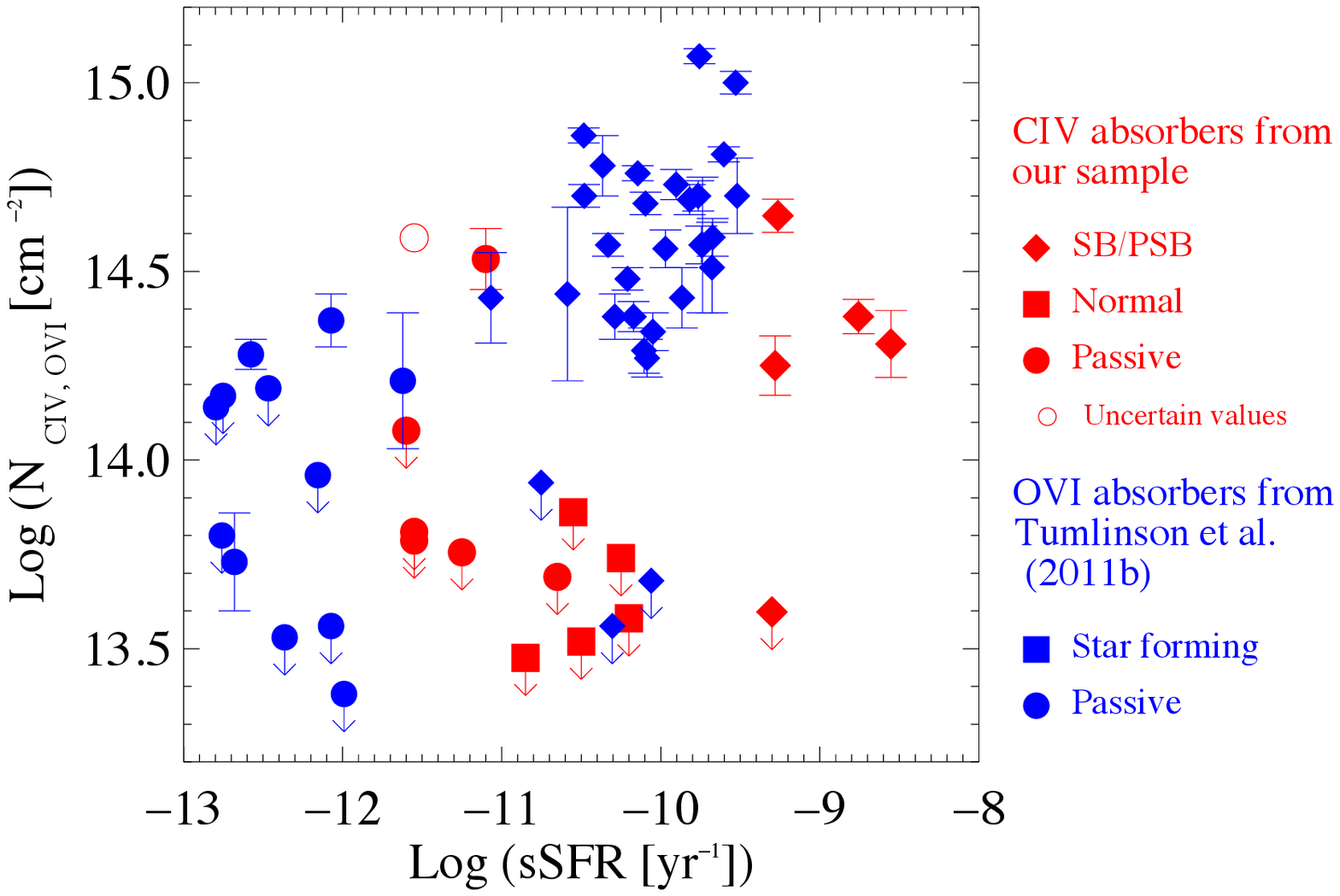}   \\
\caption{Plots showing the strength of \ion{C}{4} as a function of normalized impact parameter (left) and specific SFR (right). The scaling (X/X$_*$) used by \citet{chen01a} was in terms of luminosity i.e. L/L$_*$. Since our targets are starburst galaxies with very high luminosities, we choose to adopt a stellar mass scaling which is more appropriate to  \citeauthor{chen01a}'s scaling.
Here, M is the total galaxy stellar mass and $\rm M_* = 7 \times 10^{10}$ M$_{\odot}$. 
The three galaxy classes are labeled. In green we overplot data on \ion{C}{4} from \citet{chen01a} on the right and in blue data on \ion{O}{6} from \citet{tumlinson11b} on the left.}
\label{fig-civ_ssfr}
\end{figure*}

A drop in the strength of the \Lya line can be due to either a relative lack of gas in the CGM of starburst galaxies or to a higher degree of ionization and/or temperature. A distinct difference between the two scenarios would be an enhanced presence of absorbers associated with highly ionized and/or hot (coronal-phase) gas. The best probe of such gas in our data is the \ion{C}{4} line. Our COS data provided useful information on this line in 17 cases (5 starbursts and 12 control galaxies). It is then possible to look for any correlation between the presence of a starburst and of highly ionized gas traced by \ion{C}{4}.

We detected \ion{C}{4} absorption-lines in six galaxies. As described above, we fit Voigt profiles to the doublet and this yielded typical column densities of several $\times 10^{14}$ cm$^{-2}$ (see Table~\ref{tbl-absorptiondata}). Four of the detected galaxies are starbursts and the remaining two are control galaxies. Thus, we find \ion{C}{4} absorbers in 80\% (4/5) of the SB/PSB and in only 17\% (2/12) of the control galaxies. Interestingly, the two detections of the \ion{C}{4} absorbers in the control sample are associated with passive galaxies and none with normal star forming galaxies (see Figure~\ref{fig-civ_hist}). 

Figure~\ref{fig-civ_ssfr} shows the equivalent width  as a function of impact parameter and specific SFR. By design, we do not explore a large impact parameter range and hence no trends are prominent unlike that found by \citet{chen01a} for a sample of galaxies at a median redshift of 0.4. Our data (plotted in red) is combined with that of \citeauthor{chen01a} (plotted in blue) is shown in the left panel of the figure. Following the fitted empirical relation in \citet{chen01a}, we have scaled the impact parameters for our sample by the factor $(M/M_*)^{-0.51}$. Here $M$ is the total galaxy stellar mass and $M_*$ is the fiducial value for a Schechter function fit to the galaxy stellar mass function: $M_* = 7 \times 10^{10}$ M$_{\odot}$ \citep[e.g.][]{li_white09}. This panel clearly reinforces our previous conclusion that the CGM surrounding the starbursts has unusually strong \ion{C}{4} absorption-lines at large impact parameters compared to typical galaxies. We do explore a large parameter space in terms of specific SFR and find strong absorbers in galaxies with high specific SFRs. A qualitatively similar trend was also observed by \citet{tumlinson11b} for their \ion{O}{6} absorber sample (plotted in blue in the right panel of figure~\ref{fig-civ_ssfr}). 

Recent numerical simulations have modeled CGM properties in the presence of galactic outflows using SPH-based cosmological simulations \citep{ford12} and adaptive-mesh refinement simulations for Milky-Way massed galaxies \citep{hummels12}. While \citet{ford12} found a subset of \ion{C}{4} absorbers at impact parameters of 100~kpc from their host galaxy that had column densities greater than $\rm 10^{14}~cm^{-2}$, the median value was about $\rm \sim 10^{13}~cm^{-2}$ (corresponding to a \ion{C}{4} equivalent width of 0.05~$\AA$). 
Simulations of Milky-Way-like galaxies by \citet{hummels12} predicted typical \ion{C}{4} columns at impact parameters $>$ 100 kpc of $\sim \rm 10^{12}~cm^{-2}$. Both sets of simulations are consistent with our data for typical galaxies, but fall far short of the observed \ion{C}{4} column densities of a few $\rm \times 10^{14}~cm^{-2}$ in the CGM of starbursts.
 
We note that most of the spectra in our sample have similar signal-to-noise ratios and therefore the \ion{C}{4} non-detections are not a consequence of lower data quality in these spectra. 
Figure~\ref{fig-civ_psb_con_stack} shows the stack of all the non-detections in the two control sub-samples. This was done in order to detect any weak signature of the presence of \ion{C}{4} absorption in the data by increasing the S/N of the averaged spectrum. 
The \ion{C}{4} spectrum of each galaxy is stacked at the peak velocity of associated \Lya after correcting for the intrinsic COS wavelength calibration corrections (as described in Section~\ref{sec:cos}). The wavelength corrections help against dilution of the stacked spectra. The resulting uncertainties in rest-frame velocities are $<$~100~\kms, which is lower than the resolution of the grating or the FWHM of all the \ion{C}{4} detections. 
The stacked spectra were generated by adding all the \ion{C}{4} non-detections for the two control galaxy subsamples i.e. 5 normal galaxies and 5 passive galaxies each. 
The single case of \ion{C}{4} non-detection in the starburst sample is not shown here. The limiting equivalent width of \ion{C}{4} for the normal star-forming galaxy stack is $<$118.5 m$\rm \AA ~ ( 3\sigma$), compared to values of $\sim$ 400 to 900 m$\rm \AA$ ~ in the six detections. Passive galaxies do show a slight dip, however due to undulating nature of the noise in the spectrum this feature is still within the noise. So, in neither stacked spectrum is there a significant absorption feature.

 \begin{figure}
\hspace{-0.7cm} 
\includegraphics[trim = 0mm 0mm 0mm 0mm, clip,scale=.70,angle=-0]{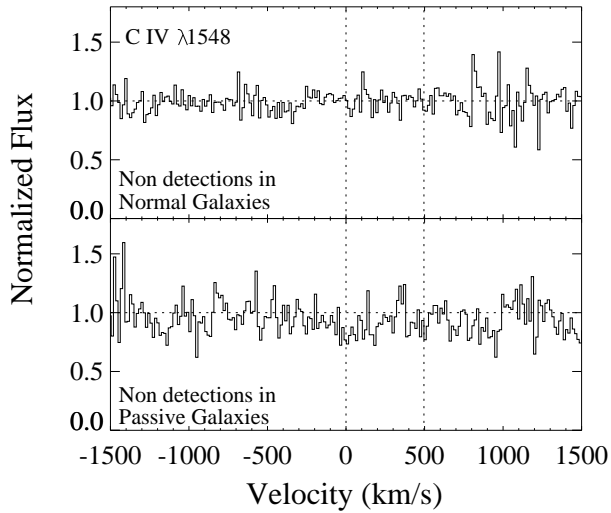}  \\
\caption{Composite rest-frame \ion{C}{4} spectra obtained by stacking all non-detections for normal star-forming and passive galaxies. The stacks contain 5 galaxies each. The dotted lines mark the expected position of the \ion{C}{4} doublets. The  $\rm \lambda 1548~\AA$ feature is at zero velocity and the  $\rm \lambda 1550~\AA$ feature is at 498~\kms. }
\label{fig-civ_psb_con_stack}
\end{figure}

\begin{figure*}
\includegraphics[trim = 00mm 0mm 0mm 0mm, clip,scale=.53, angle=-0]{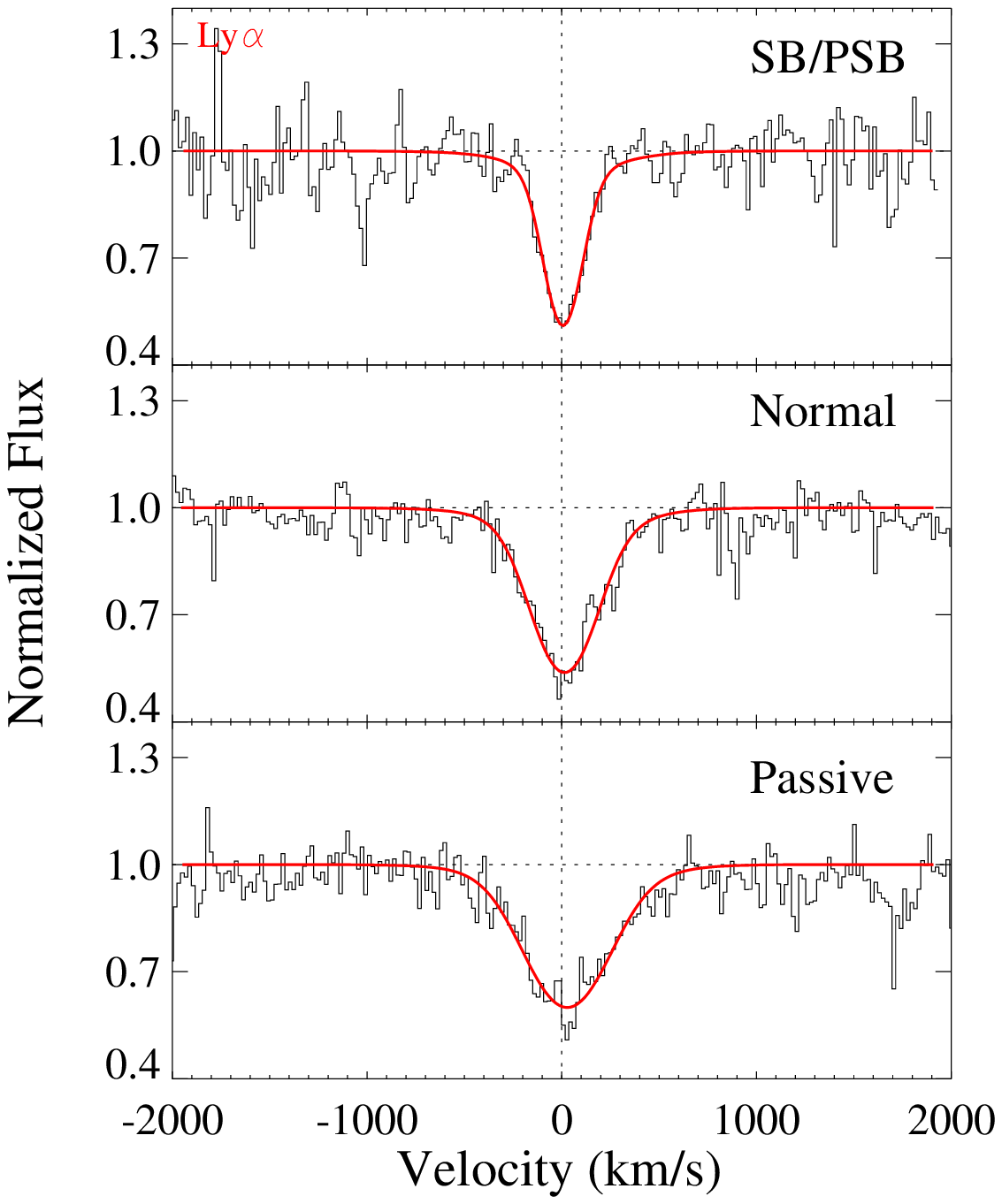}  
\includegraphics[trim = 20mm 0mm 0mm 0mm, clip,scale=.53, angle=-0]{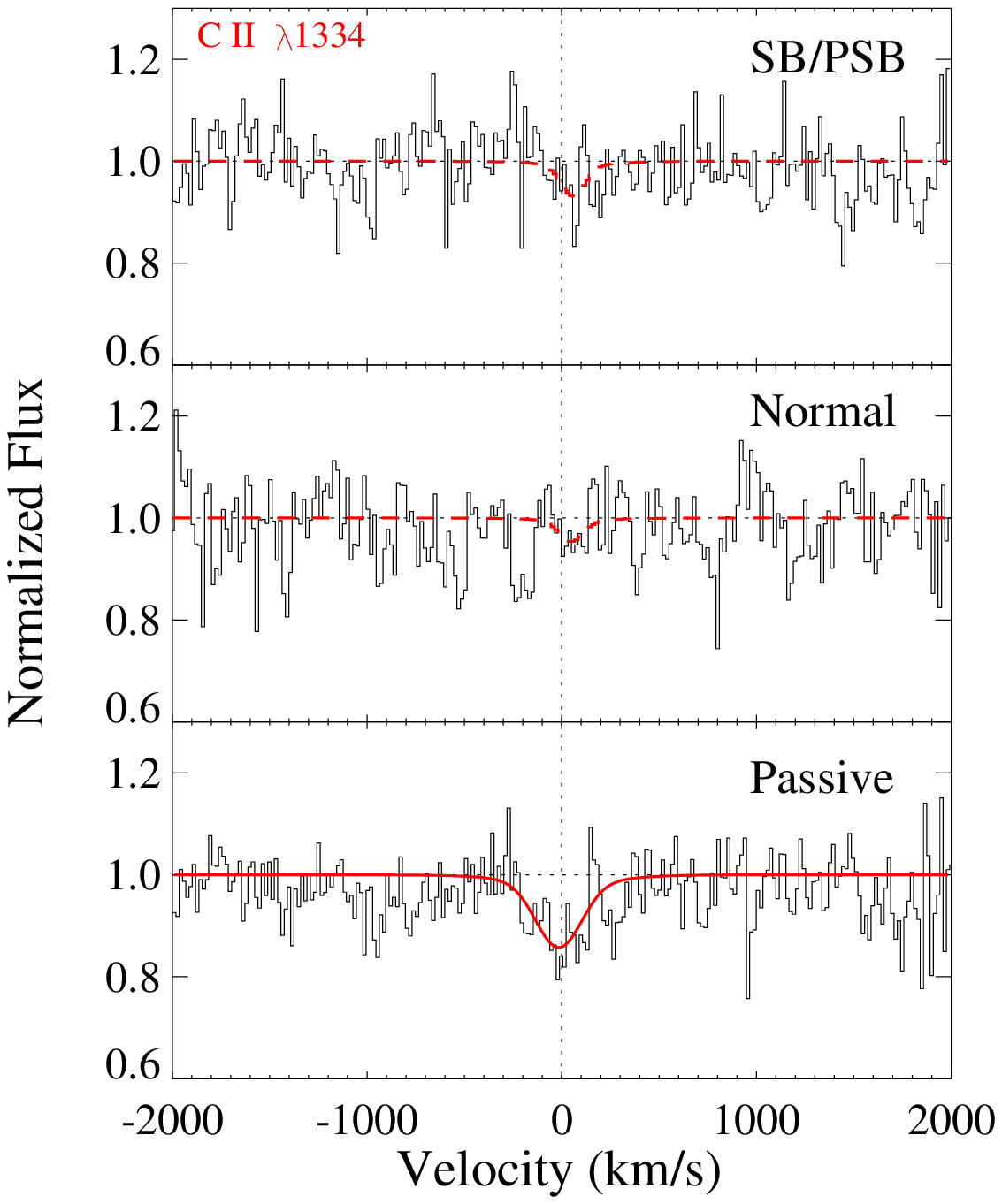}  
\includegraphics[trim = 20mm 0mm 0mm 0mm, clip,scale=.53, angle=-0]{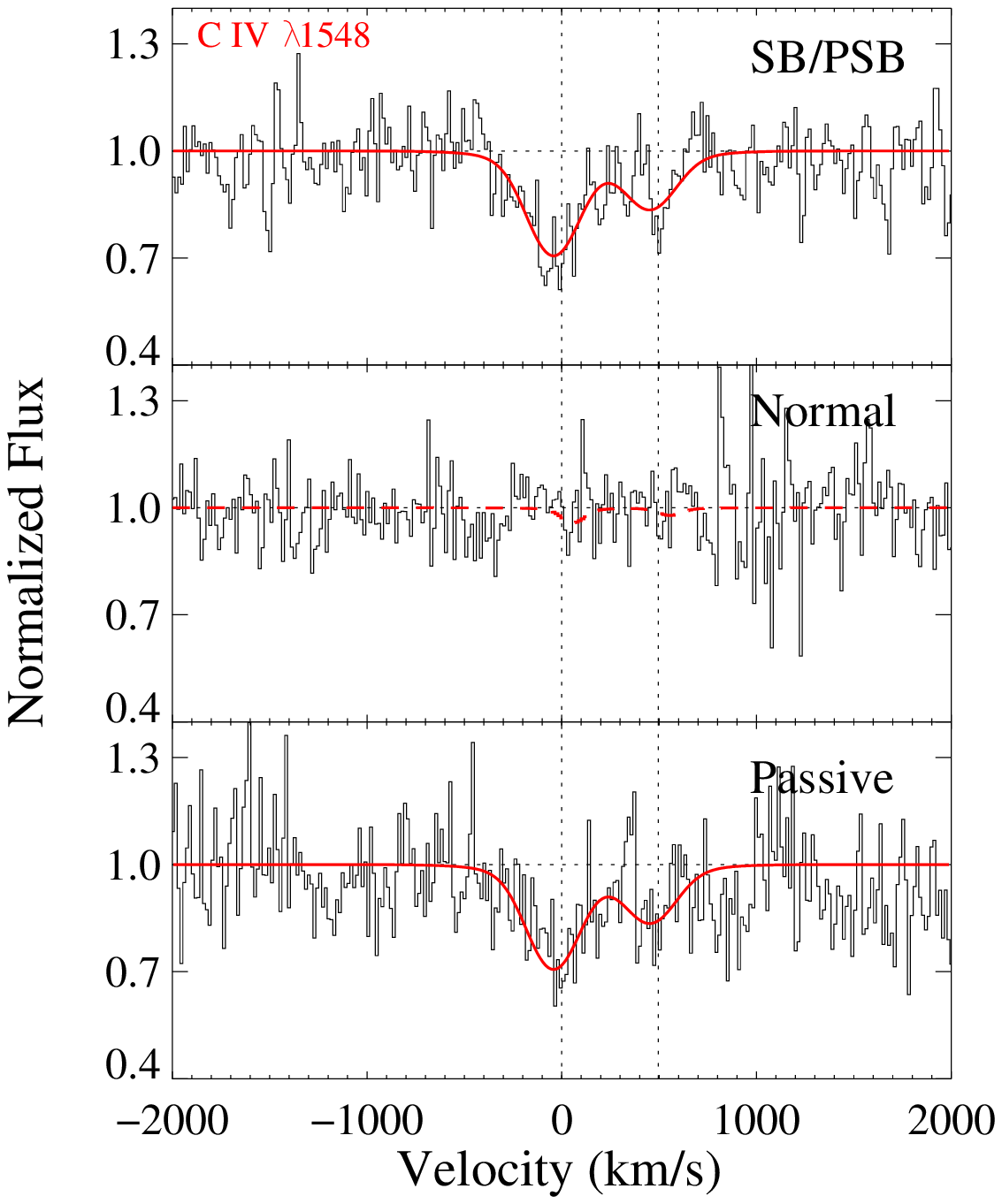} \\
\includegraphics[trim = 00mm 0mm 0mm 0mm, clip,scale=.53, angle=-0]{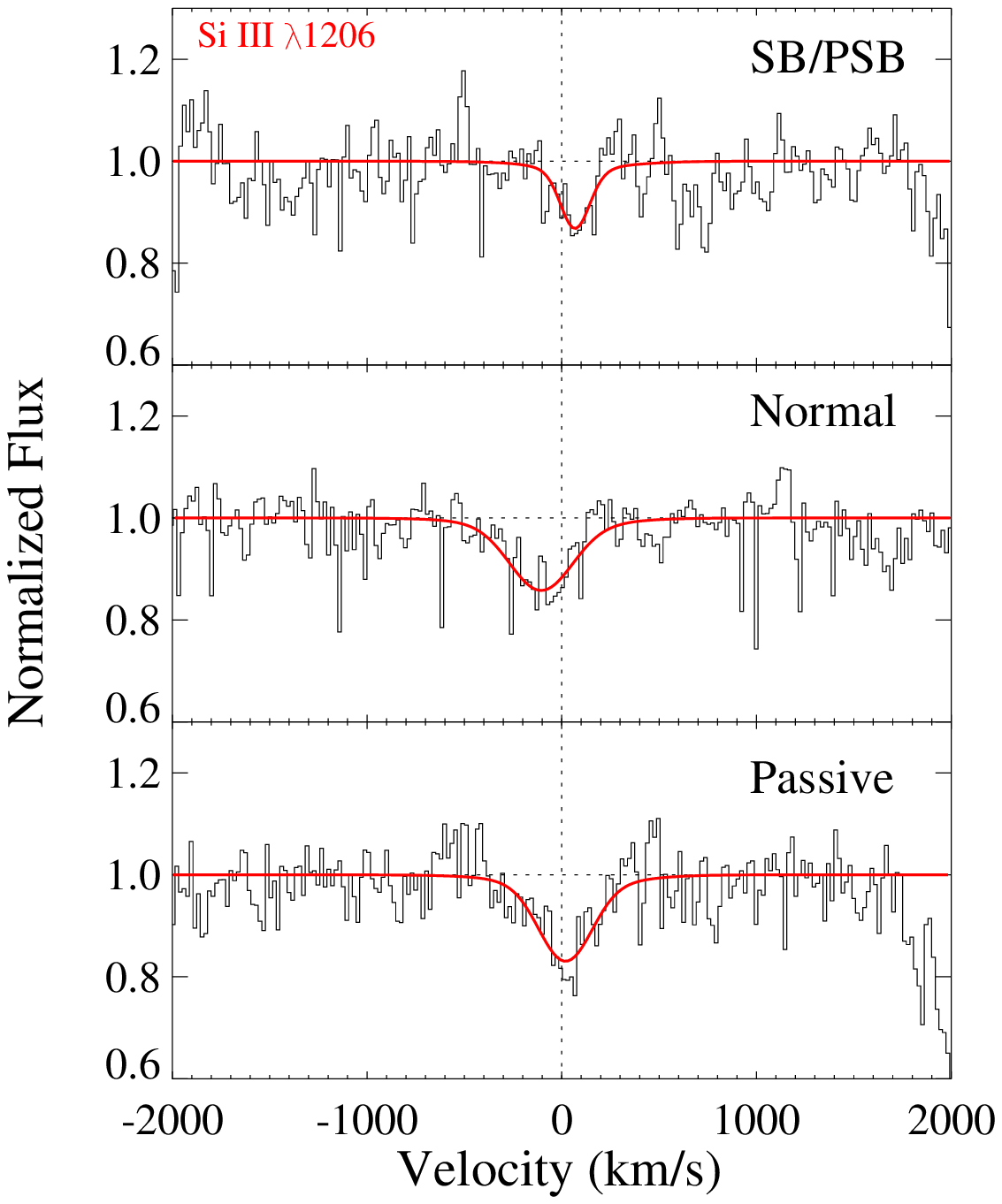}   
\includegraphics[trim = 20mm 0mm 0mm 0mm, clip,scale=.53, angle=-0]{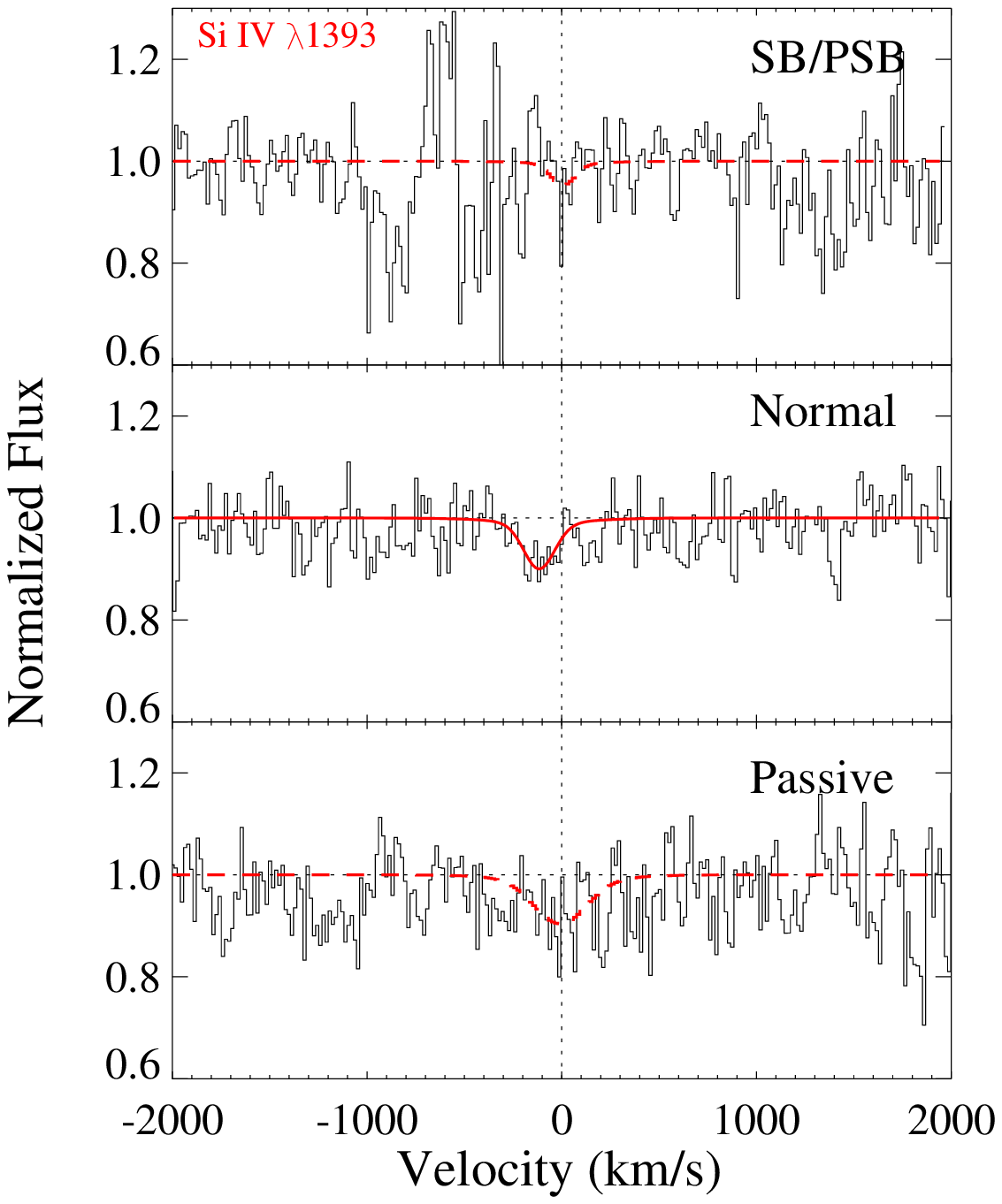}   \\
\caption{Composite rest-frame spectra of galaxies from the three classes. Each composite was obtained by stacking all the non-corrupted spectra including detections and non-detections. The panels show \Lya, \ion{C}{2}$\rm \lambda 1334$, the \ion{C}{4} doublet, \ion{Si}{3}$\rm \lambda 1206$, and \ion{Si}{4}$\rm \lambda 1393$.
The dotted line(s) marks the expected rest-frame position of the transitions. The Voigt profile fits to the spectra are shown in red solid lines. For non-detections, we plot red dashed lines showing the limiting equivalent width. }
\label{fig-civ_lya_stack}
\end{figure*}

 \begin{figure*}
\includegraphics[trim = 10mm 125mm 53mm 20mm, clip,scale=.5,angle=-0]{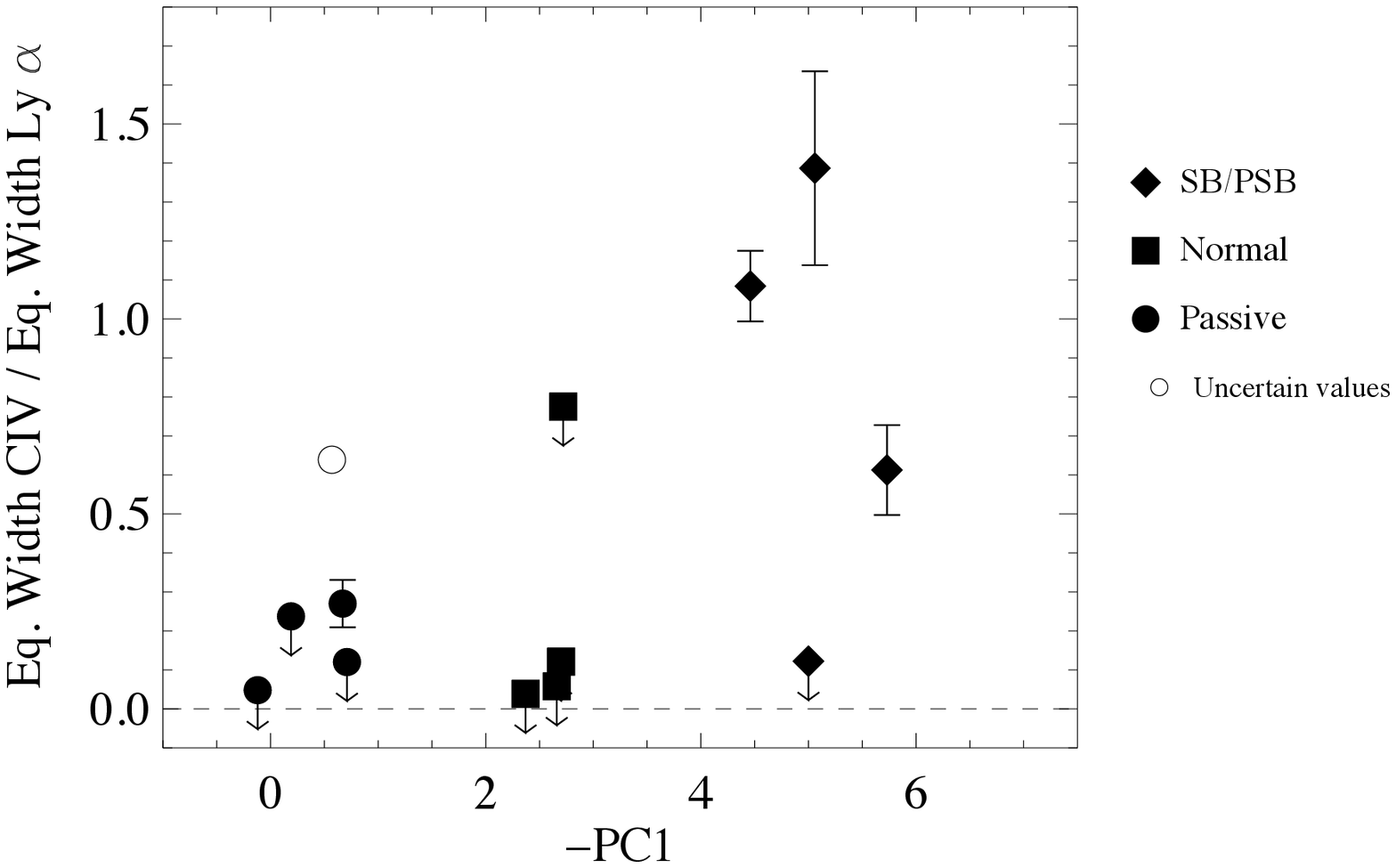}  \includegraphics[trim = 0mm 125mm 05mm 20mm, clip,scale=.5,angle=-0]{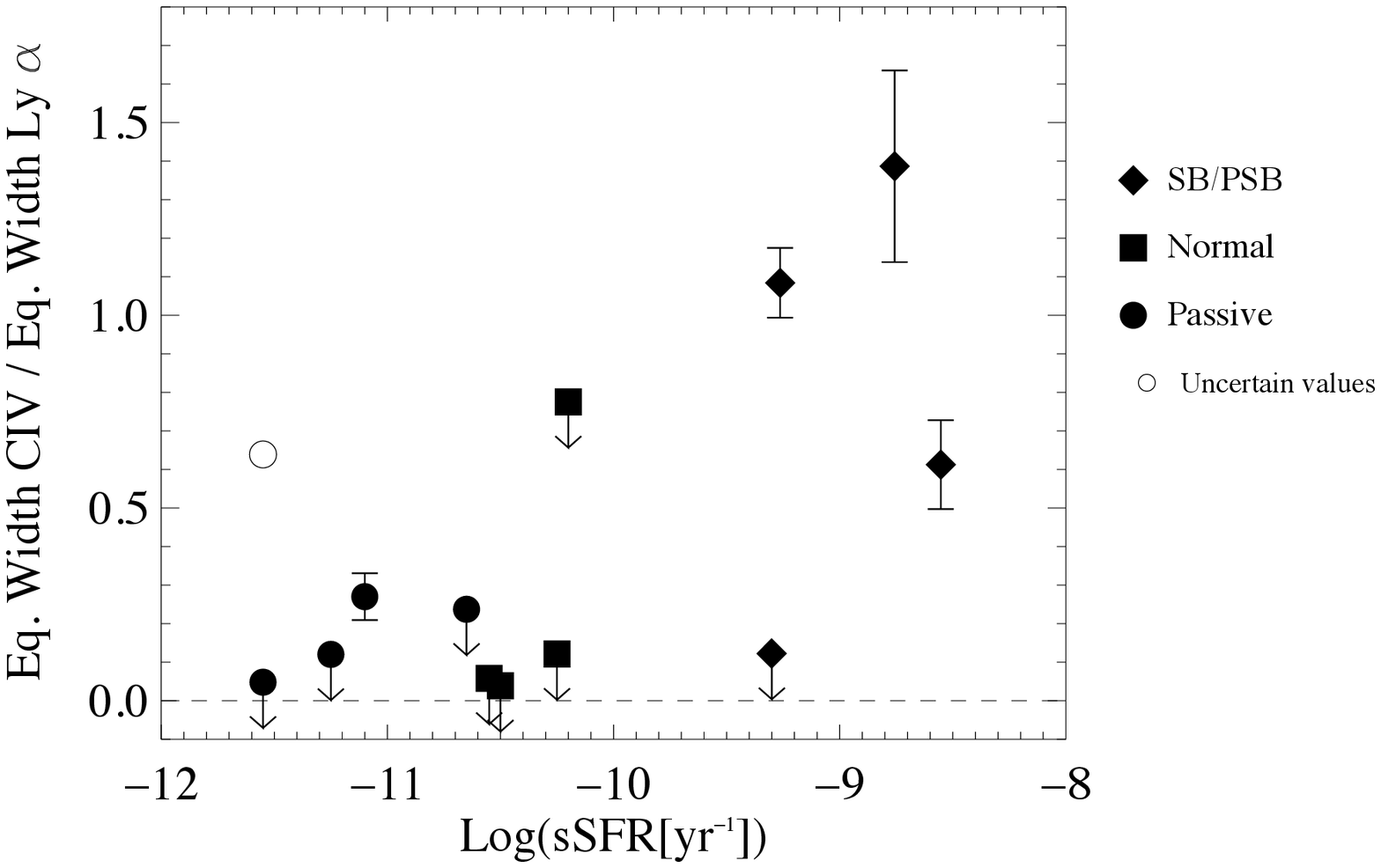}  \\
\includegraphics[trim = 10mm 125mm 53mm 20mm, clip,scale=.5,angle=-0]{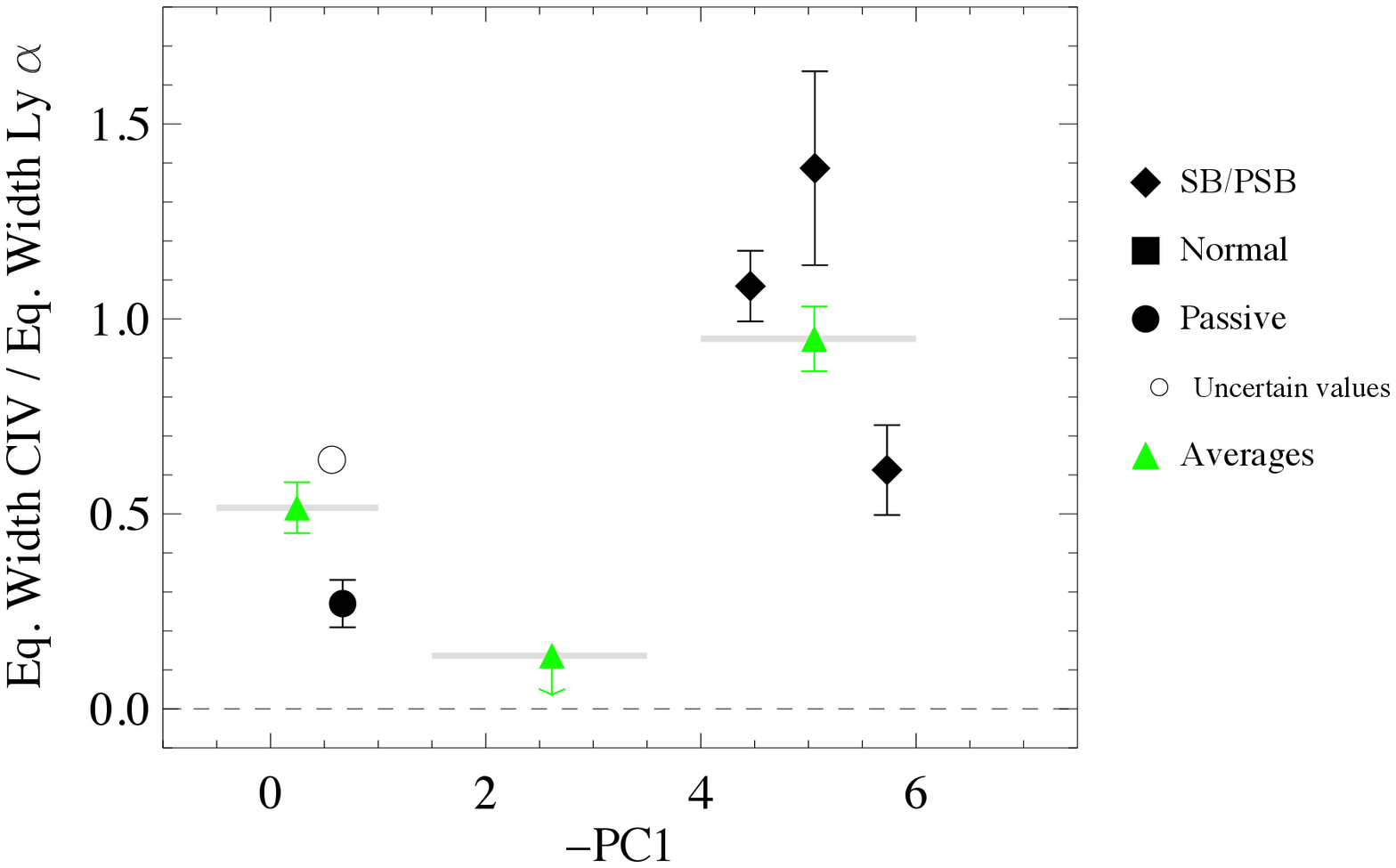}  \includegraphics[trim = 0mm 125mm 05mm 20mm, clip,scale=.5,angle=-0]{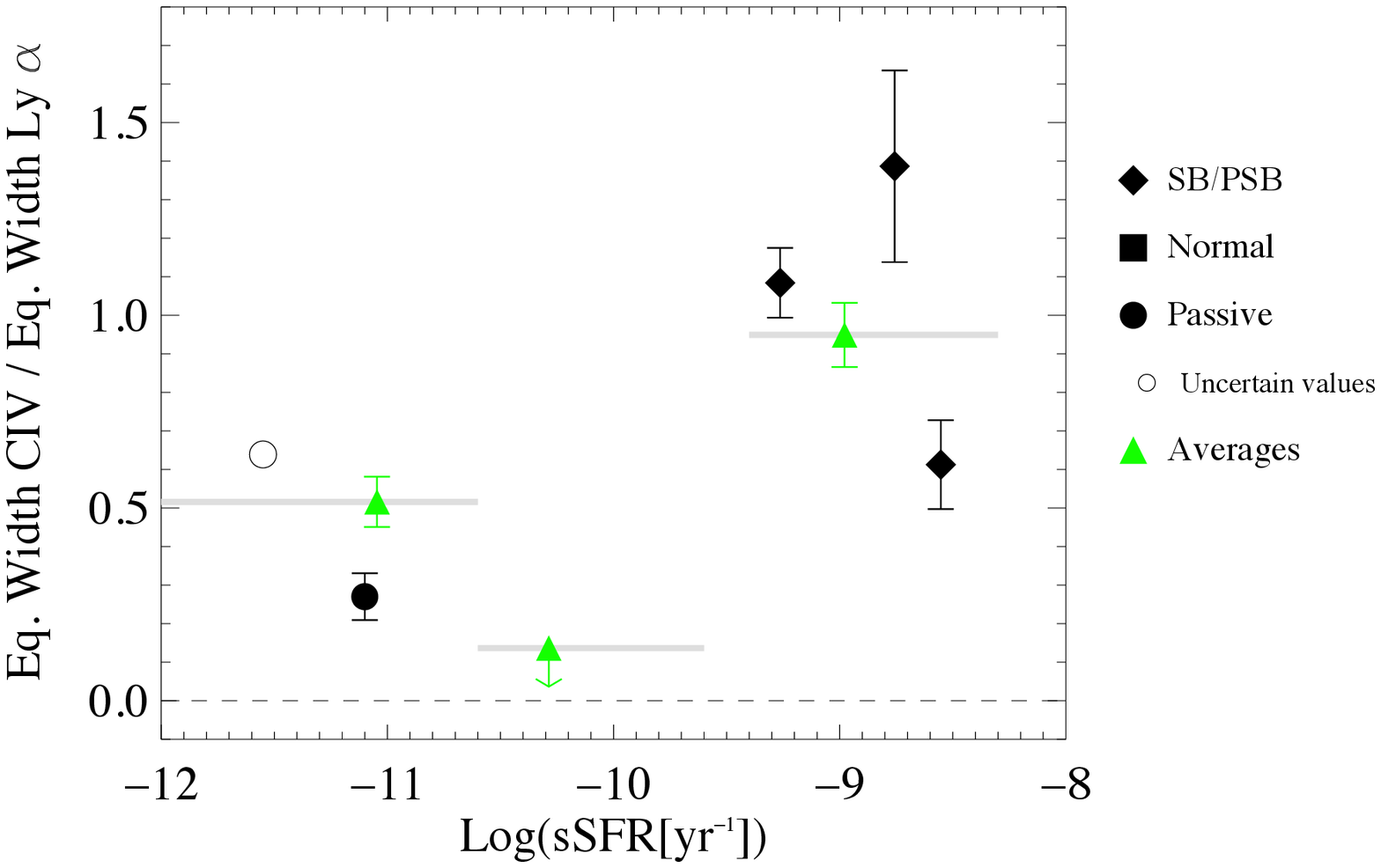}  \\
\caption{Plots showing the ratio of \ion{C}{4} to \Lya equivalent widths as a function of PC1 in the top left panel and specific SFR in the top right panel. These plots include galaxies for which we have associated \Lya absorbers. That is, galaxies with upper limits to their \Lya equivalent widths or those where we do not have measurements of either \Lya or \ion{C}{4} are not included in these plots. In the bottom panel, we omit the individual upper limits and instead plot the average for each of the classes using the composite spectra shown in Figure~\ref{fig-civ_lya_stack}.}
\label{fig-civ_lya_ratio}
\end{figure*}

There are important differences between \ion{C}{4} absorbers in starburst and control galaxies besides the detection rate. While the presence of \ion{C}{4} in the control (passive) galaxies is accompanied by the presence of strong low-ionization metal transitions, the same is not true for the starburst galaxies. 
The two passive galaxies with \ion{C}{4} detections show transitions such as \ion{Si}{2}, \ion{C}{2}, \ion{Si}{3}, \ion{C}{3}, and \ion{Si}{4}. The equivalent widths of the \Lya absorbers in each of these two systems (1.5 and 2.9~\AA) are also significantly higher than those found in the starburst galaxies (0.3 to 1.1~\AA; median of 0.5~\AA). 

To investigate this more generally, we have created stacked composite spectra for the starburst and control samples. The composite spectra for \Lya, \ion{C}{2}, \ion{C}{4}, \ion{Si}{3}, and \ion{Si}{4} for the SB/PSB sample and the two control sub-samples are presented in Figure~\ref{fig-civ_lya_stack}. These spectra were obtained by stacking both detections and non-detections together for each of the transitions - \Lya $\lambda$1216~$\rm \AA$, \ion{C}{2} $\lambda$1334~$\rm \AA$, \ion{C}{4}$\lambda\lambda$1548,1550~$\rm \AA$, \ion{Si}{3} $\lambda$1206~$\rm \AA$, and \ion{Si}{4} $\lambda\lambda$1393,1402~$\rm \AA$. The dotted lines show the rest-frame of the absorbers based on the position of the \Lya centroid.
The \Lya spectra were obtained by adding each individual spectrum corrected to the rest-frame velocity of the target galaxy. For the other transitions that are weaker than \Lyanospace, we correct the velocity to the rest-frame \Lya velocity whenever \Lya was detected. If not, we used the optical redshift of the target galaxy to set the rest-frame velocity.

\Lya was clearly detected in the stacked composite spectra in all sub-samples and \ion{C}{4} was detected for starburst and control-passive galaxies. We also detected \ion{Si}{3} in all three composite spectra at a strength of about 20-25\% of \Lyanospace. Table~\ref{tbl-stack_measurements} summarizes the equivalent widths, FWHMs, and column densities of all the detected transitions and provides upper limits for non-detections. 
Among the detections, the \Lya equivalent width as well as its velocity spread increased from the starburst to normal to passive galaxies. It is not surprising that the equivalent width and the velocity spread of the composite \Lya features are correlated. This is due to the contribution of strong (saturated) \Lya lines associated with some of the passive and normal galaxies. Since the stacked spectra were produced by averaging both saturated lines with unsaturated and even non-detections, the resulting composite may not look saturated.

Another way to study the difference in the ionization state of the CGM in the starburst halos is shown in Figure~\ref{fig-civ_lya_ratio}. Here we plot the ratio of the equivalent widths of \ion{C}{4} to \Lya as a function of PC1 and specific star formation rate. The top panels show the ratios for individual galaxies in our sample. Not all galaxies have detectable \Lya and the plot includes only those with \Lya detections. The upper limits in the plot represent cases where no \ion{C}{4} was detected. None of the four galaxies with an upper limit on the \Lya equivalent widths show \ion{C}{4} absorption. In other words, we have detected \Lya in all the galaxies where we have detected \ion{C}{4}, with the possible exception of J140502, where we have no information on the \Lya transition. 

The bottom panels of Figure~\ref{fig-civ_lya_ratio} show the subset of galaxies where we have detected \ion{C}{4} and also include the average \ion{C}{4}/\Lya ratio for the three sub-samples in green. The averages were derived from the composite spectra obtained by stacking the velocity corrected spectrum of individual galaxies. The larger mean ratio of \ion{C}{4}/\Lya in the starbursts is evident. This would suggest that the CGM has a higher overall degree of ionization in the starbursts. We believe that this is an ionization effect, and not due to higher metallicities in the CGM of the starburst galaxies. If it were just a metallicity effect, then the starburst galaxies should show strong absorption-lines from metals in the lower ionization states in addition to \ion{C}{4}. This is not the case as can be seen from our data (see Table~\ref{tbl-stack_measurements}).

\section{Discussion \label{sec:CIV_implications}}

The results presented above (section 3.2 and Tables~\ref{tbl-absorptiondata}~and~\ref{tbl-stack_measurements}) show that material with a \ion{C}{4} column density of a few $\times 10^{14}$ cm$^{-2}$ covers a large fraction ($\sim$ 80\%) of the sight-lines through the halos of starburst galaxies extending out to radii of roughly 200 kpc. 

To get a crude estimate of the mass represented by the \ion{C}{4} absorbers we follow \citet{tumlinson11b} and use a simple model-independent geometric method. We take the total number of \ion{C}{4} ions in the CGM to be given by the product of the \ion{C}{4} column density for our absorbers, the cross-sectional area of the CGM that we probe ($\pi r^2$, where r = 200 kpc), and the covering factor of the \ion{C}{4} absorbers within this radius (4/5). This implies a total mass of \ion{C}{4} within a radius of 200 kpc of roughly a few $\times 10^6$ M$_{\odot}$. For collisionally ionized gas (see below), the {\it minimum} ionization correction (which occurs for a temperature of 10$^5$ K) implies a total Carbon mass of about $\rm 10^7 ~M_{\odot}$ \citep{dopita_sutherland96}. The {\it maximum} likely metallicity of the CGM would be solar. In this case the {\it minimum} total gas mass associated with the \ion{C}{4} absorbers would be $\sim 3 \times 10^9~M_{\odot}$ (but is likely to be larger). Since the mean stellar mass of the starbursts in our sample is about $2 \times 10^{10}$ M$_{\odot}$, we conclude that the \ion{C}{4} absorbers trace a significant repository of the baryons in these galaxies. 

The results presented above also suggest that starbursts are affecting the ionization state of a significant amount of gas in the circum-galactic medium. We will first discuss whether this is physically plausible, will then use these results to estimate the properties of the absorbing material, and then briefly discuss the implications of these results.

\subsection{Photoionization}

 \begin{figure*}
\includegraphics[trim = 20mm 120mm 10mm 0mm, clip,scale=.45,angle=-0]{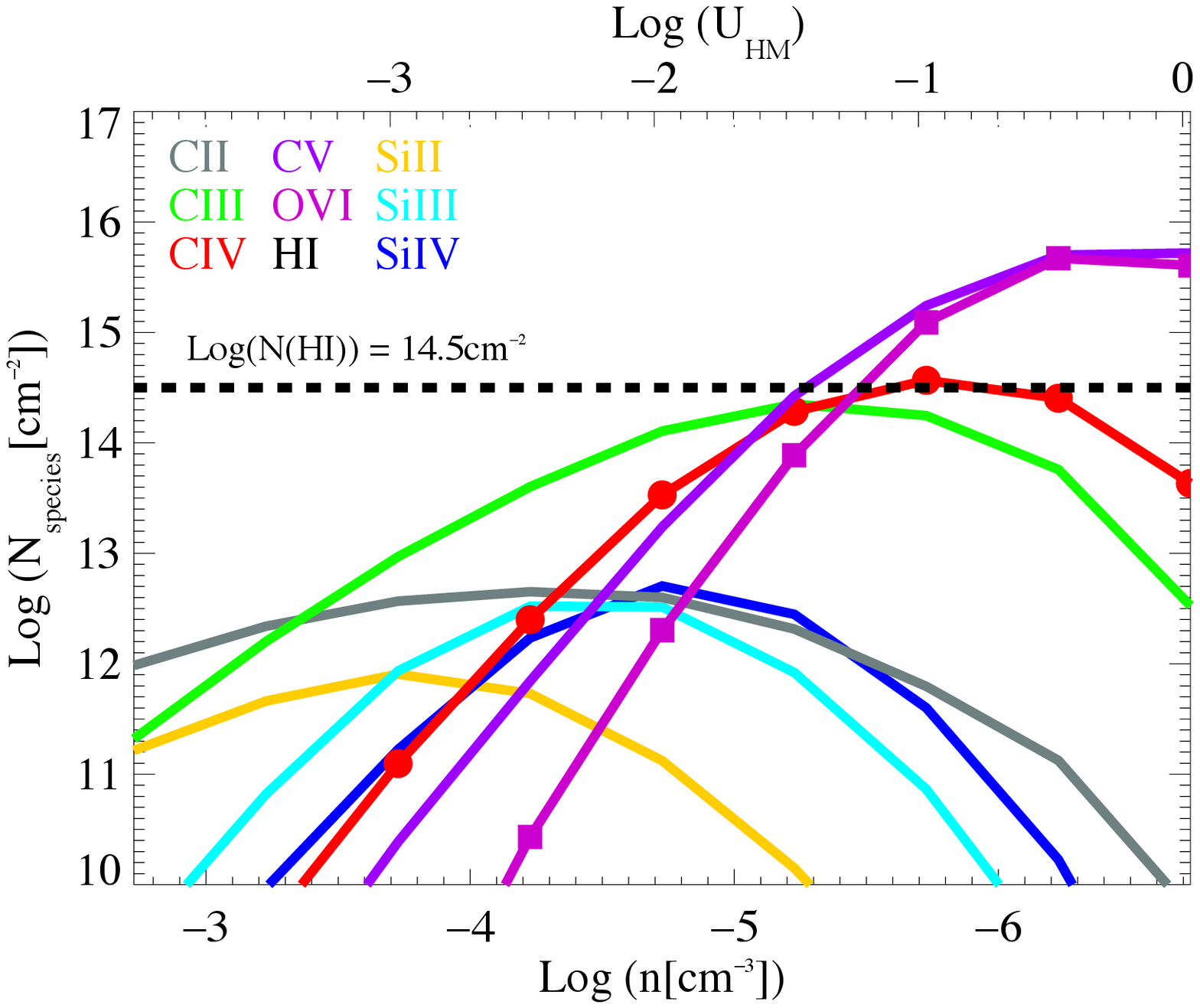}
\includegraphics[trim = 20mm 120mm 10mm 0mm, clip,scale=.45,angle=-0]{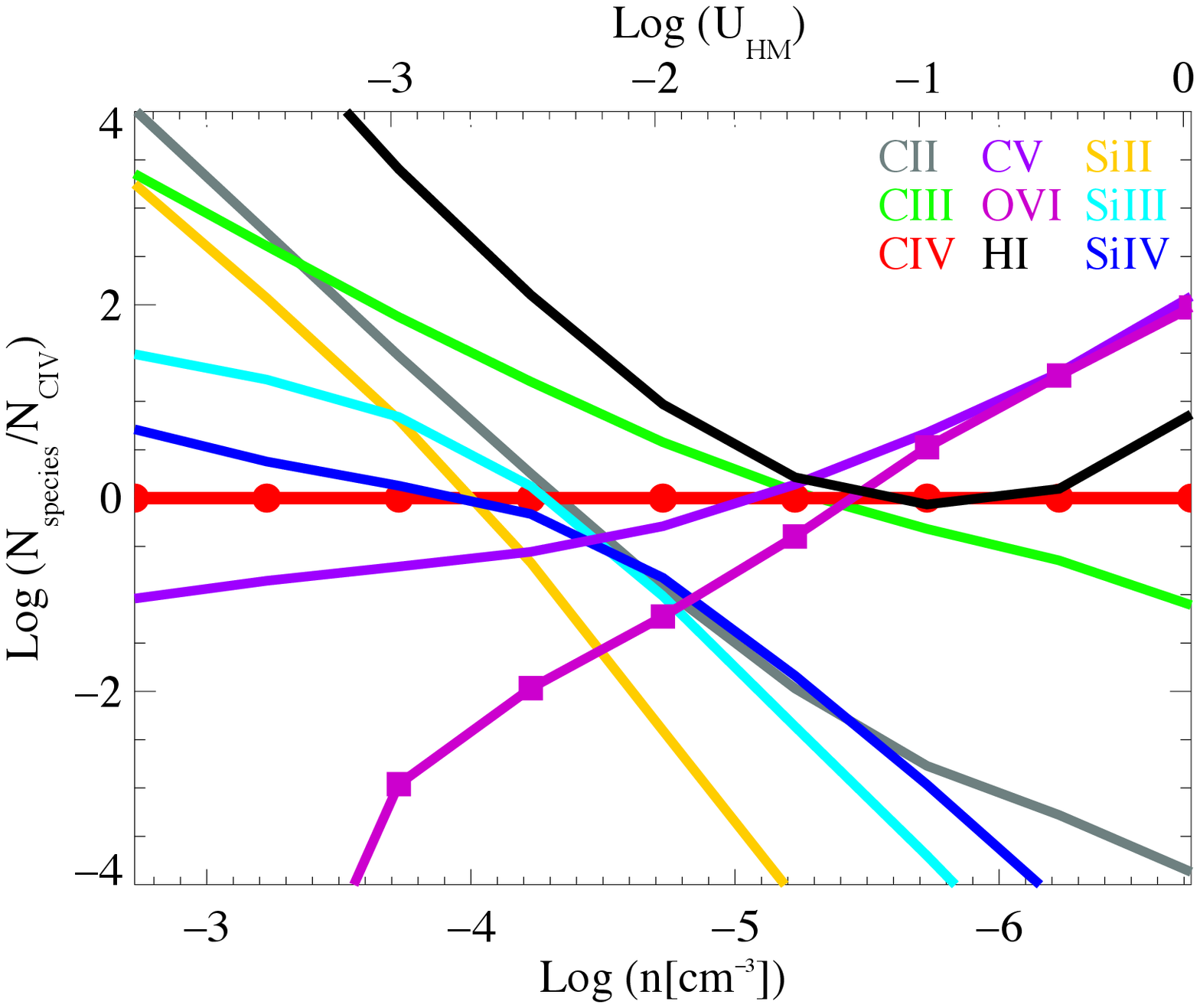}   
\caption{Photoionization models using Cloudy for a CGM cloud that is ionized by the diffuse metagalactic ionizing background. Different species are labeled using different colors. For easy viewing the \ion{C}{4} and \ion{O}{6} transitions are additionally marked with filled circles and filled squares respectively. The thick dashed line represents the \ion{H}{1} column density and was set as a termination criterion for the simulations. The right panel shows the ratio of the various species in terms of \ion{C}{4}. These ratios are independent of the termination criterion. The metagalactic flux at redshift z=0 is normalized to that described by \citet{haardt_madau12} . }
\label{fig-cloudy_hm}
\end{figure*}

 \begin{figure}
\includegraphics[trim = 20mm 120mm 10mm 20mm, clip,scale=.45,angle=-0]{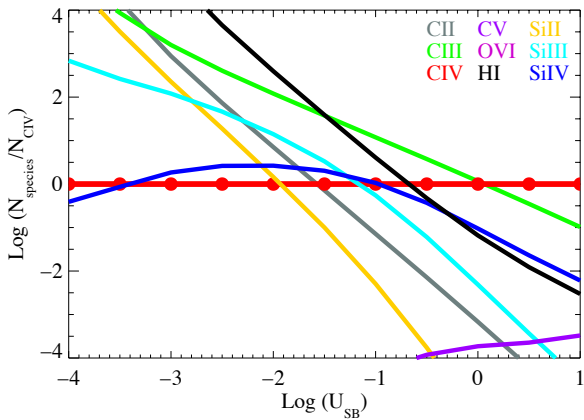}  
\caption{Photoionization models using Cloudy for a CGM cloud that is ionized by the starburst in the host galaxy. Different ionic species are color coded. The \ion{C}{4} and \ion{O}{6} transitions are additionally marked with filled circles and filled squares, respectively.
The spectrum of the starburst is modeled using Starburst99 for a 100~Myrs old continuously star forming galaxy. The plot shows the ratio of column densities of  various species with respect to column density of \ion{C}{4} as a function of the ionization parameter, U$\rm _{SB}$,  which can be expressed in terms of the physical parameters of the cloud and the starburst as described by Eq.~\ref{eq-U}. }
\label{fig-cloudy_sb}
\end{figure}

One possibility is that the \ion{C}{4} in the starburst galaxies arises in photoionized gas. The radiation field incident on a CGM cloud of density, $n$, at a distance, $r$, from the galaxy can be characterized using the ionization parameter, U, that includes contribution from both the extragalactic \citet{haardt_madau12} background and the ionizing flux from the starburst:
\begin{equation}
 U = U_{HM} + U_{SB}
\end{equation}
As described by \citet{tumlinson11a}, $U$ can be expressed as  
\begin{equation}{\label{eq-U}}
 U=\frac{1}{c~n}\big(\Phi_{HM}+ \frac{Q_* f_{esc}}{4\pi r^{2}}\big) 
\end{equation}
where $Q_*$ is the total rate of hydrogen ionizing photons produced by the starburst with an escaping fraction of $f_{esc}$, $r$ is the distance from the starburst to the cloud (which we approximate as the impact parameter) and $c$ represents the speed of light.

We present two scenarios for photoionization of a CGM cloud as modeled using the code Cloudy\citep{cloudy} for a plane-parallel slab. These two cases can be considered to bound the range of possibilities. These simulations assume the relative abundances of the different metals have solar values with no dust depletion.

In the first case, we use the Haardt \& Madau metagalactic background \citep{haardt_madau12}. We illuminate the slab from two sides in this case to account for the isotropic nature of the metagalactic background. Figure~\ref{fig-cloudy_hm} shows the ionization of the various species as a function of ionization parameter, which in turn is related to the density of the cloud (see Eq~\ref{eq-U}). The left panel shows models integrated until an \ion{H}{1} column density of 10$\rm^{14.5}~cm^{-2}$ is reached (matching our inferred values in the CGM of the starbursts). Changing this limit moves the plotted lines up and down such that their ratios remain the same. Hence, their ratios can be used as a robust matrix for estimating the ionization parameter. The plot in the right panel of Figure~\ref{fig-cloudy_hm} shows these ratios normalized to \ion{C}{4}. The strongest constraint on the ionization parameter comes from the lower limit on the ratio of the \ion{Si}{4} and \ion{C}{4} columns. This requires log(U) $>$ -1.5. We use a value for $\rm \Phi_{HM} = 2750~ photons~cm^{-2}~s^{-1}$ based on the most recent modeling of the UV background by \citet{haardt_madau12}, consistent with the value found earlier by \citet{giroux_shull97}. We then increase this flux by a factor of two to account for illumination of both sides of the photoionized slab. This yields an upper limit to the density of the cloud $n < 6 \times 10^{-6}$ cm~$^{-3}$. We find a minimum ionization correction to go from \ion{C}{4} to C is a factor of four. To produce the observed \ion{C}{4} column density would then require a cloud size 
 $>\rm 240~[Z_{\odot}/Z_{cloud}]$ kpc (where the scaling accounts for the  the cloud metallicity). For plausible (sub-solar) metallicity the implied sizes are significantly larger than the CGM. Moreover, photoionization by the metagalactic background would not naturally explain why strong \ion{C}{4} is much more common in the CGM of starbursts than in the control sample. We conclude that this model is implausible.

Next, we explore photoionization of the CGM caused by the radiation escaping from the starburst galaxies. Figure~\ref{fig-cloudy_sb} shows the distribution of various species in a CGM cloud in terms of the ionization parameter produced by a starburst. The starburst spectrum used for these runs is a model spectrum for a starburst with constant star formation for 100~Myrs as produced by Starburst99 \citep{starburst99}. Because the ionizing spectrum of a starburst is softer than that of the metagalactic background, the constraints on the ionization parameter are even more severe. The lower limit to the ratio of \ion{Si}{4} and \ion{C}{4} columns implies log(U) $>$ 0.2. We estimate $Q_*$ using the extinction corrected H$\alpha$ fluxes and the values are presented in Table~\ref{tbl-sfr}. Adopting the mean values for $Q_* \rm ~(4 \times 10^{53}~s^{-1}$) and $r$ (160 kpc) for the starburst sample, the required density is $n < 2 \times 10^{-6}~f_{esc}~\rm cm^{-3}$ and the implied cloud size is $>~\rm 130~[N_C/N_{CIV}] [Z_{\odot}/Z_{cloud}] f_{esc}^{-1}$ kpc. 
Observations of typical local starbursts yield an upper limit on the escape fraction of $f_{esc}=$0.01 or 1\% \citep[e.g.][]{grimes09}, so this scenario would yield cloud sizes even larger than in the case of the metagalactic background.

\subsection{Collisional Ionization \& Galactic Winds}

If the \ion{C}{4} does not trace photoionized gas, the alternative is that the gas is hot and collisionally ionized. This process can not be due to shocks in the CGM associated with a major galactic merger. The images of our starburst sample in Figure 2 show that they are mostly normal single galaxies. This is consistent with \citet{sanders_miradel96}, who find that only 12\% of low luminosity starbursts similar to our targets are merger driven. Since our \ion{C}{4} detection rate $\sim$80\%, we therefore conclude that any collisional ionization of the CGM is not primarily merger-driven.

Let us then explore a model in which the \ion{C}{4} traces gas clouds in the CGM that are shock-ionized by a starburst-driven wind that may or may not be escaping (galactic fountains). We first consider whether such outflows driven by starbursts with the ages and strengths corresponding to our sample would be able to affect the CGM at such large radii. The models described in \citet{wild10} applied to our sample yield average burst ages of 200 Myr, average burst masses of $\rm \sim 10^{9}~M_\odot$, average burst mass fractions of 11\%, average star-formation rates during the burst of $\sim \rm 7 M_{\odot}~yr^{-1}$.

The best-studied starburst wind is M82. Observations with Chandra \citep{strickland_heckman09} imply that the outflow speed of the wind fluid produced inside M82 is about 2000 \kms and the mass and kinetic energy outflow rates in this material are $\sim \rm 5~M_{\odot}~yr^{-1}$ and $\rm 7\times 10^{43}~erg~s^{-1}$ respectively. Our estimated burst parameters imply an average star formation rate about 70\% as high in our starbursts as in M82 so we will scale the M82 outflow rates down by a factor of two for their winds. These empirical estimates of the wind outflow rates are very similar to theoretical estimates for a starburst with a star-formation rate of $\rm 7~M_{\odot} yr^{-1}$ \citep[Starburst99; see][]{starburst99}.

The simplest model to consider is if the wind propagates freely outward at 2000 km/sec into the halo. The wind will then travel $\sim$200 kpc in 10$^8$ years (consistent with our results). More realistically, we can consider a wind that propagates into a spherically-symmetric volume-filling CGM, thereby creating an expanding wind-blown bubble. This general problem has been analyzed by \citet{dyson89} and \citet{koo_mckee92a} for power-law radial density profiles. Taking an initial radial profile for the CGM corresponding to an isothermal halo ($\rho \propto r^{-2})$, the radius of the expanding bubble is given by $r = 60 \dot{E}_{43}^{1/3} M_{CGM,10}^{-1/3}~t_8$ kpc. Here $\dot{E}$ is the rate of kinetic energy injected by the starburst  given in units of $\rm10^{43}~erg~s^{-1}$, the CGM mass out to a radius of 200~kpc is in units of $10^{10}~\rm M_{\odot}$ and the time since the starburst began is given in units of 10$^8$ years. 

The gas mass of the CGM is very uncertain, but is likely to be comparable to or a bit less than the galaxy stellar mass (e.g. Hummels et al. 2013). Since the average stellar mass of the starbursts is 2 $\times10^{10}$ M$_{\odot}$, the fiducial value for $M_{CGM,10} =$ 1 is reasonable. We take an average age of $\sim$ 200 Myr and an average kinetic energy injection rate of 5$\times 10^{42}$ erg sec$^{-1}$ for our starburst sample. This leads to an outer radius for the expanding wind-blown cavity in the halo of about 100 kpc. Given the highly idealized nature of our model and the uncertain mass of the CGM, we regard this simple estimate as only showing that it is plausible for a starburst-driven wind to affect the bulk of the CGM.

Based on the properties of our Milky Way and of QALS associated with other galaxies, we know that the real CGM contains clouds or clumps. Assuming a simple picture in which the CGM contains both clouds and a tenuous inter-cloud medium \citep[e.g.][]{kauffmann09}, the outflowing wind will then not only inflate a large bubble/cavity in the tenuous component, it will also shock-heat and accelerate the population of imbedded ambient CGM clouds. 

To produce substantial amounts of \ion{C}{4} in these clouds requires a post-shock temperature above 10$^5$ K and hence shock speeds in excess of 100~\kms ~ \citep[e.g.][]{shull_mckee79}. For a cloud shocked by a wind, the velocity of the cloud shock is related to the velocity of the wind by $v_{shock} = v_{wind} (\rho_{wind}/\rho_{cloud})^{1/2}$, 
where $\rho_{wind}$ and $\rho_{cloud}$ are the density in the wind and the pre-shocked cloud. Our adopted wind parameters give $v_{wind} =$ 2000 \kms and $\rho_{wind} = \rm 4 \times 10^{-31}~g~cm^{-3}$ at $r =$ 150 kpc. This implies that a cloud density of $\rho_{cloud} < \rm 1.8\times 10^{-28}~g~cm^{-3}$ ($n_{cloud} < \rm 1.1 \times 10^{-4} cm^{-3}$) would lead to $v_{shock} > 100$ \kms. 

Adopting the minimum corrections for ionization and metallicity discussed above, the observed \ion{C}{4} column densities imply minimum total hydrogen column densities of at least a few $\times 10^{18}$ cm$^{-2}$. Together with the cloud density above, the implied minimum cloud size would be $\sim$ 10 kpc. This is much smaller than the total path-length our line-of-sight traverses though the CGM (roughly given by the impact parameter). This size mismatch is consistent with the absorbing material having a small volume filling factor (i.e. being clumpy). The associated time for the shock to traverse the absorbing cloud (the “cloud crossing time”) would be greater than 10$^8$ years. Wind-shocked clouds can survive several cloud-crossing times before being hydrodyamically disrupted \citep[e.g.][]{pittard10}, consistent with the continued existence of the shocked clouds over a period of a few hundred Myr (the starburst duration).

These estimates can be compared to predictions for column densities behind shocks. The total column density of material cooling behind a radiative shock is independent of the gas density and is given by $N_{rad} = 10^{17.5} v_{100}^4$ cm$^{-2}$ \citep{draine_mckee93}.
Detailed shock models yield predicted \ion{C}{4} column densities of $\sim 10^{13} \rm cm^{-2}$
 for $v_{shock} \sim \rm 100 km~s^{-1}$ \citep{shull_mckee79}.
Since the observed column densities are about an order-of-magnitude larger than this, it seems likely that the \ion{C}{4} absorption we see arises from an ensemble of multiple shocks. 

The observed breadth of the \ion{C}{4} lines (FWHM $\sim$ 120 to 320 km s$^{-1}$) could be associated with wind-driven shocks. For a CGM cloud with a total gas column density of 10$^{19}$ cm$^{-2}$ located initially at rest at a radius of 150 kpc from the starburst, and then exposed to the ram pressure of our adopted wind, Equation 2 in \citet{heckman11} predicts the cloud terminal velocity would be about 200 km sec$^{-1}$.

Finally, we can consider the energetics of such a picture. If the \ion{C}{4} traces gas that is cooling radiatively, this cooling rate should not exceed the total heating rate available from the starburst wind. For the masses, densities, and temperatures discussed above the total CGM cooling rate traced by the \ion{C}{4} absorbers would be $\sim 3 \rm \times 10^{42}~erg~s^{-1}$ \citep{sutherland_dopita93}. This rate is independent of metallicity (a lower metallicity increases the total gas mass and thermal energy implied by the observed \ion{C}{4} column densities, but increases the radiative cooling time of the shocked gas by roughly the same amount). It is still a minimum rate since it assumes the gas is cooling from a temperature of $\rm 10^5$ K (the minimum needed to produce \ion{C}{4}). This cooling rate is similar to the kinetic energy outflow rate in the wind that we have estimated above based on scaling the M82 wind (5 $\rm \times 10^{42}~erg~s^{-1}$). It is possible then to provide the necessary heating, but a significant fraction of the wind kinetic energy is needed.

\subsection{Relationship to \ion{O}{6} Absorbers}

In an important paper, \citet{tumlinson11b} reported on results obtained with COS that probed the \ion{O}{6} doublet (1032,1038 \AA) along sightlines through the CGM of 42 galaxies at redshifts of 0.1 to 0.36 and with impact parameters ranging up to 150 kpc. They find a very high detection rate (90\%) of \ion{O}{6} in the normal star-forming galaxies (specific SFR $\sim 10^{-11}$ to 10$^{-9.5}$ yr$^{-1}$) and a much lower detection rate (33\%) among the passive galaxies (specific SFR $< 2 \times 10^{-12}$ yr$^{-1}$). At impact parameters similar to those we have probed ($\rho > 100$ kpc) the typical \ion{O}{6} column densities are $\sim 10^{14.5}$ cm$^{-2}$ in the CGM of the detected galaxies.

While both their results and ours link the presence of highly-ionized gas in the CGM to star formation in the central galaxy, there is an intriguing difference: we find that strong \ion{C}{4} absorbers are preferentially detected in the CGM of starbursts, but are {\it not} detected in normal star-forming galaxies. The weakness of the \ion{C}{4} absorption at impact parameters like these is also seen in the Chen et al. (2001a) sample of normal disk galaxies at $<z> \sim$ 0.4. 

Putting all these results together would imply that the ratio of \ion{O}{6} and \ion{C}{4} column densities in the CGM of normal star forming galaxies is greater than $\sim$ 10. This could imply the \ion{O}{6} in these galaxies traces exceedingly low density material that is photoionized by the metagalactic background. 
However, the Cloudy models discussed above would require densities less than $6 \times 10^{-7}$ cm$^{-3}$ in this case. The observed \ion{O}{6} column densities then lead to sizes greater than a Mpc (even for solar Oxygen abundances). This is not tenable for the CGM.
 
Alternatively, the large ratio of \ion{O}{6} to \ion{C}{4} could indicate collisionally ionized gas significantly hotter than $T = 10^5$ K. We have argued that the \ion{C}{4} in the starburst CGM arises in collisionally ionized gas as well. Our results that \ion{C}{4} is much stronger in the CGM of starbursts might suggest that the \ion{O}{6} lines would also be stronger in the CGM of starbursts than in normal star forming galaxies. Currently, the data to establish that result is lacking. Observations of \ion{O}{6} along site lines directly into starbursts similar to those in our sample show high column densities of $\rm \sim 10^{15}~cm^{-2}$ \citep{grimes09}. However it is not clear if this gas is actually located in the CGM (rather than close to the starburst). Future observations of \ion{O}{6} in the CGM of starbursts with HST/COS would be most instructive.

\section{Summary \& Implications \label{sec:conclusion}} 

The circum-galactic medium (CGM) comprises of gas located in the halo of a galaxy extending out to the virial radius. As such, it is the interface between a galaxy and the inter-galactic medium (IGM). The effect of winds driven by intense star formation on gas in the CGM may be crucial to the evolution of both galaxies and the IGM. For example, can such negative feedback shut down the delivery of fresh gas flowing through the CGM into a galaxy and thereby quench future star formation? If so, does this play a role in the on-going migration of galaxies from the blue star-forming population to the red passive one? Can metals and energy be transported through the CGM and ultimately be delivered to the IGM? Answering these questions requires detailed observations of galactic winds. The best-studied such winds are driven from starbursts in the nearby universe. While existing data demonstrate that such starburst-driven winds exist, they provide scant information on their properties beyond radii of a few tens of kpc and so do not probe the bulk of the CGM.

In order to probe the effect of starburst-driven winds on the CGM we conducted a COS UV spectroscopic study using background QSOs to probe the CGM (in absorption) around foreground galaxies. This study probed sight-lines out to radii of 200 kpc for 20 galaxies. We targeted starburst galaxies and control galaxies consisting of normal star forming and passive galaxies. We compared the properties of CGM in these three classes and found the following:

1. Most (83\%) of the target galaxies in our sample showed \Lya absorbers in the CGM. The dispersion in \Lya equivalent width was larger for passive galaxies, with some absorbers having values of 3~$\rm \AA$ and others with upper limits whose values were an order-of-magnitude lower. This is unlike the starburst sample where \Lya was detected in all cases but had lower strengths (0.4 to 1.1~$\rm\AA$).

2. Highly ionized gas traced by \ion{C}{4}, was predominantly found in the CGM of starburst galaxies. The \ion{C}{4} line was detected in 80\% of the starburst galaxies (4/5) compared to only 17\% (2/12) of the control galaxies. These lines are resolved (intrinsic FWMH $\sim$ 100 to 320 km s$^{-1}$ and are mildly saturated with implied \ion{C}{4} column densities of a few $\times 10^{14}$ cm$^{-2}$.

3. Multiple transitions from low-ionization species such as \ion{Si}{2} and \ion{C}{2} as well as intermediate-ionization transitions like \ion{Si}{3} and \ion{C}{3} were detected in the two passive galaxies where \ion{C}{4} was detected. This was not the case for the four starbursts with detected \ion{C}{4}. This implies that the CGM in the starbursts is more highly ionized than the CGM in the passive galaxies, even those where \ion{C}{4} is detected.
This inference is further supported by the result that the ratio of the equivalent widths of the \ion{C}{4} and \Lya lines is larger in the starburst galaxies than in either the passive or normal star forming galaxies.

4. It is very unlikely that the \ion{C}{4} absorbers in the starbursts are photoionized by either the metagalactic background or the stellar radiation from the starburst. CLOUDY models show that high ionization state of the gas requires cloud densities that are very small. To explain the observed column densities requires path-lengths through the absorbing gas that are significantly larger than the size of the CGM. 

5. We have shown that a starburst with properties like those we infer for our sample members (average burst ages of $\sim$200 Myr and average burst mass fractions of $\sim$10\% of the stellar mass) can plausibly drive an outflow that can affect the bulk of the CGM. In such a model the \ion{C}{4} would arise in pre-existing clumpy clouds or filaments in the CGM that are being shock-ionized and accelerated by the ram pressure of the wind. The ensemble of such clouds must have a large areal covering factor ($\sim 80$\%). 

6. Even for the most conservative assumptions about ionization and metallicity corrections, the \ion{C}{4} absorbing clouds would constitute a significant baryon repository when compared to the galaxy stellar mass (at least $\sim$10\% of the $\rm M_*$).

While our understanding of how starbursts affect the circum-galactic gas is far from complete, our new observations show for the first time that starburst-driven outflows in the nearby universe can have a major impact on physical conditions in the CGM. This might hold the key to understanding whether such winds may quench star formation by reducing the supply of infalling cool gas and help us to better understand how the intergalactic medium is chemically enriched and heated. 

\vspace{.5cm}
\acknowledgements 
We are thankful to Mike Shull for his in-depth comments and suggestions.
We also thank Jason Tumlinson, Andrew Fox, Chris Thom, Guinevere Kauffmann, Xavier Prochaska, Hsiao-Wen Chen, and Guangtun Zhu for useful discussions.
This work is based on observations with the NASA/ESA Hubble Space Telescope, which is operated by the Association of Universities for Research in Astronomy, Inc., under NASA contract NAS 5-26555. The data analysis was supported by grant HST GO 111728.1.

This project also made use of SDSS data. Funding for the SDSS and SDSS-II has been provided by the Alfred P. Sloan Foundation, the Participating Institutions, the National Science Foundation, the U.S. Department of Energy, the National Aeronautics and Space Administration, the Japanese Monbukagakusho, the Max Planck Society, and the Higher Education Funding Council for England.  The SDSS Web Site is http://www.sdss.org/. 
The SDSS is managed by the Astrophysical Research Consortium for the Participating Institutions. The Participating Institutions are the American Museum of Natural History, Astrophysical Institute Potsdam, University of Basel, University of Cambridge, Case Western Reserve University, University of Chicago, Drexel University, Fermilab, the Institute for Advanced Study, the Japan Participation Group, Johns Hopkins University, the Joint Institute for Nuclear Astrophysics, the Kavli Institute for Particle Astrophysics and Cosmology, the Korean Scientist Group, the Chinese Academy of Sciences (LAMOST), Los Alamos National Laboratory, the Max-Planck-Institute for Astronomy (MPIA), the Max-Planck-Institute for Astrophysics (MPA), New Mexico State University, Ohio State University, University of Pittsburgh, University of Portsmouth, Princeton University, the United States Naval Observatory, and the University of Washington.

{\it Facilities:}  \facility{Sloan ()} \facility{COS ()}

\bibliographystyle{apj}	        
\bibliography{myref_bibtex}		

\clearpage
 \begin{landscape}

\begin{deluxetable}{lrrrl rlrrr crrcr cccc}  
\tabletypesize{\scriptsize}
\tablecaption{Our QSO-Galaxy Sample with Information on the Foreground Galaxy and the Background QSO. \label{tbl-sample}}
\tablewidth{0pt}
\tablehead{
\colhead{Foreground Galaxy} & \colhead{RA} & \colhead{Dec} & \colhead{$z$} & \colhead{Class} &\colhead{PC1}   &\colhead{PC2}& \colhead{$\rm M_{*_{total}}^{a}$}  & \colhead{$\rm M_{*_{fiber}}^{a}$}   & \colhead{$\Theta^{b}$} & \colhead{QSO}  &\colhead{RA} & \colhead{Dec} & \colhead{$z$} & \colhead{$\rho$}\\
\colhead{} & \colhead{} & \colhead{} &\colhead{} &  \colhead{}  & \colhead{} & \colhead{} & \colhead{(Log~(M$_{\odot}))$}& \colhead{(Log~(M$_{\odot}))$} & \colhead{} & \colhead{} & \colhead{} & \colhead{} & \colhead{} & \colhead{(kpc)} }
\startdata
J075622.09+304329.0 &  119.092 &  30.725 &  0.0796 &  Normal &  -2.49$\pm$0.08
 &  -0.05$\pm$0.09 &   11.02 &  10.32  &         0 &  J075620.07+304535.4 &  
119.08 &  30.76 &  0.236 &  193.9\\
J084356.12+261855.3 &  130.984 &  26.315 &  0.1128 &  SB/PSB &  -5.00$\pm$0.08
 &  0.19$\pm$0.09 &   10.55 &  10.22  &        41 &  J084349.75+261910.7 &  
130.96 &  26.32 &  0.257 &  178.5\\
J085252.73+031320.4 &  133.220 &  3.222 &  0.0954 &  Passive &  -0.19$\pm$0.22
 &  0.61$\pm$0.24 &   10.92 &  9.99  &         0 &  J085259.22+031320.6 &  
133.25 &  3.22 &  0.297 &  171.9\\
J085254.99+030908.3 &  133.229 &  3.152 &  0.0346 &  Normal &  -1.96$\pm$0.09
 &  0.68$\pm$0.10 &   10.32 &  9.62  &        60 &  J085259.22+031320.6 &  
133.25 &  3.22 &  0.297 &  179.1\\
J085301.26+031425.1 &  133.255 &  3.240 &  0.1292 &  Passive &  -0.57$\pm$0.16
 &  -0.07$\pm$0.16 &   10.90 &  10.50  &        72 &  J085259.22+031320.6 &  
133.25 &  3.22 &  0.297 &  164.6\\
J092844.89+602545.7 &  142.187 &  60.429 &  0.1541 &  Passive &  -0.67$\pm$0.15
 &  0.51$\pm$0.15 &   10.95 &  10.57  &        11 &  J092837.98+602521.0 &  
142.16 &  60.42 &  0.295 &  152.1\\
J100801.20+500915.6 &  152.005 &  50.154 &  0.0523 &  Normal &  -2.66$\pm$0.15
 &  0.86$\pm$0.15 &   10.39 &  9.55  &         0 &  J100744.54+500746.6 &  
151.94 &  50.13 &  0.212 &  186.6\\
J101008.85+300252.5 &  152.537 &  30.048 &  0.0874 &  Passive &  -0.71$\pm$0.20
 &  0.77$\pm$0.22 &   10.41 &  10.04  &         0 &  J101000.68+300321.5 &  
152.50 &  30.06 &  0.256 &  179.8\\
J102846.43+391842.9 &  157.194 &  39.312 &  0.1135 &  SB/PSB &  -4.46$\pm$0.13
 &  0.58$\pm$0.15 &   10.50 &  9.94  &        80 &  J102847.00+391800.4 &  
157.20 &  39.30 &  0.473 &  88.7\\
J120314.43+480316.4 &  180.810 &  48.055 &  0.0629 &  Passive &  0.12$\pm$0.12
 &  0.04$\pm$0.12 &   10.78 &  10.20  &         7 &  J120329.84+480313.5 &  
180.87 &  48.05 &  0.817 &  187.4\\
J122115.76-020009.6 &  185.316 &  -2.003 &  0.0625 &  SB/PSB &  -5.06$\pm$0.08
 &  0.89$\pm$0.09 &   10.15 &  9.43  &        85 &  J122115.32-020253.3 &  
185.31 &  -2.05 &  0.785 &  197.2\\
J122534.26-025029.1 &  186.393 &  -2.841 &  0.0673 &  SB/PSB &  -5.73$\pm$0.04
 &  0.88$\pm$0.04 &   10.13 &  9.69  &         0 &  J122534.79-024757.2 &  
186.40 &  -2.80 &  0.195 &  196.1\\
J132107.48+295615.3 &  200.281 &  29.938 &  0.0723 &  SB/PSB &  -3.79$\pm$0.11
 &  0.16$\pm$0.12 &   10.48 &  9.97  &        33 &  J132059.41+295728.1 &  
200.25 &  29.96 &  0.206 &  175.5\\
J132150.89+033034.1 &  200.462 &  3.509 &  0.0816 &  Normal &  -2.70$\pm$0.09
 &  0.62$\pm$0.09 &   10.81 &  10.13  &        16 &  J132144.97+033055.7 &  
200.44 &  3.52 &  0.269 &  140.2\\
J140502.20+470525.9 &  211.259 &  47.091 &  0.1452 &  SB/PSB &  -5.02$\pm$0.13
 &  -0.19$\pm$0.14 &   10.43 &  10.08  &         0 &  J140505.77+470441.1 &  
211.27 &  47.08 &  1.240 &  146.9\\
J151136.53+402852.6 &  227.902 &  40.481 &  0.0945 &  Passive &  0.04$\pm$0.14
 &  0.35$\pm$0.14 &   10.67 &  10.22  &         0 &  J151140.33+402721.5 &  
227.92 &  40.46 &  0.558 &  177.0\\
J154527.12+484642.2 &  236.363 &  48.778 &  0.0752 &  Normal &  -2.37$\pm$0.19
 &  0.12$\pm$0.20 &   10.50 &  9.87  &        10 &  J154530.23+484608.9 &  
236.38 &  48.77 &  0.399 &  64.7\\
J161708.92+063822.2 &  244.287 &  6.640 &  0.1526 &  Passive &  0.26$\pm$0.11
 &  0.50$\pm$0.12 &   11.39 &  10.81  &         0 &  J161711.42+063833.4 &  
244.30 &  6.64 &  0.229 &  103.2\\
J161913.50+334146.8 &  244.806 &  33.696 &  0.1326 &  Normal &  -2.23$\pm$0.15
 &  0.24$\pm$0.16 &   11.04 &  10.50  &        12 &  J161916.54+334238.4 &  
244.82 &  33.71 &  0.471 &  150.9\\
J230842.91-091112.8 &  347.179 &  -9.187 &  0.1902 &  Normal &  -2.72$\pm$0.11
 &  0.19$\pm$0.11 &   11.09 &  10.69  &         2 &  J230845.60-091123.9 &  
347.19 &  -9.19 &  0.193 &  131.0\\
\enddata
\tablenotetext{a}{From the latest calculation by Jarle Brinchman using the method published by \citet{kauffmann03} and \citet{salim07}.}
\tablenotetext{b}{Orientation is the angle between the QSO sightline and the major axis of the galaxy. The major axis was identify using position angle from SDSS. These values were not accompanied by individual errors and therefore, we estimated a mean error of 10$^{\circ}$ in orientation .}
\end{deluxetable}

  \clearpage
  \end{landscape}

\begin{deluxetable}{ccccc ccccc crrcr cccc}  
\tabletypesize{\scriptsize}
\tablecaption{Table Presenting Star Formation Related Properties of Galaxies in Our Sample. \label{tbl-sfr}}
\tablewidth{0pt}
\tablehead{
\colhead{Foreground Galaxy} & \colhead{H$\alpha$ Flux$^a$} & \colhead{H$\alpha$ Luminosity$^a$} & \colhead{Q$_*$} & \colhead{Avg. sSFR$^{b}$}   & \colhead{sSFR$_{Fib}^{c}$} \\ 
\colhead{} & \colhead{($\rm erg~s^{-1}~cm^{-2}$)} & \colhead{($\rm erg~s^{-1}$)}& \colhead{($\rm s^{-1}$)} & \colhead{($\rm Log~(yr^{-1}$))} & \colhead{($\rm Log~(yr^{-1}$))}  
 } 
\startdata
J075622 &  $-$ &  $-$&$-$&$-$&-10.70\\
J084356 &  2.3$\times 10^{     -14}$ &  7.7$\times 10^{      41}$&
5.6$\times 10^{      53}$&-9.30&-9.30\\
J085252 &  7.7$\times 10^{     -17}$ &  1.8$\times 10^{      39}$&
1.3$\times 10^{      51}$&$-$&-10.65\\
J085254 &  1.1$\times 10^{     -15}$ &  3.0$\times 10^{      39}$&
2.2$\times 10^{      51}$&$-$&-10.85\\
J085301 &  3.4$\times 10^{     -16}$ &  1.5$\times 10^{      40}$&
1.1$\times 10^{      52}$&$-$&-11.55\\
J092844 &  $-$ &  $-$&$-$&$-$&-11.10\\
J100801 &  1.8$\times 10^{     -15}$ &  1.2$\times 10^{      40}$&
8.7$\times 10^{      51}$&$-$&-10.55\\
J101008 &  1.9$\times 10^{     -17}$ &  3.5$\times 10^{      38}$&
2.6$\times 10^{      50}$&$-$&-11.25\\
J102846 &  7.6$\times 10^{     -15}$ &  2.5$\times 10^{      41}$&
1.8$\times 10^{      53}$&-9.26&-9.80\\
J120314 &  2.9$\times 10^{     -15}$ &  2.8$\times 10^{      40}$&
2.0$\times 10^{      52}$&$-$&-11.55\\
J122115 &  1.5$\times 10^{     -14}$ &  1.4$\times 10^{      41}$&
1.0$\times 10^{      53}$&-8.76&-9.75\\
J122534 &  5.1$\times 10^{     -14}$ &  5.5$\times 10^{      41}$&
4.1$\times 10^{      53}$&-8.55&-9.40\\
J132107 &  2.7$\times 10^{     -14}$ &  3.5$\times 10^{      41}$&
2.5$\times 10^{      53}$&-9.73&-9.45\\
J132150 &  4.2$\times 10^{     -15}$ &  7.0$\times 10^{      40}$&
5.1$\times 10^{      52}$&$-$&-10.25\\
J140502 &  1.3$\times 10^{     -14}$ &  7.2$\times 10^{      41}$&
5.3$\times 10^{      53}$&-9.28&-9.05\\
J151136 &  $-$ &  $-$&$-$&$-$&-11.55\\
J154527 &  1.8$\times 10^{     -15}$ &  2.5$\times 10^{      40}$&
1.8$\times 10^{      52}$&$-$&-10.50\\
J161708 &  $-$ &  $-$&$-$&$-$&-11.60\\
J161913 &  3.3$\times 10^{     -15}$ &  1.5$\times 10^{      41}$&
1.1$\times 10^{      53}$&$-$&-10.10\\
J230842 &  3.5$\times 10^{     -15}$ &  3.6$\times 10^{      41}$&
2.6$\times 10^{      53}$&$-$&-10.20\\
\enddata
\tablenotetext{a}{Extinction corrected flux and luminosity values as derived from MPA-JHU catalog using prescription published by \citet{wild07} }
\tablenotetext{b}{Average specific SFR derived based on the age of the new stellar population and burst mass fraction as described by \citet{wild10}. }
\tablenotetext{c}{Based on SDSS fiber spectra from \citet{brinchmann04}. These were determined from H$\alpha$ and/or D$_n$(4000) depending on availability of H$\alpha$ and H$\beta$ measurements that are uncontaminated by AGN emission.}
\end{deluxetable}

\begin{deluxetable}{lccccccccccc}  
\tabletypesize{\scriptsize}
\tablecaption{\Lya and \ion{C}{4} Absorption Features Associated with Foreground Galaxies. \label{tbl-absorptiondata}}
\tablewidth{0pt}
\tablehead{
\colhead{Galaxy} & \colhead{W$\rm _{Ly\alpha_{rest}}$ $^{a,b}$} &   \colhead{FWHM$\rm _{Ly\alpha}^c$} & \colhead{Log~N$\rm _{Ly\alpha}$}  & \colhead{W$\rm _{CIV1548_{rest}}$}  &  \colhead{FWHM$\rm _{CIV1548}^c$} &  \colhead{W$\rm _{CIV1550_{rest}}$} & \colhead{Log~N$\rm _{CIV}$}  \\
\colhead{} & \colhead{$\rm (m\AA)$}   &   \colhead{(\kms)} &  \colhead{$\rm (Log~(cm^{-2}))$} &  \colhead{$\rm (m\AA)$}  &   \colhead{(\kms)} &  \colhead{$\rm (m\AA)$} &   \colhead{$\rm (Log~(cm^{-2}))$} }
\startdata
J075622 &  $<$     146$^b$&$-$&$-$&$-$&$-$&$<$     255&$-$\\
J084356 &       437$~\pm~$      43&$\rm  104.1^{+21.0}_{-17.5}$&14.11$~\pm~$0.05
&$<$     160&$-$&$<$     215&$-$\\
J085252 &       279$~\pm~$      23&$\rm <29.8^d$&$>$13.71$^d$&$<$     199&$-$&
$<$     256&$-$\\
J085254 &  $-$&$-$&$-$&$<$     121&$-$&$<$     114&$-$\\
J085301 &      1452$~\pm~$      47&$\rm  186.6^{+21.3}_{-19.1}$&$>$15.05$^e$&
     928$~\pm$      86$^{f}$&$\rm 162.0^{+56.0}_{-41.6}$&$-$&
14.59$~\pm~$0.06$^{f}$\\
J092844 &      2885$~\pm~$     132&$\rm  515.3^{+20.6}_{-19.8}$&$>$15.08$^e$&
     779$~\pm$     172&$\rm 129.0^{+41.4}_{-31.4}$&     505 $~\pm$     104&
14.53$~\pm~$0.08\\
J100801 &      1694$~\pm~$     168&$\rm  417.9^{+25.8}_{-24.3}$&14.69$~\pm~$0.02
&$<$     295&$-$&$-$&$-$\\
J101800 &       640$~\pm~$      84&$\rm  449.6^{+83.5}_{-70.4}$&14.13$~\pm~$0.05
&$<$     231&$-$&$<$     221&$-$\\
J102846 &      1129$~\pm~$      67&$\rm  182.8^{+24.2}_{-21.4}$&14.69$~\pm~$0.07
&    1224$~\pm$      72&$\rm 279.7^{+47.4}_{-40.5}$&     741 $~\pm$      92&
14.65$~\pm~$0.04\\
J120314 &      1815$~\pm~$     129&$\rm  363.4^{+26.0}_{-24.3}$&14.79$~\pm~$0.03
&$<$     261&$-$&$<$     279&$-$\\
J122115 &       497$~\pm~$      85&$\rm  143.0^{+93.3}_{-56.4}$&14.12$~\pm~$0.10
&     689$~\pm$      37&$\rm 152.5^{+42.2}_{-33.0}$&     409 $~\pm$      56&
14.38$~\pm~$0.05\\
J122534 &       898$~\pm~$      44&$\rm  199.2^{+21.7}_{-19.6}$&14.46$~\pm~$0.03
&     550$~\pm$     100&$\rm 115.8^{+115.8}_{-57.9}$&     332 $~\pm$      61&
14.31$~\pm~$0.09\\
J132107 &       275$~\pm~$      75&$\rm  398.6^{+662.0}_{-248.8}$&
13.73$~\pm~$0.20&$-$&$-$&$-$&$-$\\
J132150 &       613$~\pm~$      47&$\rm  161.3^{+16.4}_{-14.9}$&14.23$~\pm~$0.02
&$<$     223&$-$&$<$     234&$-$\\
J140502 &  $-$&$-$&$-$&     627$~\pm$     105&$\rm 319.8^{+107.6}_{-80.5}$&
     336 $~\pm$      57&14.25$~\pm~$0.08\\
J151136 &  $<$     127$^b$&$-$&$-$&$<$     248&$-$&$<$     289&$-$\\
J154527 &      1154$~\pm~$      59&$\rm  281.9^{+16.0}_{-15.2}$&14.53$~\pm~$0.02
&$<$     134&$-$&$<$     125&$-$\\
J161708 &  $<$     199$^b$&$-$&$-$&$<$     485&$-$&$<$     486&$-$\\
J161913 &      1847$~\pm~$      45&$\rm  310.0^{+8.0}_{-7.8}$&14.90$~\pm~$0.01&
$-$&$-$&$<$      94&$-$\\
J230842 &        66$~\pm~$      21&$\rm <12.5^d$&$>$13.09$^d$&$<$     154&$-$&
$-$&$-$\\
\enddata
\tablenotetext{a}{The error represented here takes into account both the statistical error as well as error in continuum fitting.}
\tablenotetext{b}{The limiting \Lya equivalent width represents the 3~$\sigma$ error in the equivalent width over 200 \kms i.e. $\pm$100~\kms in the rest frame of the foreground galaxy.  For \ion{C}{4} we estimate limiting equivalent width around regions of $\pm$100~\kms from the \Lya line center.}
\tablenotetext{c}{FWHM takes into account the instrumental line spread function.}
\tablenotetext{d}{~Lines are unresolved and the column densities presented here were derived assuming the transitions to be optically thin.}
\tablenotetext{e}{The lines are saturated and the quoted column density is a lower limit.}
\tablenotetext{f}{The data coverage of the line is partial as it lies at the edge of the grating. Consequently, there could be much larger systematic uncertainty in the measurement than estimated in the error.}
\end{deluxetable}

\begin{deluxetable}{lccrccccccc}
\tabletypesize{\scriptsize}
\tablecaption{Transitions Associated with Two strongest \Lya absorbers Associated with Galaxies J085301 and J092844. \label{tbl-J085301_J092844}}
\tablewidth{0pt}
\tablehead{
\colhead{Galaxy} & \colhead{Redshift} & \colhead{Transition} &\colhead{$\lambda_{rest}$} & \colhead{$\rm W_{rest}$}  &\colhead{FWHM$^b$}                    &\colhead{Log~N}  \\
\colhead{}           & \colhead{}              & \colhead{}                 & \colhead{$\rm (\AA)$}       & \colhead{$\rm (m\AA)$}   &\colhead{($\rm km~s^{-1}$)} &\colhead{($\rm Log~ cm^{-2}$)} }
\startdata
J085301 & 0.1292 &  \HI Ly$\alpha$    & 1215.67  & 1452   $\pm$  47   & $\rm187~^{+21}_{-17}$ & $>$15.05$^c$  \\
              &             &  \HI Ly$\gamma$ $^d$ &  972.54   & 1023  $\pm$  154   & $-$  & $-$   \\
              &             &   \ion{Si}{2}           & 1190.42   &  215   $\pm$   23 & \multirow{5}{*}{$\Big] \scriptsize \rm 82~^{+10}_{-9}$}  & \multirow{5}{*}{$\Big ]$\scriptsize $>$13.8$^e$} \\  
              &             &   \ion{Si}{2}           & 1193.29   &  345   $\pm$   23   \\
              &             &   \ion{Si}{2}           & 1260.42   &  524   $\pm$   47   \\
              &             &   \ion{Si}{2}           & 1304.37   &  165   $\pm$  36   \\
              &             &   \ion{Si}{2}           & 1526.71   &  311   $\pm$  63   \\
              &             &   \ion{Si}{3}           & 1206.50   &  790   $\pm$  46   & 171~$\rm ^{+21}_{+19}$& 13.81  $\pm$   0.03 \\
              &             &   \ion{Si}{4}           & 1393.76   &  423   $\pm$   37   & \multirow{2}{*}{$\Big] \rm 92^{+13}_{-11}$}  & \multirow{2}{*}{$\Big]$\scriptsize $>$13.7$^e$}  \\ 
              &             &   \ion{Si}{4}           & 1402.77   &  240   $\pm$   50   &    \\
              &             &   \ion{C}{2}            & 1334.53   &  633   $\pm$   54   & $\rm 120~^{+26}_{-22}$ & 14.68  $\pm$   0.05\\
              &             &   \ion{C}{3}  $^d$  &  977.02     & 1031$^c\pm$ 126  &  $-$    & $-$\\
              &             &   \ion{C}{4}            & 1548.20   &  928   $\pm$   86   &  $\rm 162~^{+56}_{-42}$    & 14.59  $\pm$   0.06\\      
              \\  
J092844 & 0.1541 &  \HI Ly$\alpha$    & 1215.67  & 2987   $\pm$  132   & $\rm 515~^{+21}_{-20}$ & $>$15.08$^c$ \\
              &             &  \HI Ly$\gamma^d$ &  972.54   & 1063   $\pm$  117    &  $-$  & $-$ \\           
              &             &   \ion{Si}{2}           & 1190.42   &  424   $\pm$   61 & \multirow{5}{*}{$\Big] \rm 73~^{+8}_{-7}$}   & \multirow{5}{*}{$\Big]$\scriptsize $>$14.13$^e$} \\
              &             &   \ion{Si}{2}           & 1193.29   &  526   $\pm$   80  \\
              &             &   \ion{Si}{2}           & 1260.42   &  647   $\pm$   86  \\
              &             &   \ion{Si}{2}           & 1304.37   &  382   $\pm$   63  \\
              &             &   \ion{Si}{2}           & 1526.71   &  578   $\pm$   95  \\              
              &             &   \ion{Si}{3}           & 1206.50   &  611   $\pm$   84   & $\rm 82~^{+42}_{-28}$  &  $>$13.45$^e$  \\ 
              &             &   \ion{Si}{4}           & 1393.76   &  738   $\pm$   77   & $\rm 243~^{+46}_{-39}$ &  14.03  $\pm$   0.05\\
              &             &   \ion{C}{2}            & 1334.53   &  777   $\pm$  68   & $\rm 106~^{+24}_{-20}$ &   $>$14.56$^e$  \\
              &             &   \ion{C}{3}$^d$    &  977.02   & 1063   $\pm$  118  &     $-$ &     $-$    \\
              &             &   \ion{C}{4}            & 1548.20   &  779   $\pm$  172  &  \multirow{2}{*}{$\Big ] \rm \scriptsize 126~^{+41}_{-31}$}  & \multirow{2}{*}{$\Big ]$\scriptsize 14.53  $\pm$   0.08}\\
              &             &   \ion{C}{4}            & 1550.77   &  505   $\pm$  104    
\enddata
\tablenotetext{a}{The error takes into account both the statistical error as well as error in continuum fitting.}
\tablenotetext{b}{The FWHMs were obtained after correcting for the appropriate line spread function of grating G140L.}
\tablenotetext{c}{Saturated lines.}
\tablenotetext{d}{The transition lies in the Segment B of the G140L grating and wavelength calibration is highly uncertain in this part of the spectrum. Therefore, we do not use these transition for fitting Voigt profile.}
\tablenotetext{e}{Lines are unresolved and the column densities presented here were derived assuming the transitions to be optically thin.}
\end{deluxetable}

\begin{deluxetable}{lcccccccccc}  
\tabletypesize{\scriptsize}
\tablecaption{Absorption-line Measurements from Stacks for the Three Sub-samples. \label{tbl-stack_measurements}}
\tablewidth{0pt}
\tablehead{
\colhead{Galaxy} & \colhead{Transition}&\colhead{$\lambda_{rest}$}& \colhead{$\rm W_{rest}^{a,b}$} & \colhead{FWHM $^c$}  &\colhead{Log~N$^d$}  \\
\colhead{} &\colhead{} & \colhead{$\rm (\AA)$} & \colhead{$\rm (m\AA)$} &\colhead{$\rm (km~s^{-1})$} &\colhead{($\rm Log~(cm^{-2})$)}  }
\startdata
SB$/$PSB &  \HI Ly$\alpha$        & 1215.67   & 590.4       $\pm$    30.0   &$\rm 171~^{+18}_{-16}$     &  $>$14.19$^e$  $\pm$    0.03   \\
              &   \ion{C}{2}                & 1334.53   &  $<$60.1                          &   $-$                                     &  $<$13.5                          \\
              &   \ion{C}{4}                & 1548.20   &  559.6       $\pm$   42.3   &$\rm 278~^{+40}_{-35}$  &  14.20 $\pm$    0.04   \\ 
              &   \ion{Si}{3}                & 1206.50   &  125.2       $\pm$   19.5  &$\rm  205~^{+87}_{-51}$  &  12.85    $\pm$    0.12  \\ 
              &   \ion{Si}{4}                & 1393.76   &  $<$44.0                         &    $-$                                     &  $<$12.71                     \\
\hline
\\
Control  &  \HI Ly$\alpha$         & 1215.67   & 881.3        $\pm$   21.5  & $\rm 331~^{+12}_{-12}$   & $>$14.32$^e$   $\pm$   0.01    \\
(Normal) &   \ion{C}{2}                & 1334.53   & $<$40.1                          &    $-$                                  & $<$13.32                                    \\
              &   \ion{C}{4}                & 1548.20   & $<$39.5                          &    $-$                                   & $<$13.00                           \\
              &   \ion{Si}{3}                & 1206.50   &  255.1   $\pm$  18.7      & $\rm 332~^{+62}_{-78}$  &   13.10     $\pm$   0.05  \\ 
              &   \ion{Si}{4}                & 1393.76   &  111.2  $\pm$   20.0      &  $\rm 146~^{+108}_{-62}$   &  13.12     $\pm$   0.10\\
\hline
\\
Control   &  \HI Ly$\alpha$         & 1215.67   &  946.2    $\pm$  31.7   & $\rm 425~^{+23}_{-22}$  &  $>$14.33$^e$        $\pm$  0.02    \\
(Passive) &   \ion{C}{2}                & 1334.53   &  213.5    $\pm$  31.7   & $\rm 383~^{+88}_{-65}$  &  14.06        $\pm$  0.08    \\  
              &   \ion{C}{4}                & 1548.20   &  487.5    $\pm$  56.0   & $\rm 282~^{+70}_{-57}$  &  14.13       $\pm$  0.06   \\  
              &   \ion{Si}{3}                & 1206.50   &  264.0    $\pm$  28.7   & $\rm 274~^{+83}_{-63}$ &  13.12       $\pm$   0.06   \\
              &   \ion{Si}{4}                & 1393.76   & $<$167.7                     &    $-$                                     & $<$13.30               
\enddata
\tablenotetext{a}{The lower limits were derived using data within $\pm$100\kms of the expected line center. }
\tablenotetext{b}{The error represented here takes into account both the statistical error as well as error in continuum fitting.}
\tablenotetext{c}{The FWHMs were obtained after correcting for the appropriate line spread function of grating G140L.}
\tablenotetext{d}{Column densities for the non detections were derived by assuming a Doppler parameter,$\rm b=60$\kms.}
\tablenotetext{e}{Some of the absorption features that were included for making the stacks were saturated. Therefore, the column density represented here is a lower limit.}
\end{deluxetable}

\end{document}